\documentclass[aps,twocolumn,prd,superscriptaddress]{revtex4-2}

\usepackage{dcolumn}
\usepackage{bm}
\usepackage{bbold}

\usepackage[utf8]{inputenc}
\usepackage{babel}

\usepackage{mathtools}
\usepackage{amsfonts}
\usepackage{mathrsfs}
\usepackage{bbm}
\usepackage{slashed}
\usepackage{tensor}

\usepackage{graphicx}
\usepackage{color}
\usepackage[dvipsnames]{xcolor}
\usepackage{array}
\usepackage[abs]{overpic}

\usepackage{placeins}
\usepackage{makecell}
\usepackage{caption}
\usepackage{subcaption}

\usepackage{xspace}
\usepackage{siunitx}
\usepackage{xfrac}
\usepackage{hyperref}
\usepackage[nameinlink]{cleveref}
\usepackage{appendix}

\usepackage{xifthen}
\usepackage{xcolor}
\hypersetup{
	colorlinks,
	linkcolor={red!75!black},
	citecolor={blue!75!black},
	urlcolor={blue!75!black}
}

\usepackage{booktabs}
\usepackage{multirow}

\newcolumntype{C}{>{$}c<{$}}
\AtBeginDocument{
	\heavyrulewidth=.08em
	\lightrulewidth=.05em
	\cmidrulewidth=.03em
	\belowrulesep=.65ex
	\belowbottomsep=0pt
	\aboverulesep=.4ex
	\abovetopsep=0pt
	\cmidrulesep=\doublerulesep
	\cmidrulekern=.5em
	\defaultaddspace=.5em
}

\newcommand{\sumint}{\int\hspace{-4.8mm}\sum}
\newcommand{\suminttext}{\int\hspace{-3.7mm}\sum}

\newcommand{\tinytext}[1]{\text{\tiny{#1}}}
\renewcommand{\Re}{\operatorname{Re}}

\def\num#1{\numx#1}\def\numx#1e#2{{#1}\mathrm{e}{#2}}

\graphicspath{{./figures/}{figures/QMY}{figures/QM}}

\renewcommand{\Re}{\operatorname{Re}}

\newcommand{\gettitle}{Towards quantitative precision for QCD at large densities}

\newcommand{\getHeidelbergAffiliation}{\affiliation{Institut f{\"u}r Theoretische Physik, Universit{\"a}t Heidelberg, Philosophenweg 16, 69120 Heidelberg, Germany}}
\newcommand{\getEMMIAffiliation}{\affiliation{ExtreMe Matter Institute EMMI, GSI, Planckstr. 1, 64291 Darmstadt, Germany}}
\newcommand{\getDarmstadtAffiliation}{\affiliation{Institut für Kernphysik, Technische Universität Darmstadt, D-64289 Darmstadt, Germany}}

\hypersetup{
	pdftitle={\gettitle},
	pdfauthor={},
	pdfkeywords={discontinuous galerkin}
	{functional renormalization group} {effective potential}
	{phase transition} {numerical methods} {phase structure},
	bookmarksopen=true,
	bookmarksopenlevel=2,
	bookmarksnumbered=true
}

\begin{document}

\title{\gettitle}

\author{Friederike Ihssen}\getHeidelbergAffiliation
\author{Jan M. Pawlowski}\getHeidelbergAffiliation\getEMMIAffiliation
\author{Franz R. Sattler}\thanks{sattler@thphys.uni-heidelberg.de}\getHeidelbergAffiliation
\author{Nicolas Wink}\getDarmstadtAffiliation

\begin{abstract}
	QCD at large density reveals a rich phase structure, ranging from a potential critical end point and inhomogeneous phases or moat regimes to color superconducting ones with competing order effects. Resolving this region in the phase diagram of QCD with functional approaches requires a great deal of quantitative reliability, already for a qualitative access. In the present work, we systematically extend the functional renormalisation group approach to low energy QCD by setting up a fully self-consistent approximation scheme in a low energy effective quark-meson theory. In this approximation, all pointlike multi-scattering events of the mesonic pion and the sigma mode are taken into account in terms of an effective potential as well as all higher quark--antiquark--mesonic scattering orders. As a first application we compute the phase structure of QCD including its low temperature - large chemical potential part. The quantitative reliability of the approximation and systematic extensions are also discussed.     	
\end{abstract}

\maketitle

\section{Introduction}
\label{sec:intr}

Understanding the rich phase structure of quantum chromodynamics has been a long-standing challenge. Amongst the many phenomena under investigation are the potential existence, location and experimental signatures of a critical endpoint (CEP) and of inhomogeneous phases or moat regimes, as well as QCD phases like color superconductivity at low temperatures and high densities including the highly demanding resolution of competing order effects.

From recent calculations with the functional renormalisation group (fRG) \cite{Fu:2019hdw} and Dyson-Schwinger equations \cite{Gao:2020fbl, Gao:2020qsj, Gunkel:2021oya, Bernhardt:2022mnx}, a coherent understanding of the phase structure up to a baryon chemical potential over temperature ratio of $\mu_\tinytext{B} / T \lesssim 4$ has emerged, for recent reviews see \cite{Fischer:2018sdj, Dupuis:2020fhh, Fu:2022gou}. Beyond this regime, further progress is impeded by the emergence of competing order effects in the multi-quark sector and the emergence of potential first-order phase transitions. While the former is directly related to possible new phases in the phase structure and necessitates the use of more sophisticated truncations in functional approaches, the latter is related to the formation of discontinuities in field space and its resolution asks for advanced numerical methods.

In this paper, we improve on existing state-of-the-art truncations of the mesonic sector of QCD. This is done by considering a low energy effective theory of QCD, the quark-meson (QM) model, within the self-consistent variant of the local potential approximation (LPA) set up in \cite{Pawlowski:2014zaa}: We consider all quark--antiquark--$n$-meson scattering processes for arbitrary $n$, in addition to all $n$-meson scatterings. After integrating out the quarks, it is the combination of these scatterings that contribute to the effective mesonic potential.
We call this approximation \textit{self-consistent} LPA (scLPA).  Accordingly, dropping the quark--antiquark--$n$-meson scatterings potentially leaves out a large part of the local scattering processes in the potential and is not self-consistent. 
For related work in scalar-Yukawa systems see \cite{Jakovac:2015kka, Gies:2017zwf, Held:2018cxd, Fejos:2020lli}, for respective UV-fixed point analyses see \cite{Zanusso:2009bs, Vacca:2015nta}, for various results on the QCD phase structure at finite density with the QM model see e.g.~\cite{Schaefer:2004en, Kamikado:2012cp, Tripolt:2013zfa, Mitter:2013fxa, Pawlowski:2014zaa, Jiang:2015xqz, Zhang:2017icm, Tripolt:2017zgc, Resch:2017vjs, Yin:2019ebz, Otto:2019zjy, CamaraPereira:2020xla, Otto:2020hoz, Otto:2022jzl}, for the embedding in QCD, for extensions including confinement and the relations to other low energy effective theories  see the recent reviews \cite{Dupuis:2020fhh, Fu:2022gou}.

As the balance between purely mesonic scatterings and quark--antiquark--meson scatterings is successively changed with increasing finite chemical potential, the use of such an approximation becomes increasingly more important for the reliability of results. At temperatures and chemical potentials in the regime of the potential critical end point, inhomogeneous phases or moat regimes and color superconducting phases, even the qualitative resolution of the phase structure within this advanced approximation requires computational progress in two directions:\\[-2ex]

Firstly, the stability and quantitative precision of the field space discretisation has to be improved. This allows for a full resolution of $n$-meson scatterings for arbitrary $n$ in the quark-meson model at high densities for the first time. This is done using an extension of the Discontinuous Galerkin field space discretisation investigated already in \cite{Ihssen:2022xkr, Grossi:2019urj, Grossi:2021ksl}.\\[-2ex]

Secondly, we investigate the scale-dependence of the hadronisation process of the scattering of a quark--antiquark pair into the pion and the sigma mode.
We find that the momentum transfer between quarks and mesons can be improved by using different scale-dependences for both particle species. This is implemented by introducing a relative shift in the cutoff-scales. Within self-consistent LPA, this adjustment of the RG-flow is necessary for convergence and we find a stable regime for an intermediate range of the shift parameter.\\[-2ex]

Together, these improvements allow us to access the low-density regime within this low-energy effective theory self-consistently for the first time.

This paper is structured as follows:
In \Cref{sec:LE-QCD} we introduce the quark-meson model within the self-consistent local potential approximation.
We also discuss there the relations between quark and composite (i.e.~bosonic) cutoff scales and possible criteria for an optimal momentum transfer.
\Cref{sec:LDG} introduces the Local Discontinuous Galerkin scheme, which is used for the numerical computation of the full mesonic field dependences. Finally, we present and discuss our results in \Cref{sec:results}.

\section{Low energy QCD}
\label{sec:LE-QCD}

Low energy effective theories (LEFTs) and their respective effective actions $\Gamma_\textrm{LEFT}$ are obtained from QCD when the fundamental high energy degrees of freedom, here those of quarks and gluons, are integrated out above an (infrared) momentum scale $k^2$. In most cases, the glue fluctuations are fully integrated out, while the fermions are only integrated out down to the ultraviolet cutoff scale $k = \Lambda_\textrm{UV}$ of the LEFT at hand. This leads us to
\begin{align}
	\Gamma_{\textrm{\tiny{LEFT}},k}[\bar q, q]=\Gamma_{\textrm{\tiny{QCD}},k}[\bar q, q , \langle A\rangle ]\,,
	\label{eq:GammaLEFT}
\end{align}
where $q,\bar q$ are the quark and antiquark fields respectively and $\langle A\rangle $ is the solution of the gluonic equation of motion (EoM). We have 
\begin{align}
\left. \frac{\delta \Gamma_{\textrm{\tiny{QCD}},k}}{\delta A}\right|_{A=\langle A\rangle }=0\,, 
\label{eq:EoMA}
\end{align}
where the solution to the equation of motion $\langle A\rangle$, i.e. the expectation value, depends on the other fields of the effective action. 

\subsection{Emergent composites} 
\label{sec:EmergentComposites} 

In the functional renormalisation group (fRG) approach to QCD in a covariant gauge, this task is naturally performed by integrating down QCD from large infrared cutoff scales $k \ll 10$\,GeV to infrared cutoff scales of $k= \Lambda_\textrm{UV}\approx 1$\,GeV. At about this scale the gluon degrees of freedom decouple due to the gluon mass gap of about 1\,GeV. While the glue background is still present, gluon off-shell fluctuations decouple. Below this scale, full QCD is well-described by an effective action as in \labelcref{eq:GammaLEFT} with dynamical quarks, if the degrees of freedom of the matter sector are sufficiently well captured in the effective action. For a detailed discussion of this decoupling in the phase structure of QCD see \cite{Fu:2019hdw}, for recent reviews see \cite{Dupuis:2020fhh, Fu:2022gou}. 

After the gluons are integrated out, the remaining infrared regularisation is implemented within the fRG approach by a change of the kinetic term in the action that suppresses the propagation of either four-momentum modes $p$ with $p^2\lesssim k^2$ or spatial momenta $\boldsymbol{p}$ with $\boldsymbol{p}^2\lesssim k^2$. 

However, such an effective action and its flow have to incorporate the effects of any resonant interaction channels of the four-quark scattering vertex. Foremost this is the scalar-pseudoscalar channel, and the full four-quark term in the effective action is given by 
\begin{align} \nonumber 
\Gamma_{4q,k}=&\, -\int \prod_{i=1}^4  \left(
 \frac{d^{4} p_i}{(2\pi)^{4}} \right)\, (4 \pi)^4 \delta(p_1+p_2+p_3+ p_4) \\[1ex]
&\, \hspace{.6cm}\times \lambda_{q,k}(p_1,...,p_4) (\bar{q}\tau_0 q + \bar{q}\boldsymbol{\tau} q)^2+\cdots \,,
\label{eq:G4q}
\end{align}
with 
\begin{align}
\tau =(\tau_0, {\boldsymbol{\tau}}) = \bigg(\frac{\mathbb{1}}{\sqrt{2 N_f}},\, i \gamma_5\,\frac{\boldsymbol{\sigma}}{2} \bigg)\,,
\label{eq:tau+boldp}
\end{align}
where $\boldsymbol{\sigma}$ are the Pauli-matrices, mixing the quarks in flavor-space.
All quark and antiquark fields in \labelcref{eq:G4q} carry different momenta $p_i$ with $i=1,...,4$. The dots refer to further terms in a Fierz-complete set-up with momentum-independent tensor structures. In QCD, the dressing $\lambda_q$ of the four-quark scattering vertex is resonant ($t$-channel), and in the chiral limit a pole at $p=0$ emerges at the chiral symmetry breaking scale $k_\chi$ with 
\begin{align}
	k<k_\chi: \quad \lim_{ {\boldsymbol{p}}\to 0 } \frac{1}{\lambda^{(\pi)}_{q,k}({\boldsymbol{p}}) }\to 0\,, 
	\label{eq:kchi}
\end{align}
where the superscript ${}^{(\pi)}$ stands for the projection on the pseudoscalar part of the four-quark interaction \labelcref{eq:G4q}. \Cref{eq:kchi} signals the presence of a massless mode in the system. This leaves us with a computationally challenging situation: for cutoff scales $k\lesssim k_\chi$ the system describes the propagation of massless modes in the four-quark vertex, while the propagation of the quark itself is suppressed by the regulator below the momentum scale $k$. While such a system can be implemented in functional approaches, its description requires an advanced approximation of the effective action in order to accommodate the momentum transfer in the diagrams between momentum scales $p^2 \propto 0$ and $p^2 \propto k^2$. For a respective discussion see \cite{Fu:2022uow}.

This challenge can be resolved if the resonant channel is also infrared regularised with a suitable infrared cutoff. In order to describe such a cutoff procedure within the standard fRG, we use dynamical hadronisation with emergent composites. It has been applied in the $\pi-\sigma$ channel in \cite{Gies:2002hq, Braun:2009ewx, Mitter:2014wpa, Braun:2014ata, Cyrol:2017ewj, Fu:2019hdw}, also including the vector channel in \cite{Rennecke:2015eba}, for further developments towards the inclusion of all general hadrons see \cite{Fukushima:2021ctq}. Here, we briefly describe the ingredients relevant for the present work. 

To begin with, for momenta above the chiral symmetry breaking scale $k_\chi$, where the resonance is formed, the coupling of the scalar-pseudoscalar channel can be parametrized by
\begin{align}
k>k_\chi: \quad  \lambda_{q,k}(p) = \frac{h_k^2} {Z_{\phi, k}(p^2)( p^2 +m_{\phi,k}^2)}\,, 
\label{eq:ResonantChannel}
\end{align}
with the $t$-channel momentum $p$. Here, we fully describe the momentum dependence in the $t$-channel of the four-quark coupling in terms of the wavefunction renormalization $Z_{\phi}$ and mass $m_{\phi}$ of the emergent composite. $h$ is later introduced as Yukawa coupling between quarks, anti-quarks and the composite.

At a given momentum scale, the resonant $t$-channel can now be incorporated with a momentum-dependent Hubbard-Stratonovich transformation,  
\begin{align}\nonumber 
& \int_x m_q \bar q q -\int_p  \lambda_q(p) \Bigl[(\bar{q}\tau_0 q + \bar{q}\boldsymbol{\tau} q)^2\Bigr](p) = \int_p h_k\,\Biggl[ \bar{q}\, \tau\phi\, q\\[1ex] 
& +\frac12 Z_{\phi,k}(p^2) \phi(-p) \left( p^2 +m_{\phi,k}^2 \right) \phi(p)\Biggr]- c_\sigma \int_x \sigma \,,
\label{eq:HS}
\end{align}
where we have dropped a field-independent term. The matrix-valued vector $\tau$ is defined in \labelcref{eq:tau+boldp} and $\phi$ is evaluated in the solution $\langle \phi\rangle$ of the quadratic equations of motion of the emergent scalar--pseudo-scalar fields, 
\begin{align}
	\phi = (\sigma, \boldsymbol{\pi})\,,\quad \boldsymbol{\pi}=(\pi_1,...,\pi_{N_f^2-1})\,, \quad 	\rho = \frac{\sigma^2 + \boldsymbol{\pi}^2}{2}\,.
	\label{eq:PiSigma}
\end{align}
\Cref{eq:HS} also includes the current quark mass term $\int_x m_q \bar q q$ on the right hand side in the isospin-symmetric approximation $m_u=m_d=m_q$. This approximation will be used throughout this work. We also used the abbreviations 
\begin{align}
\int_x =\int d^4 x\,,\qquad  \int_p =\int \frac{d^4 p}{(2 \pi)^4}\,. 
\end{align}
Using \Cref{eq:PiSigma} the solution of the EoM of $\phi$ on the right hand side of \labelcref{eq:HS} is obtained as 
\begin{align}
	\langle \phi\rangle(p) = \frac{h_k} {Z_{\phi, k}(p^2)( p^2 +m_{\phi,k}^2)}\,\Bigl(c_\sigma (2\pi)^4 \delta(p)-\bar q \tau q(p) \Bigr)\,, 
	\label{eq:EoMphi}
\end{align}
where the term proportional to $c_\sigma$ is a shift in the zero-momentum mode $\langle \int_x \sigma\rangle$. With \labelcref{eq:ResonantChannel} and the choice 
\begin{align}
	c_\sigma = 2 m_q\, \frac{Z_\phi(0) m_\phi^2}{h_k(0)}\,, 
	\label{eq:csigmaPhys}
\end{align} 
the right hand side of \labelcref{eq:HS} agrees with the left hand side up to a field-independent term. 

In the remainder of this work we use the derivative expansion to the lowest order. This entails the limit $p\to~0$ in the above relations, and the reduction to constant solutions of the mesonic equations of motion. 

The field $\boldsymbol{\pi}$ carries the same quantum numbers as the pseudo-scalar pion, while $\sigma$ is the $\sigma$-mode which overlaps with the $\sigma$-meson but also with the $f_0(500)$. The pole masses of both fields are given by the lowest lying hadron pole with these quantum numbers, whereas the field $\rho$ is the radial meson field entering the classical and effective potential of the LEFT. 

While the above transformation is done at a given scale $k$, dynamical hadronisation allows us to perform the respective steps continuously during the flow, which consequently eliminates the $t$-channel of the vertex $\lambda_{q,k}(p_1,...,p_4)$, keeping only a remnant of the vertex; see in particular \cite{Mitter:2014wpa, Cyrol:2017ewj, Fukushima:2021ctq} for detailed discussions and developments of the formalism.

This leads us to the representation of the low energy effective action of QCD in terms of a LEFT with only quark and meson degrees of freedom. In the present work we consider an approximation of this effective action which includes the momentum-dependent ($t$-channel) approximation of the four-quark vertex in terms of the exchange of emergent scalar-pseudoscalar fields $\phi$ as defined in \labelcref{eq:PiSigma} which carry the quantum numbers of the $\sigma$-mode and the pions.
Moreover, we include an effective potential $V( \rho)$ and hence all orders of self-scatterings of the resonant scalar-pseudoscalar channel. Note also that a part of the multi-scatterings in this channel is carried by the quark--antiquark--meson scatterings and hence requires a field-dependent Yukawa coupling $h(\rho)$ which includes all orders of these scatterings. 

In conclusion, the low energy physics of the resonant scalar-pseudoscalar channel is self-consistently taken into account by the effective action 
\begin{align}\nonumber 
	\Gamma_k[\bar q, q, \phi] = &\, \int_0^{1/T}d \tau \int d^3 x\,\Biggl[
	\bar{q}(\slashed{\partial} - \mu_q\gamma_0)q
	+\frac{1}{2}(\partial_\mu\phi)^2 \\[1ex]
	&\hspace{-.2cm}+ h_k(\rho) \bar{q}(\tau_0 \sigma + \boldsymbol{\tau}\,\boldsymbol{\pi})q+V_k(\rho)-c_\sigma \sigma
	\Biggr]\,. 
	\label{eq:QMAction}
\end{align}
In \labelcref{eq:QMAction} we have already introduced a quark chemical potential $\mu_q$ and finite temperature $T\neq 0$. \Cref{eq:QMAction} makes explicit that the explicit chiral symmetry breaking can be rephrased as a source term for the zero mode $\int_x \sigma$ which accordingly does not take part in the dynamics. Consequently, the full dynamics of the theory is governed by that in the chiral limit. 

The effective action \labelcref{eq:QMAction} is augmented with infrared regularisations for both the quarks and the emergent mesonic degrees of freedom with 
\begin{align} \nonumber 
	\Delta S_k[q,\bar q,\phi] =&\, \sumint_p\Biggl[\bar q(-p) R_q(p) q(p) \\[1ex] 
	& \hspace{2cm}+ \frac12 \phi(-p) R_\phi(p) \phi(p)\Biggr]\,. 
\label{eq:DeltaS}
\end{align} 
At finite temperature the four-momentum integration turns into $\suminttext_p$, which comprises a spatial momentum integral and a thermal Matsubara sum. The regulator functions $R_q(p^2)$ and $R_\phi(p^2)$ suppress infrared momentum fluctuations below the cutoff scale $k$ and decay sufficiently fast for momenta larger than $k$. The choices $R_q, R_\phi$ used in the present work are detailed in \Cref{app:thfkt}.

The fRG setup is supplemented by the equation of motion of the $\sigma$-mode. Its solution at $k=0$ determines the physical value $\langle\sigma\rangle$ of the order parameter in dependence of the explicit symmetry parameter $c_\sigma$ in \labelcref{eq:csigmaPhys}. The EoM is defined as the generalisation of \labelcref{eq:HS,eq:EoMphi}, 
\begin{align}
	\left. \frac{\delta \Gamma_k}{\delta \sigma}\right|_{\sigma=\langle \sigma\rangle } = 0\,,
	\label{eq:EoMphiGen}
\end{align}
evaluated at $q,\bar q=0$ and $\boldsymbol{\pi}=0$. As already discussed above, the dynamics of the theory is that of the chiral limit even in the presence of explicit chiral symmetry breaking with $c_\sigma \neq 0$. Hence, the effective potential and other coupling parameters such as the field-dependent Yukawa coupling will show non-analyticities at the solution of the equation of motion in the chiral limit, 
\begin{align}
	\langle\sigma\rangle_\chi = \lim_{c_\sigma\to 0} \langle\sigma\rangle\,. 
	\label{eq:chiral_order}
\end{align}
Below, we will discuss the reliability of this approximation as well as the potential impact of missing terms: first of all, we have dropped the momentum dependence of the quark propagator and that of the meson field, for a respective analysis see \cite{Helmboldt:2014iya}. Furthermore, we have implicitly restricted the momentum dependence of the four-quark scattering vertex to the $t$-channel of the scalar-pseudoscalar tensor structure. From investigations in full functional QCD we know that we do not have to consider the full momentum-dependence of the propagators for chemical potentials $\mu_\tinytext{B} / T \lesssim 3$ as they are well covered by scale-dependent wave functions.
Moreover, the analysis of functional QCD in vacuum in \cite{Mitter:2014wpa, Cyrol:2017ewj} has revealed that the $t$-channel and symmetric-point approximation of the four-quark scattering vertex are sufficient for regulator-independent results. This holds true as long as the momentum transfers in the systems are not deformed too much. 

However, if aiming for quantitative precision at large chemical potentials, one has to take into account the increasing difference between integrating out momentum fluctuations about the Fermi surface (quarks and antiquarks) and integrating out momentum fluctuations about $p=0$. Therefore, a detailed analysis of the required momentum dependences of propagators and vertices is needed in this regime of the model.

\subsection{Physical mass gaps and mesonic and quark cutoff scales} 
\label{sec:PhysGap+Cutoff}

The present low-energy approximation of \labelcref{eq:QMAction} lacks the momentum dependences that are required for quantitative reliability, in particular at large chemical potentials. A necessary condition for quantitative reliability is the regulator independence of the results. At finite temperature and vanishing chemical potential the quark-meson model in the present approximation shows a subleading regulator dependence for relevant observables, such as the critical temperature, and hence the condition is satisfied. 
Note that this regulator independence is deeply rooted in the procedure of adjusting the model parameters in the vacuum at the ultraviolet scale $k=~\Lambda$. This leads to a compensation of some of the regulator dependences by the regulator dependence of the initial conditions required for the correct vacuum physics.
However, at finite chemical potential, the balance between quark and mesonic quantum fluctuations is shifted and the regulator dependence is rising. In particular, some regulator choices may even destabilise the model in specific regimes of temperature and chemical potential. 

This situation calls for optimised regulator choices that minimise the momentum transfer between the different modes during the flow, as large transfers cannot be accommodated within the approximation \labelcref{eq:QMAction}, see \cite{Pawlowski:2005xe, Pawlowski:2015mlf}. 
The quark and mesonic regulators $R_q,\,R_\phi$ in \labelcref{eq:DeltaS} carry two potentially different infrared cutoff scales 
\begin{align} 
	\textrm{quarks:} \ k_q=k\,,\qquad \textrm{mesons:} \ k_\phi\,.
\end{align}
The respective propagators in the quark-meson model $G_\phi(p)$, $G_q(p)$ are gapped with 
\begin{align}\nonumber 
	m_{\phi,\textrm{gap}}^2 =&\, \min_{p,\sigma>\langle \sigma\rangle_\chi } \textrm{spec}\, G_\phi^{-1}(\phi,p)\,,\\[1ex] 	m_{q,\textrm{gap}}^2 =&\, \min_{p,\sigma>\langle \sigma\rangle_\chi }  \textrm{spec}\, G_q^{-2}(\phi,p)\,,
\label{eq:mgap}
\end{align}
evaluated at $q,\bar q=0$ and $\boldsymbol{\pi}=0$. The restriction $\sigma>\langle \sigma\rangle_\chi$ in \labelcref{eq:mgap} with $\langle \sigma\rangle_\chi$ defined in \labelcref{eq:chiral_order} is introduced for avoiding the unphysical regime of the theory including non-analyticities in the definition of the effective mass gap for $k_{q,\phi}\neq 0$. 

\Cref{eq:mgap} defines the \textit{physical} mass gap or infrared cutoff scale in the quark and meson sectors with the minimal spectral values of the inverse propagators. In the broken phase the mesonic sector has two mass gaps, that of the $\sigma$-mode and that of the pion. 

\Cref{eq:mgap} entails, that the physical cutoff $m_\textrm{gap}$ of the different fields may differ substantially from the running cutoff scale $k$. Accordingly, by lowering $k$ one integrates out momentum fluctuations at the physical cutoff scale $p^2\approx m_\textrm{gap}^2$ and not at $p^2\approx k^2$, even though the momentum range of the loop integrations is cut off by the latter. Consequently, a large difference between the different physical cutoffs asks for a significantly better momentum resolution of the effective action in order to capture the same physics as with flows with the same physical cutoff scale. This has been investigated in the context of functional optimisation in \cite{Pawlowski:2005xe}, for applications see also \cite{Pawlowski:2015mlf, Diehl:2007xz, BuschSchaefer2023}.  

It follows from this analysis that fermionic fluctuations are best integrated out in terms of their distance to the Fermi surface, while bosonic fluctuations (of bosons with vanishing fermion number such as the mesons) are best integrated out in terms of their distance to their curvature mass. In conclusion, the quantitative access to these systems with sizeable fluctuations in both types of fields necessitates a non-trivial momentum dependence of both vertices and propagators, or at least a fully optimised infrared regularisation. In QCD this requirement certainly holds true for the color-superconducting regime at large chemical potentials. It may also play a role for the access to inhomogeneous phases or a moat regime as well as for the liquid-gas transition at low temperatures and large chemical potentials ($\mu_B \approx 936$\,MeV). 

In any case, the optimal choice $k_\phi(k_q)$ is related to 
\begin{align}
m_{\phi,\textrm{gap}}^2 (k_q,k_\phi)\approx m_{q,\textrm{gap}}^2 (k_q,k_\phi)\,,
\label{eq:OptimalChoice} 
\end{align}
for cutoff scales that exceed the physical scales of the system. At lower physical cutoff scales the quarks and the $\sigma$-mode decouple at about 200\,MeV and only the pion remains. \Cref{eq:OptimalChoice} can be readily implemented in an adaptive procedure. In the present work we only use the variation of the relative cutoff scale for a systematic error estimate of the approximation. This is done using
\begin{align}
	 k_q =k\,, \qquad k_\phi &= \Lambda \left( \frac{k}{\Lambda}\right)^\gamma\,, \quad \gamma\in (\gamma_\textrm{min}\,,\, \gamma_\textrm{max})\,, 
\label{eq:kqkphi}
\end{align}
with the UV cutoff scale for the LEFT of $\Lambda\approx 650$\,MeV. With \labelcref{eq:kqkphi} both cutoff scales agree in the UV and, in the absence of approximations, a variation of the cutoff scales does not affect the initial condition for the effective action. Moreover, for $\gamma>1$ the bosonic cutoff is lagging behind and the symmetry breaking scale $k_\chi$ can be delayed after the decoupling scale of the quark. Let us now discuss the limits $\gamma\to 1$ and $\gamma\to0$: \\[-2ex]

For $\gamma\to 1$ we obtain identical cutoff scales $k = k_q = k_\phi$ and hence in the broken phase with $k\leq k_\chi$ we have 
\begin{align}
	m_{\phi,\textrm{gap}}^2 \approx 0 \ll m_{q,\textrm{gap}}^2\,. 
	\label{eq:WorstChoice}
\end{align}
These very light modes in a given system we call \textit{quasi-massless}. This may happen during the flow as described above or in the physical system at $k=0$. Indeed, the physical pion with its mass of about 140\,MeV is a quasi-massless mode in QCD. 

The physical cutoff scale of the pion drops to zero at the chiral symmetry breaking scale $k_\chi\approx 650$\,MeV, and for $k\leq k_\chi$ the effective action has to accommodate momentum transfers between $p^2\approx 0$ and $p^2 \approx k^2$. This requires fully momentum-dependent propagators and vertices.\\[-2ex]

In turn, for $\gamma\to 0 $, we first integrate out the quarks and an O(4) model remains. While this is also not doing full justice to the underlying dynamics of infrared QCD, it is a qualitative or even semi-quantitative limit, as the quarks decouple anyway at their physical mass scale, the constituent quark mass of $m_{q,\textrm{const}} \approx 350$\,MeV. 

Finally we remark that so far we have counted the mesonic momentum and hence the mesonic mass gap identical to the quark momentum and the quark mass gap, see e.g.~\labelcref{eq:OptimalChoice}. However, this relation depends on the details of the dynamical hadronisation procedure. The default choice in literature is that also used in the present work: both momenta are counted equally, leading to \labelcref{eq:OptimalChoice}. We note that while this choice is not particularly important at vanishing chemical potential, it is most likely not the optimal choice. For example, naïvely a factor of two is suggestive if the quark momenta are, at least partially, collinear. 

We close this Section with a summary of our findings: We have discussed the emergence of the quark-meson (QM) model as the emergent theory of low energy QCD and the reliability of the approximation \labelcref{eq:QMAction} for describing the dynamics of QCD at finite temperature and density. We have argued that a self-consistent lowest order of the derivative expansion requires the inclusion of a field-dependent Yukawa coupling. Moreover, the latter is pivotal for describing the physical decoupling of the quark degrees of freedom. This is in particular crucial for the access of QCD at larger density and low temperatures. Moreover, we have discussed that an optimal cutoff choice requires \labelcref{eq:OptimalChoice} and is essential for the present approximation, which lacks the full momentum-dependences of propagators and vertices.

\subsection{Flow equations for the effective potential $V(\rho)$ and the Yukawa coupling $h(\rho)$}
\label{sec:FlowsVh}

In the following we derive the flow for the effective action \labelcref{eq:QMAction} in the self-consistent LPA  \cite{Pawlowski:2014zaa} described in the introduction. This requires solving the coupled set of flow equations for the effective potential $V_k(\rho)$ and $h_k(\rho)$. We apply a high-order spatial DG discretisation, in  contradistinction to \cite{Pawlowski:2014zaa}, where mainly high order Taylor expansions have been used for both $V_k$ and $h_k$. Importantly, a high-order spatial DG discretisation can capture non-analytic features that cannot be dealt with in Taylor expansions or other methods that might wash out sharp features, such as finite difference methods.
Moreover, it also provides overall better quantitative precision. In \cite{Pawlowski:2014zaa} also cutoff-dependent wave functions $Z_q, Z_\phi$ have been included and their impact has been studied there. We refrain from including them here, as in a future work we will include field-dependent ones within the set-up \cite{Ihssen:2023nqd}. 

We use three-dimensional spatial flat or Litim regulators, that are detailed together with their threshold functions $l_0^{(F/B,d)}$ in \Cref{app:thfkt}. This leads us to the flow equation of the meson potential $V_k$ with 
\begin{align} 
  \partial_t V_k(\rho) &= \frac{k_\phi^4}{4\pi^2}\bigg[
      (N_f^2-1)l_0^{(B,4)}(m_\pi^2;T)
      + l_0^{(B,4)}(m_\sigma^2;T)\bigg]\ \notag\\[1ex]
      &+ \frac{k_q^4}{4\pi^2}\bigg[- 4 N_c N_f l_0^{(F,4)}(m_q^2;T,\mu_q)
    \bigg]\,, 
\label{eq:QMflow}
\end{align}
where we introduced the RG time $t=\log \frac{k}{\Lambda}$. The parameters in the effective potential as well as the choice of $h_k$ at the UV scale $\Lambda$ are inferred from the physical effective potential and Yukawa coupling at $k=0$, where observables such as the quark mass, the pion mass and the chiral condensate or rather $\langle \sigma\rangle  \propto \langle \bar q q \rangle$ are fixed to their physical values. In the present approximation with constant fields, the equation of motion \labelcref{eq:EoMphiGen} for $\sigma$ takes the form 
\begin{align}
	\frac{\partial V_k(\rho)}{\partial \sigma}\bigg\vert_{\sigma=\langle \sigma\rangle  } = \sigma \frac{\partial V_k(\rho)}{\partial \rho}\bigg\vert_{\sigma=\langle \sigma\rangle  } = c_\sigma\,, 
	\label{eq:EoMsigma}
\end{align}
with $\boldsymbol{\pi}=0$ and hence $\rho = \sigma^2/2$. 

The flow of the Yukawa coupling $h_k(\rho)$ is obtained from the following  projection of the flow of the effective action of the quark-meson model, 
\begin{align}
	\partial_t h_k(\rho) = \frac{\sqrt{2 N_f}}{4 N_c N_f \, \sigma } \Re\Big[
		\text{tr} \,\Gamma^{(2)}_{q\bar{q}}(p) 
	\Big]_{\hspace{-1mm}\substack{\phi(x) = \phi,\\[.1ex]\hspace{7.mm} p = p_\textrm{low}}}\,, 
\label{eq:h_projection}
\end{align}
with constant field $\phi$ and $p_\textrm{low} = (\pi\,T,\,\boldsymbol{0})$ with the lowest-lying fermionic Matsubara frequency. The trace $\text{tr}$ sums over Dirac, color and flavor indices, and $\Gamma^{(2)}_{q\bar{q}}(p)$ is the momentum-diagonal kernel of the two-point function in the presence of constant fields,  
\begin{align}
	\frac{\delta^2\Gamma_k}{\delta q(r)\delta\bar q(p)} = \Gamma^{(2)}_{q\bar{q}}(p) \,\, (2\pi)^4\delta(p+r)\,.
\end{align}
Instead of the flow \labelcref{eq:h_projection} of the Yukawa coupling we shall use that of the quark mass 
\begin{align}
	m_q^2(\rho) = 2\rho \, h_k(\rho)^2\,,
\end{align}
as this flow is more accessible numerically. Furthermore, we obtain a simple boundary condition for the quark mass,
\begin{align}
	m_q^2(\rho = 0) = 0\,,
	\label{eq:mqboundary}
\end{align} 
due to the flow of $h_k(\rho)$ being finite even at $\rho = 0$. The quark mass at $\rho = 0$ can critically influence the entire interaction potential and decide whether or not symmetry breaking occurs. Thus, it is essential to have a good hold on the boundary condition. 

Using the projection \labelcref{eq:h_projection}, the flow of $m_q^2(\rho)$ is derived as
\begin{align}\label{eq:flow_mq}
	\partial_t m_q^2 =  
	& - \nu_3 k_\phi^2 \, \partial_{\rho} m_q^2 \Big(\left(N_f^2-1\right) l_1^{(B,4)}\left(m_\pi^2;T\right)
	\notag\\
	& \hspace{99pt}+l_1^{(B,4)}\left(m_\sigma^2;T\right)\Big) \notag\\[1ex]
	& + \nu_3 \, \rho\,\big(\partial_{\rho } m_q^2\big)^2 \Big(k_\phi^2 l_1^{(B,4)}\left(m_\sigma^2;T\right)
	\notag\\
	& \hspace{49pt}+2 m_q^2 L_{(1,1)}^{(4)}\left(m_q^2,m_\sigma^2;T,\mu_q\right)\Big) \notag\\[1ex]
	& - \nu_3 k_\phi^2 \, 2\rho\,\partial_{\rho }^2 m_q^2 l_1^{(B,4)}\left(m_\sigma^2;T\right) \notag\\[1ex]
	& - \nu_3 \left(N_f^2-1\right) \frac{m_q^2}{\rho} \Big(2 m_q^2 L_{(1,1)}^{(4)}\left(m_q^2,m_\pi^2;T,\mu_q\right)
	\notag\\
	&\hspace{90pt} -k_\phi^2 l_1^{(B,4)}\left(m_\pi^2;T\right)\Big)\,,
\end{align}
with $\nu_3 = \frac{1}{4\pi^2}$.

\section{Discontinuous Galerkin in the functional renormalisation group}
\label{sec:LDG}

In the following we discuss the numerical intricacies of the current setup in detail. The less technically interested reader may skip ahead to the results in \Cref{sec:results}, which can be understood with little knowledge of the numerical details of the present Section.

The self-consistent LPA \cite{Pawlowski:2014zaa} leads to coupled system of partial differential equations (PDEs) for $V_k(\rho)$ and $h_k(\rho)$. Due to its non-linearity, solving this system of PDEs requires the precise computation of higher order derivatives. In recent years, this task has been approached by employing a variety of numerical methods for solving increasingly complex models and approximations. The application of numerical fluid dynamics to fRG equations has so far been investigated in~\cite{Grossi:2019urj, Wink:2020tnu, Grossi:2021ksl, Koenigstein:2021syz, Koenigstein:2021rxj, Steil:2021cbu, Stoll:2021ori, Ihssen:2022xjv, Ihssen:2022xkr, Ihssen:2023qaq, Murgana:2023xrq}.

The dynamics of spontaneous symmetry breaking and convexity restoration in dimensions $d>2$ have been tackled by introducing direct Discontinuous Galerkin methods (dDG) in the large-$N$ limit \cite{Grossi:2019urj}.
An application of the dDG method to the quark-meson model at finite $N$ has provided results for the phase structure at high temperatures and around the critical point \cite{Grossi:2021ksl}. Moreover, development of shock-waves in the RG-flows has been shown at high densities, indicating the necessity of using inherently discontinuous numerical methods. It was also found that the dDG, and hence all derived finite element methods, fail to solve the RG-flows at high densities.

In the following we will work with the pion mass
\begin{align}
	u = m_\pi^2 = \partial_{\rho}V \,.
\end{align}
Its flow can be obtained by taking a $\rho$-derivative of the flow \labelcref{eq:QMflow} and reads schematically
\begin{align} \label{eq:orig_diffeq}
  \partial_t u + \partial_{\rho} F(u, \partial_\rho u, \rho) = s(\rho)\,,
\end{align}
which structurally resembles a convection-diffusion equation, where $t$ has been chosen like above as $t=~\log \frac{k}{\Lambda}$.

Formally, this is a second-order PDE with a non-linear flow. As such, it has the potential to generate large convective and diffusive contributions due to $F(u, \partial_\rho u, \rho)$ having poles at $u =-k_\phi^2$ and $u+ 2\rho \partial_\rho u =-k_\phi^2$. 

The dDG discretisation fails if the non-linear diffusive flow is dominating, since a naive computation of $\partial_\rho u$ may yield errors of $O(1)$. This can also lead to apparent convergence in less sensitive discretisations, as been shown explicitly in case of the heat equation \cite{dDGanalysis}. Caution is advised at any point in the numerical evaluation of the flow where the diffusive pole $u+ 2\rho \partial_\rho u =-k_\phi^2$ is driving the dynamics, i.e.~where $\partial_\rho u<0$. In the present application this concerns primarily the low-$T$, high-$\mu_q$ area of the phase diagram, where a precise control of higher derivatives is vital.

We overcome these difficulties by using a so-called local discontinuous Galerkin Method (LDG).
Specifically, we choose a LDG method using left- and right-sided derivatives as detailed in \cite{FENG201668}. The stability and correctness of the solution has been checked in \cite{FENG201668} on the example of various higher-order convection-diffusion equations. It is a variation of the standard LDG method put forward in \cite{yan_local_2002}, which has already been successfully applied to the fRG in \cite{Ihssen:2022xkr}.
In contrast to the standard LDG method, the current method is specifically designed to stabilise second order derivatives. This allows for the computation of large diffusive flows which may be highly non-linear also in the derivatives. Furthermore, it allows us to discretise structurally more complex equations.

The main idea of this method is to replace the higher-order object $F(u, \partial_\rho u,\rho)$ with a mathematical operator $\mathcal{F}(u, g_\eta,w_{\eta\xi},\rho)$ which behaves in a more controlled way. We introduce the right and left derivatives $g_1$ and $g_2$ via the respective numerical flux. From $g_1$ and $g_2$ we can then reconstruct $\partial_\rho u$ in order to get a well-conditioned derivative. Similarly we use the four second-derivative terms from the left and right, $w_{\eta\xi}$, for the second derivative and also to correct errors in $\partial_\rho u$.
The following subsection will explain this numerical setup in detail.

\subsection{Spatial discretisation}

We start the spatial discretisation by dividing the computational domain (i.e.~field space) $\Omega = \cup_k D_k$ into cells $D_k$. Furthermore, we also define a local function space $V_h$ on each cell. Then, all functions are expanded locally within each cell in terms of basis functions $v^{(i)} \in V_h$, which can be for example Legendre polynomials of some finite order.

This means that within the $k$-th cell in the domain we have a decomposition
\begin{align}
	u_{h,k} = \sum_i u_{h,k}^{(i)} \, v^{(i)}\,,
\end{align}
where we require that all cells $D_k$ share the same function space $V_h$. The subscript $h$ is used to distinguish any numerically approximated quantity, such as $u_h$, from the exact solution $u$.

Due to the locality of the function spaces $V_h$, the equations can be solved separately within each cell. 
The cell-wise discontinuity of this approach still has to be stabilised. Hence, in the following we couple the cell-wise equations on the interfaces with all nearest neighbour cells.

From here on we use the usual DG notation and define for any variable $r$ at some cell interface at position $\rho$ the inner and outer limits
\begin{align}
	r^-(\rho) = \lim\limits_{y\rightarrow \rho, \, y<\rho}r(y),\notag \\[1ex]
	r^+(\rho) = \lim\limits_{y\rightarrow \rho, \, y>\rho}r(y).
\end{align}
Furthermore, we define the shorthand expressions for averages and jumps at a cell interface,
\begin{align}
	[\![ r ]\!] &:= \hat{n} \, (r^- - r^+), \notag\\[1ex]
	\{\!\!\{ r \}\!\!\} &:= \frac{r^- + r^+}{2},
\end{align}
where $\hat{n}$ is the outward-facing normal vector orthogonal to the cell border. As $u_h$ is double-valued at cell interfaces, the above quantities are important for treating cell-boundary terms.

\Cref{eq:orig_diffeq} is second order in $\rho$, which is not directly discretisable by the DG method. In a LDG approach we wish to reduce this to a first-order system of equations and apply the DG method to each equation. These equations are again highly local and can be solved on a per-cell basis, giving rise to the name local discontinuous Galerkin. 

Following this rationale, we introduce additional stationary equations for new variables $g_\eta$ and $w_{\eta\xi}$ accompanying the instationary one,
\begin{align}\label{eq:LDG_cont_system}
	\partial_t u + \partial_\rho \mathcal{F} = 0\,, \quad 
	g_\eta = \partial_\rho u \,, \quad 
	w_{\eta\xi} = \partial_\rho g_\eta \,,
\end{align}
with $\eta, \xi \in \{1,2\}$ and the yet to be specified operator $\mathcal{F}$. We choose $g_\eta$ and $w_{\eta\xi}$ to be projections of first and second order derivatives of $u$ respectively. The aim is then to build projections which are in analogy to right- and left-side derivatives and use them to reconstruct the (continuum) derivative by averaging.

Next, we apply the standard DG spatial discretisation to \labelcref{eq:LDG_cont_system}, i.e.~we integrate \labelcref{eq:LDG_cont_system} against local test functions (i.e.~the basis functions) $v^{(i)}$ on each cell and perform a partial integration.
From this we obtain the set of equations
\begin{widetext}
	\begin{align}
			M_{ij}\partial_t u_{h,k}^{(j)}(t)
			- \int_{D_k} \mathcal{F}(u_{h,k},\,(g_{\eta})_{h,k}) \partial_\rho v^{(i)}
			+ \int_{\partial D_k}
			\widehat{\mathcal{F}}(u^-_h,u^+_h) \cdot\hat{n}_k\, v^{(i)} &= \int_{D_k} s(\rho) v^{(i)}
		\,,\notag\\[1ex]
			M_{ij} (g_\eta)_{h,k}^{(j)}
			- \int_{D_k} u_{h,k} \partial_\rho v^{(i)}
			+ \int_{\partial D_k}\widehat{g}_\eta(u^-_h,u^+_h)\cdot\hat{n}_k\, v^{(i)} &= 0
			\,,\hspace*{20pt} \eta \in \{1,2\}
		\,,\notag\\[1ex]
			M_{ij} (w_{\eta\xi})_{h,k}^{(j)}
			- \int_{D_k} (g_{\eta})_{h,k} \partial_\rho v^{(i)}
			+ \int_{\partial D_k}\widehat{w}_{\eta\xi}((g_{\eta})^-_h,(g_{\eta})^+_h)\cdot\hat{n}_k\, v^{(i)} &= 0
			\,,\hspace*{20pt} \eta,\xi \in \{1,2\}\,,
		\label{eq:residual}
	\end{align}
\end{widetext}
for each grid cell $D_k$ separately.
Here, we have introduced the outward-facing normal vector $\hat{n}_k$ on the cell surface $\partial D_k$ and the value of $u_h$ on the cell surface from the inside $u_h^-$ and outside $u_h^+$. The mass matrix $ M_{ij}$ is defined as
\begin{align}
	M_{ij} = \int_{D_k} v^{(i)} v^{(j)}\,.
\end{align}
Importantly, we have introduced the numerical fluxes $\widehat{\mathcal{F}}$, $\widehat{g}_\eta$, and $\widehat{w}_{\eta\xi}$, which connect solution variables across cell boundaries as discussed before.
They constitute a crucial ingredient to formulating a consistent and stable DG method.

We want the auxiliary variables $g_\eta$ and $w_{\eta\xi}$ to be numerical right- and left-derivatives, which is achieved by using up- and downwind fluxes respectively for $\widehat{g}_\eta$. The choice of fluxes for $\widehat{w}_{\eta\xi}$ is analogously, 
\begin{align}
	\widehat{g}_1(u^-_h,u^+_h) &= u^+_h
	\,,\notag\\[1ex]
	\widehat{g}_2(u^-_h,u^+_h) &= u^-_h
	\,,\notag\\[1ex]
	\widehat{w}_{\eta 1}((g_\eta)^-_h,(g_\eta)^+_h) &= (g_\eta)^+_h 
	\,,\notag\\[1ex]
	\widehat{w}_{\eta 2}((g_\eta)^-_h,(g_\eta)^+_h) &= (g_\eta)^-_h
	\,.
\end{align}
With this, $g_1$ is a right-sided derivative and $g_2$ a left-sided one, whereas the $w_{\eta\xi}$ give all combinations of two either right- or left-sided derivatives.
In doing this, we provide the instationary equation with information on all possible jumps in derivatives between cell borders. We can now go on to devise a scheme on how to introduce this information to the flow in a consistent, stabilising manner.

For the collective flux of $\mathcal{F}$, i.e.~$\widehat{\mathcal{F}}$, we utilise the standard Lax-Friedrichs flux,
\begin{align}
	\widehat{\mathcal{F}} = \{\!\!\{F\}\!\!\} + \frac{c}{2}\,[\![u]\!]\,,
\end{align}
where $c$ is the local wavespeed in the system, i.e., the largest eigenvalue of $\frac{\partial F}{\partial u}$, which allows for stable convection and also introduces a moderate artificial diffusion.

It is left to define $\mathcal{F}$. We set it to
\begin{align}\label{eq:operator_def}
		\mathcal{F} = &F(u,\,\frac{g_1 + g_2}{2},\,\frac{w_{12} + w_{21}}{2},\,\rho) \notag \\[1ex] 
					  & \quad+\,\alpha \big(w_{11} - w_{12} - w_{21} + w_{22}\big) \,.
\end{align}
This choice of derivative $\partial_\rho u \simeq \frac{g^+ + g^-}{2}$ is elaborated on in \cite{FENG201668}, where one sees that it can be viewed as analogous to the central difference derivative operator on a Cartesian grid.
The convergence and stability of this method has been shown and tested in \cite{yan_local_2002} for a set of toy problems.
Also, note that $\mathcal{F} \rightarrow F$ in the continuum limit.

The correction term
\begin{align}
	\label{eq:addTerm}
	\alpha \big(w_{11} - w_{12} - w_{21} + w_{22}\big) \,,
\end{align}
that has been added to \labelcref{eq:operator_def} in order to define the numerical operator can be viewed as
\begin{align}
    \alpha h^2 \frac{\big(w_{11} - w_{12} - w_{21} + w_{22}\big)}{h^2} \rightarrow \alpha h^2 \partial_\rho^4 u \,,
\end{align}
in the limit of small $h$.
Effectively, this source term added to the flow of $u$ exerts a force that pushes the second derivatives $w_{\eta\xi}$ onto each other, reducing oscillations in the first and second derivatives of $u$.

Framed differently, we are approximating our parabolic problem by a fourth-order problem which reduces to the original equation in the limit of $h\rightarrow0$. This has been shown to work well for certain classes of non-linear second-order equations \cite{Feng2011TheVM}, stabilising their high-order contributions in a controlled manner.
Additionally, \labelcref{eq:addTerm} implements a penalty for jumps in $g^+, \,g^-$, thus stabilising the solution at the cell borders.

The choice of $\alpha$ should be such that $\alpha > \lvert \frac{\partial F}{\partial (\partial^2_\rho u)} \lvert$. We will however define it to be constant in time and by choosing it heuristically: One wants it to be as small as possible, but large enough to suppress oscillations in the derivatives. We comment that a choice of $\alpha = c \lvert \frac{\partial F}{\partial (\partial^2_\rho u)} \lvert$ with some constant $c>1$ is possible, but works worse in numerical experiments than choosing $\alpha$ globally constant.

We further note that \labelcref{eq:operator_def} allows us to directly discretise the flow of $m_q^2(\rho)$ in \labelcref{eq:flow_mq}, which would not be possible in the standard DG formulation due to the nonlinear dependences on higher derivatives of $m_q^2(\rho)$. This can be done by going through the above set of definitions without a flux term $F$ but rather using only a source term. This source is then similarly replaced by an operator defined as above.

\subsection{Time discretisation}

It has been shown in \cite{Ihssen:2023qaq} that in the presence of symmetry breaking, and thus convexity restoration, the use of implicit timestepping methods is necessary to avoid stability issues and exponentially small timesteps. Furthermore, \texttt{BDF} methods have proven to be very favorable for fRG applications due to both high precision and computational efficiency.

Therefore, we use the implicit scheme provided by the \texttt{IDA} timestepper of the \texttt{SUNDIALS} suite for time integration. In our application it is equivalent to the variable order (1--5) backwards differentiation schemes of \texttt{CVODE\_BDF}, which has been found to be highly successful for the time integration within the broken phase in \cite{Ihssen:2023qaq}.

\subsection{Numerical implementation}
\label{sec:NumericalImpl}

\subsubsection{Effective potential}
\label{sec:NumericsPotential}

For the numerical implementation of the flow of the effective potential $V(\rho)$, we take a $\rho$-derivative of \labelcref{eq:QMflow}. This leads to a flow equation for the field-dependent (curvature) pion mass
\begin{equation}
	\partial_{\rho} V (\rho)= m_\pi^2(\rho)\,. 
\end{equation}
The flow equation for $ m_\pi^2$ has the shape of a conservation equation, 
\begin{equation}
	\partial_{t} m_\pi^2 = \partial_{\rho}F(m_\pi^2,\,\partial_{\rho}m_\pi^2,\,\rho)\,.
	\label{eq::numQM}
\end{equation}
This allows us to directly apply the LDG formulation as discussed in \Cref{sec:LDG}.

For boundary conditions we use inflow/outflow flux conditions wherever a conservative flux formulation applies; that is, effectively, we simply set $\widehat{F} = F((m^2_\pi)^-,\,(\partial_{\rho}m^2_\pi)^-,\,\rho)$ at the boundary, which transports information simply in or out depending on the sign of $\partial_{m_\pi^2}F$.

\begin{figure*}[t]
	\centering
	\begin{minipage}{.48\textwidth}
		\centering
		\includegraphics[width=1\linewidth]{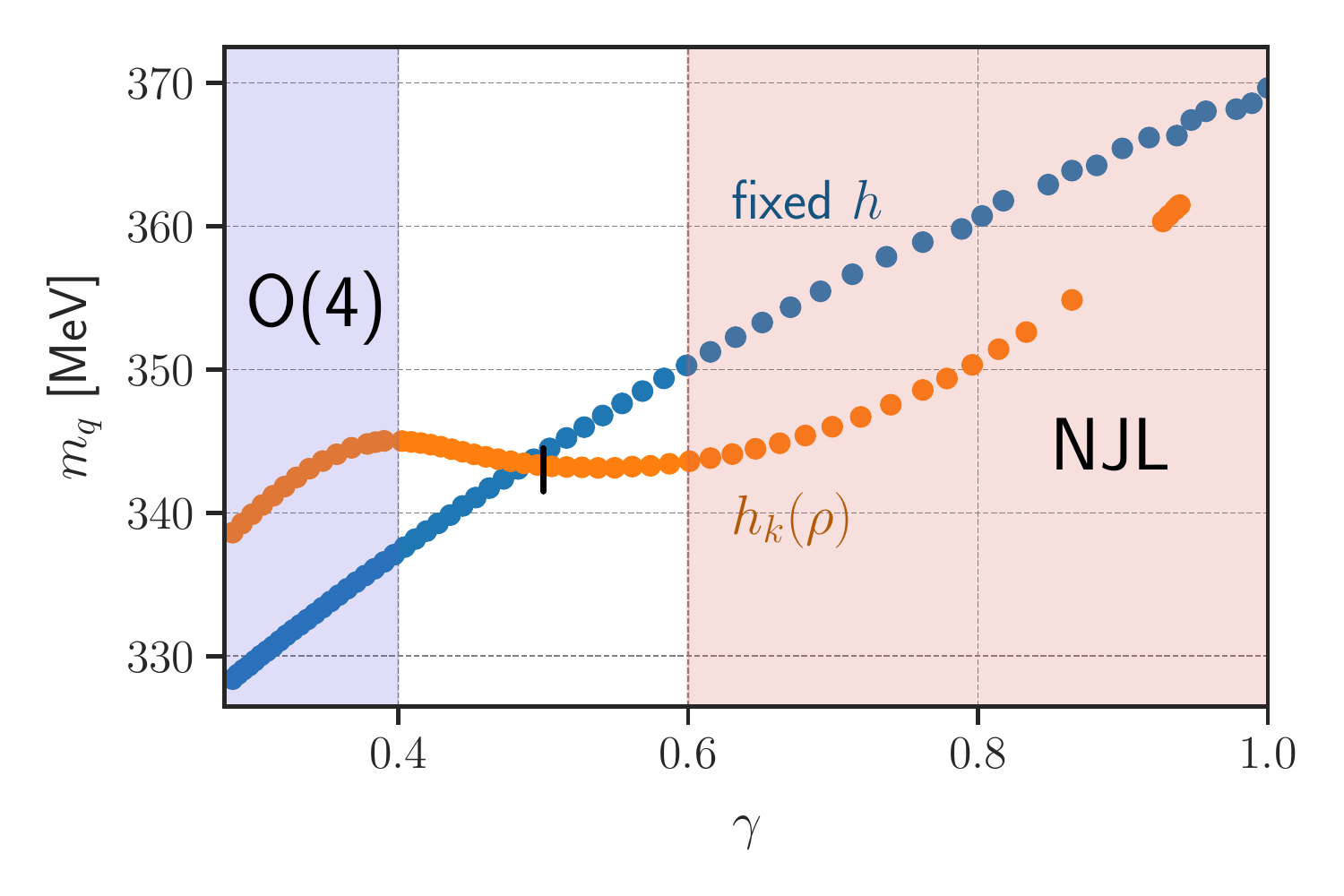}
		\caption{The quark mass close to vacuum as a function of the relative cutoff parameter $\gamma$. For $\gamma\to\infty$ we arrive at a bosonised NJL-type setup with a quasi-massless mode in the broken phase. For $\gamma\to0$ the model approaches an O(4) model in LPA. In between, one sees the  relatively stable region for intermediate $\gamma$ in the case of a self-consistent LPA (scLPA).}
		\label{fig:gammadep}
	\end{minipage}%
	\begin{minipage}{.04\textwidth}
		\hfill
	\end{minipage}%
	\begin{minipage}{.48\textwidth}
		\centering
		\includegraphics[width=1\linewidth]{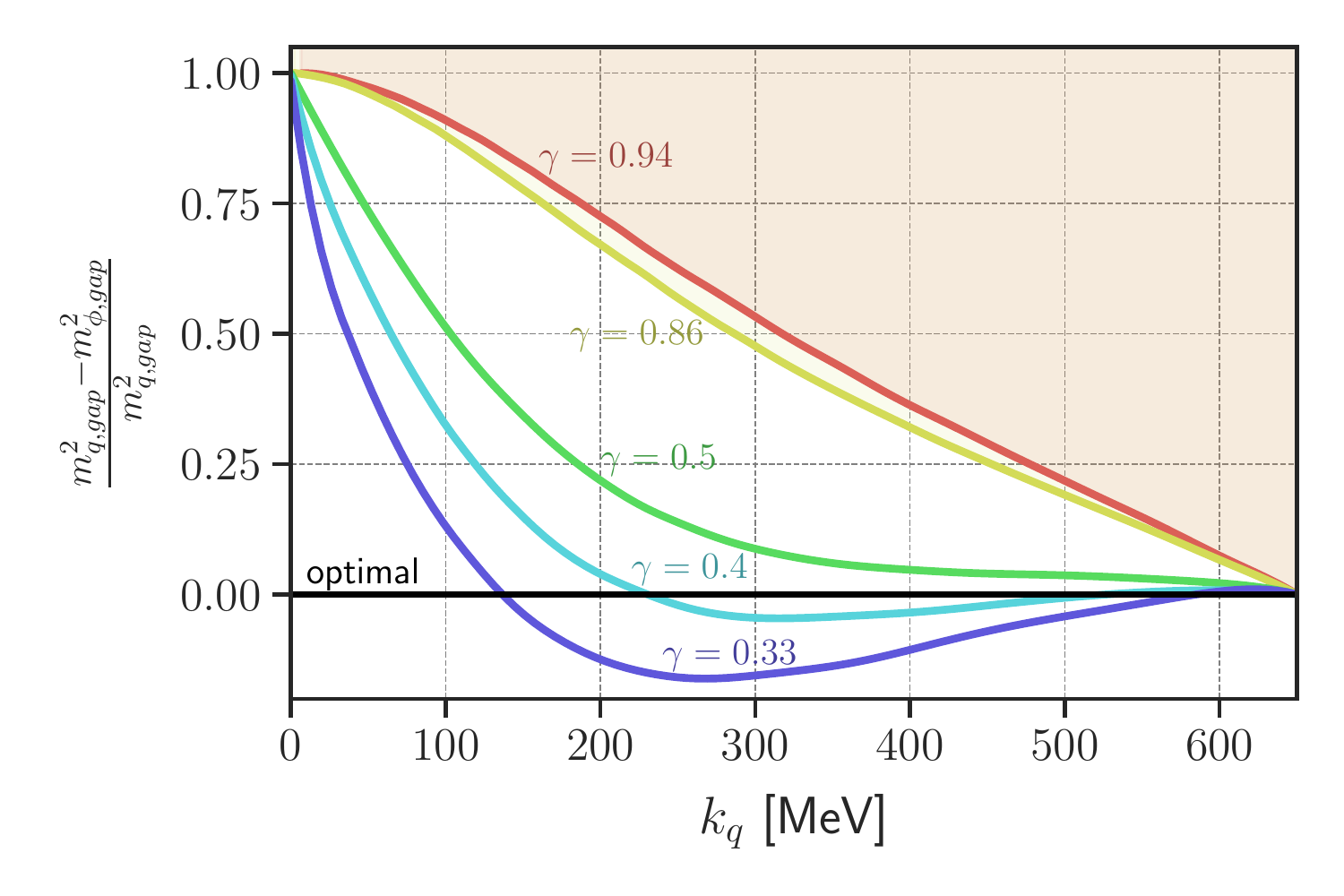}
		\caption{Difference between the physical gaps for quarks and mesons at various $\gamma$. The gaps are evaluated at the flowing EoM with no explicit symmetry breaking. In the vacuum, the region above $\gamma = 0.94$ is forbidden as the quark mass diverges there. Similarly, $\gamma=0.86$ is the same bound at $\mu_q = 280\,\text{MeV}$.\vspace{4mm}}
		\label{fig:gamma_gaps}
	\end{minipage}
\end{figure*}
%

\subsubsection{Yukawa coupling}
\label{sec:NumericsYukawa}

Unfortunately, no reformulation is known for the flow of the Yukawa coupling \labelcref{eq:flow_mq} which casts the flow equation into a conservative form. Therefore, we discretise it directly. If we write \labelcref{eq:flow_mq} in a shorthand manner as
\begin{align}
	\partial_t m_q^2(\rho) =  G(m_q^2,\,\partial_{\rho} m_q^2,\,\partial^2_{\rho}m_q^2)\,,
\end{align}
we can once again replace $G$ by an operator $\mathcal{G}$, which is defined analogously to \labelcref{eq:operator_def} as
\begin{align}\label{eq:operator_def_Yukawa}
	\mathcal{G} =\, &G(m_q^2,\,\frac{b_1 + b_2}{2},\,\frac{s_{12} + s_{21}}{2},\,\rho) \notag \\[1ex] 
	& \quad+\,\alpha_H \big(s_{11} - s_{12} - s_{21} + s_{22}\big) \,.
\end{align}
Here, $b_\eta$ and $s_{\eta\xi}$ have been defined as the first and second derivatives, respectively, once again from left and right.

Note that \labelcref{eq:operator_def_Yukawa} is implemented as a source term for the flow $\partial_t m_q^2(\rho)$. Furthermore, we fix the boundary condition $m_q^2(0)=0$ at $\rho = 0$, as we do not expect a divergence of $h_k(\rho)$ at finite $k$.

\section{Results}
\label{sec:results}

In \Cref{sec:InitialCondOptimisedk} we derive the parameters of the UV effective action. This goes hand in hand with the optimisation of the relative cutoff scales discussed in \Cref{sec:PhysGap+Cutoff}. We emphasise that the self-consistent LPA  is the minimal one that leads to results in the quark-meson model that are stable against deformations of the regulators and the cutoff scales for sufficiently small chemical potentials.

Within this setup we compute the phase structure of the quark-meson model and discuss it in \Cref{sec:results:QMY}. While this model lacks some important QCD physics at large densities, i.e.~the diquark channel that is relevant for color-superconducting phases, we consider the present setup a qualitative step towards stable functional QCD results in this regime.

\subsection{Initial conditions and optimised relative cutoff scales}
\label{sec:InitialCondOptimisedk}

The conceptual advances in \Cref{sec:PhysGap+Cutoff} are now put to work: We consider the self-consistent LPA in the quark-meson model with the flows for the effective potential, \labelcref{eq:QMflow}, and the field-dependent Yukawa coupling or quark mass term, \labelcref{eq:flow_mq}. The relative cutoff scale is adjusted by the parameter $\gamma$ in \labelcref{eq:kqkphi}. 

Evidently, for $\gamma\to0$ we first integrate out the quarks and then the mesonic degrees of freedom. Effectively this leaves us with an O(4) model, which lacks the interplay between quarks and mesons that is the important advantage of the quark-meson model. 

In turn, for $\gamma\gtrsim1$ we deal with a setup where the pion becomes quasi-massless very quickly in the symmetry-broken regime.
This amounts to a setup with maximal momentum transfer and all the difficulties discussed in \cite{Fu:2022uow}. 

Without approximation, we should arrive at a unique effective action at $k=0$ for all $\gamma$, if initiating the flow at a given UV cutoff scale $\Lambda$: the regulators $R_{q,\Lambda}$ and $R_{\phi,\Lambda}$ are $\gamma$\nobreakdash-independent at the initial scale, because $k_\phi=\Lambda$ for $k_q=\Lambda$, see \labelcref{eq:kqkphi}.
We expect a significant $\gamma$\nobreakdash-dependence in the asymptotic regimes $\gamma\to 0$ and $\gamma\to~\infty$ and a plateau around the optimal $\gamma_\textrm{opt}$ for a sufficiently advanced approximation.
Therefore, a minimised $\gamma$-dependence of our results in the vacuum for a fixed initial condition entails the stability of the approximation.

Implementing the above setup we observe that a plateau is reached in the self-consistent LPA with a field-dependent quark mass, see \Cref{fig:gammadep}, while no such plateau is seen in the standard LPA. 
We conclude that the higher-order Yukawa scatterings contained in $h_k(\rho)$ lead to non-trivial dynamics. These stabilise the relative scale dependence in the regime of maximal back-coupling with the meson scattering potential $V_k(\rho)$. 
In the asymptotic regimes, the evolutions of the couplings are decoupled, as either all mesons or all quarks are integrated out first and thus no coupled quark-meson dynamics is included.

The initial conditions used for \Cref{fig:gammadep} and in the remainder of the work are summarised in \labelcref{eq:InitialParameters} unless specified otherwise. For the initial effective potential at the UV scale $k=\Lambda$, we have a vanishing meson mass and only an initial $\phi^4$-coupling given by the parameter $\lambda_\Lambda$,
\begin{align}
	V_\Lambda(\rho) = \frac{\lambda_\Lambda}{4}\rho^2\,, 
\end{align}
and the explicit values
\begin{align}
	\Lambda &= 650\,\text{MeV}\,,\quad m_\phi^2 = 0\,,\\ 
	\lambda_\Lambda &= 58.7\,,\hspace{32pt}
	h_\Lambda = 6.6 \,.
	\label{eq:InitialParameters}
\end{align}
These initial conditions lead to physical values for the quark, pion and sigma masses on the physical point.

From \Cref{fig:gammadep} we infer the plateau regime as 
\begin{align}
	\gamma \in (\gamma_\textrm{min} \,,\gamma_\textrm{max}  )  \,,\qquad \gamma_\textrm{min} \approx 0.4\,,\quad \gamma_\textrm{max} \approx 0.6\,.
	\label{eq:gammaPlateau}
\end{align}
This supports our evaluation of self-consistency, while also significantly improving the reliability of the predictions of the quark-meson model. Moreover, the plateau in \Cref{fig:gammadep} entails that we can take any $\gamma$ in this regime without changing the physics. A full analysis will be presented elsewhere. Here, we simply pick a $\gamma$ in the middle of this regime: we expect this regime to shrink at finite temperature and density and, while this may not happen symmetrically, it will happen on both sides (c.f.~\Cref{app:mu_dep_gamma}). We choose 
\begin{align}
	\gamma_\textrm{opt} = 0.5 \,, 
	\label{eq:gammaOpt}
\end{align}
as an estimate for the optimal $\gamma$ in the vacuum. We remark again that $k_\phi(k_q)$ could be in principle chosen fully adaptively, implementing \labelcref{eq:OptimalChoice}. However, here we opt for a more pragmatic approach which is easier accessible: we simply use the choice $\gamma_\textrm{opt}$ as identified above.

In \Cref{fig:gamma_gaps} we also show the relative gap difference for several choices of $\gamma$ at zero chemical potential and $T=~4\,\textrm{MeV}$. It is clearly visible that a choice of $\gamma \gtrsim 1$ leads to a quick buildup of a large difference in the physical gaps, whereas our optimal choice $\gamma_\textrm{opt} = 0.5$ keeps the gaps relatively close to each other in the regime before the quark decouples. The latter has to happen eventually, so keeping the gaps close to each other for all $k$ is neither possible nor wanted, but certainly preferable for scales before the eventual decoupling.

\subsection{Momentum transfer at small temperatures and large chemical potentials}
\label{sec:results:relative_cutoffs}

At both small temperatures and increasing chemical potential we face the problem of large momentum transfers in the flows, as discussed in \Cref{sec:EmergentComposites}, leading to qualitative regulator dependences. Consequently, even the semi-quantitative reliability of the present approximation is at stake in the low temperature and large chemical potential regime of the phase diagram. Moreover, dropping significant momentum transfers in each flow step can destabilise the system. 

Indeed, we detect both signatures of a lack of reliability in the present approximation: For field-independent Yukawa coupling the phase structure shows an increasing regulator dependence for larger chemical potentials and low temperatures. Even worse, in fully self-consistent LPA the system cannot be solved for general $\gamma$ in \labelcref{eq:kqkphi}, including the standard choice $\gamma = 1$. 
More specifically, the quark mass increases exponentially at small $\rho \approx 0$ as soon as $k < k_\chi$. This fast divergence also makes the boundary condition \labelcref{eq:mqboundary} impossible to implement and strongly deforms the phase diagram. The divergence gets worse at higher $\mu_q$ and is triggered by the additional momentum-dependence of the Fermi-surface. With the standard choice $\gamma=~1$, the difference between the physical cutoff scales \labelcref{eq:mgap} are increasing with increasing $\mu_q$. Moreover, the pion becomes quasi-massless in the broken phase for $k\leq k_\chi$ as discussed below \labelcref{eq:WorstChoice}. This maximises the difference between physical cutoff scales and the momentum transfer towards $p=0$, which is relevant for the field dependence of the Yukawa coupling. 
\begin{figure}[t]
	\includegraphics[width=1\linewidth]{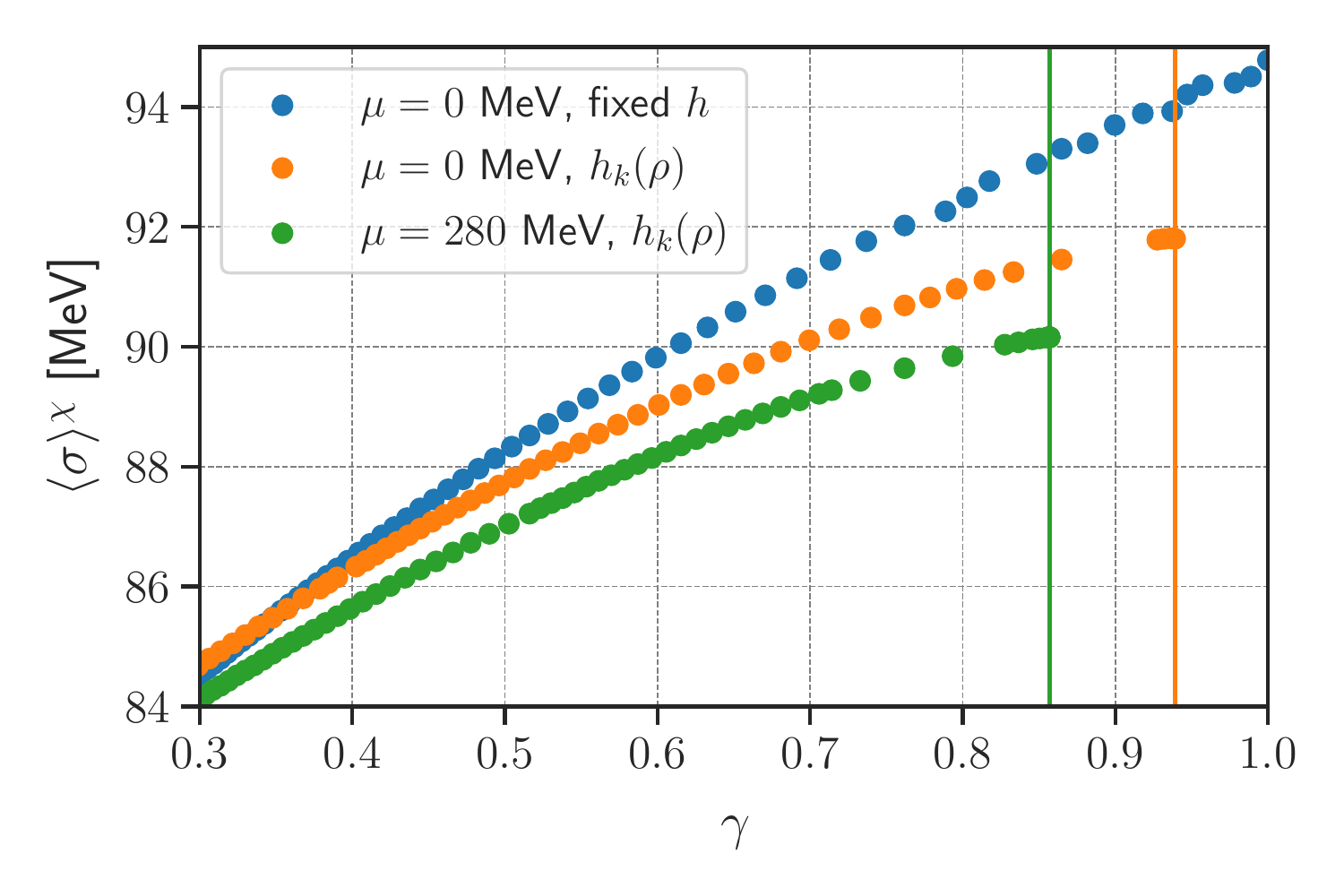}
	\caption{We compare the order parameter for different $\gamma$. The colored vertical lines show the values of $\gamma$ where one is able to obtain results for running $h_k(\rho)$, i.e.~when \labelcref{eq:subtract_mq} does not diverge.}
	\label{fig:gamma_scan}
\end{figure}
\begin{figure*}[t]
	\begin{subfigure}{0.5\textwidth}
		\includegraphics[width=1\textwidth]{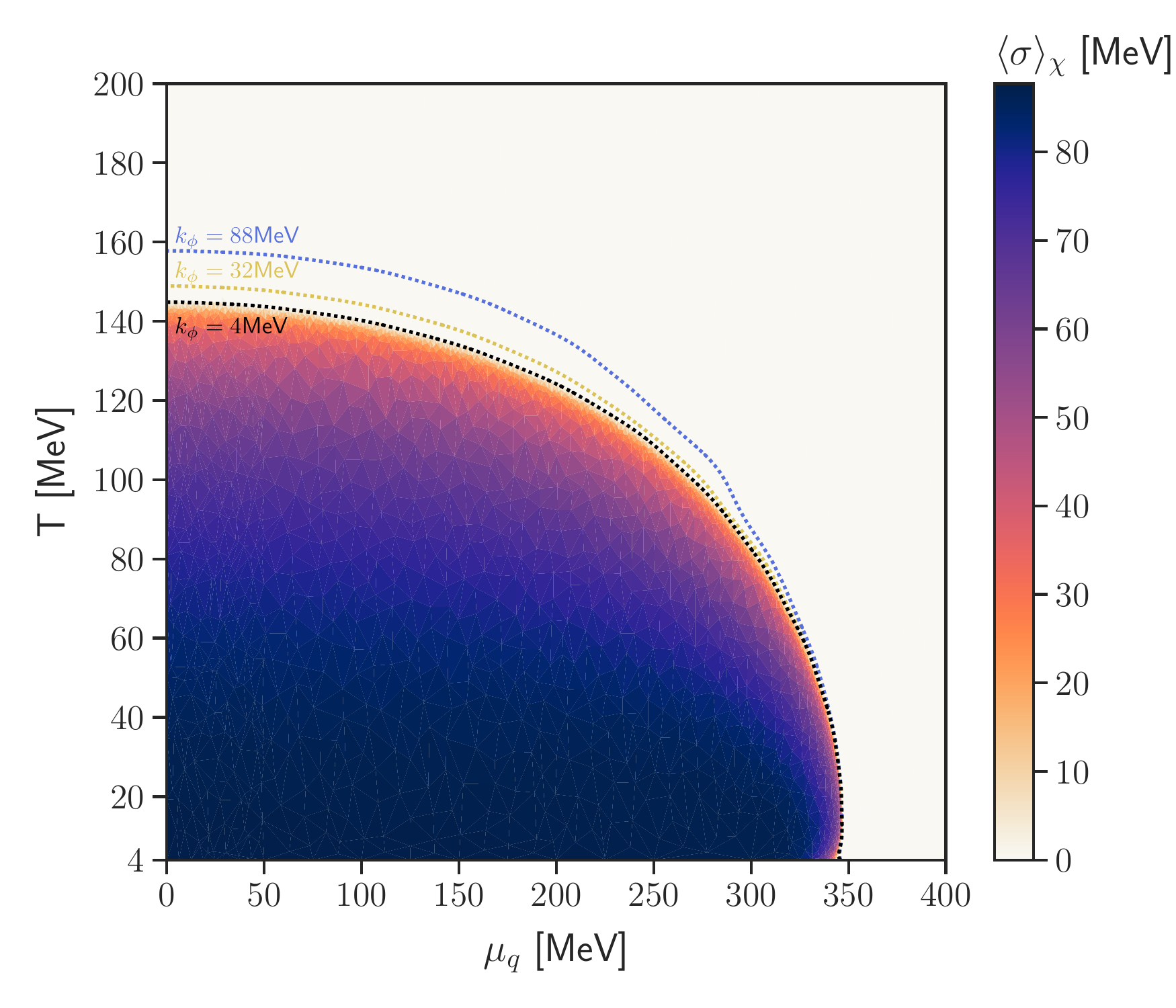}
		\caption{$c_\sigma=0$.}
	\end{subfigure}\hfill%
	\begin{subfigure}{0.5\textwidth}
		\includegraphics[width=1\textwidth]{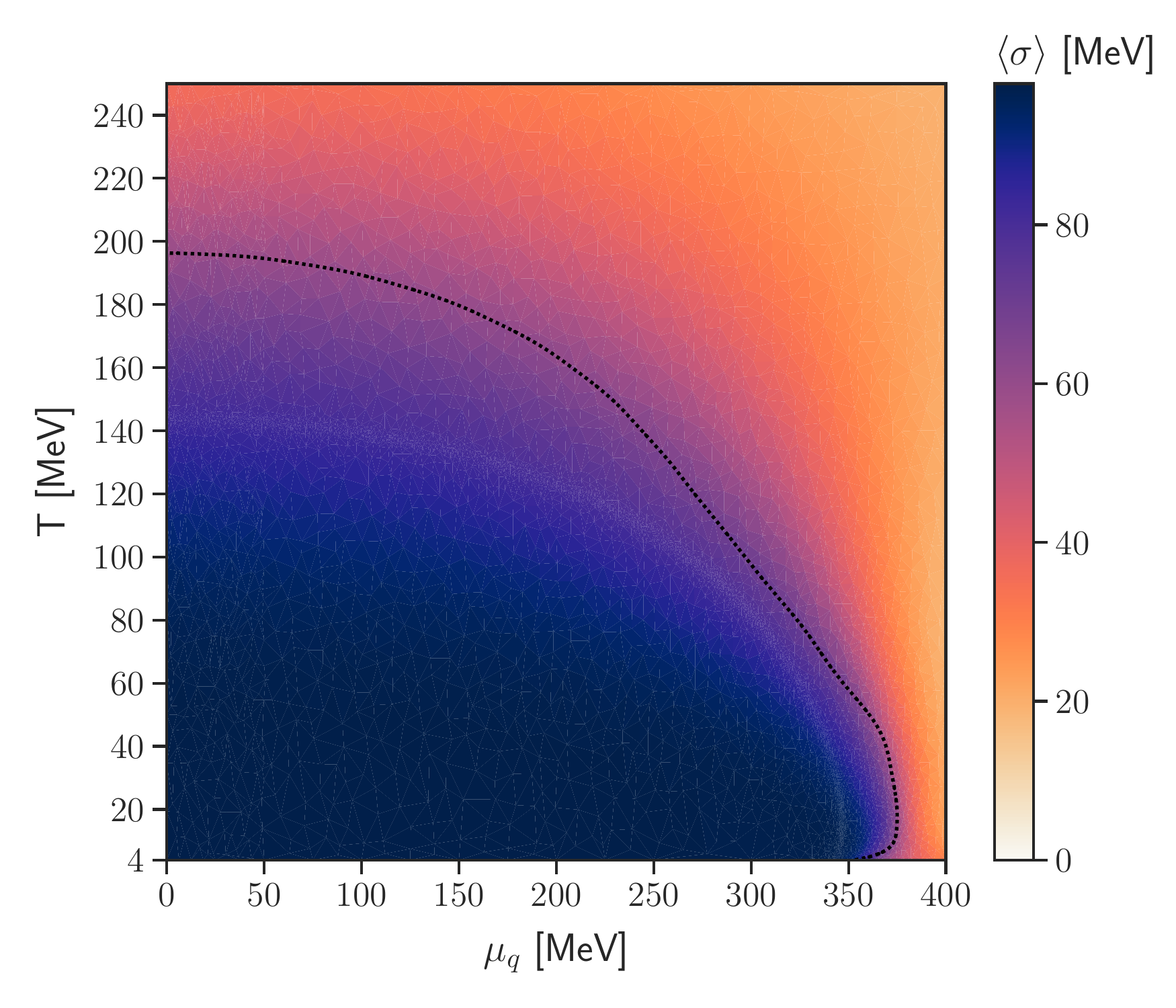}
		\caption{$c_\sigma=1.319\cdot10^{-3}\,\text{GeV}^3$.}
	\end{subfigure}
	\caption{The phase diagram of the quark-meson model with higher quark-meson scatterings (scLPA) in the chiral case $c_\sigma = 0$ in (a) and with explicit chiral symmetry breaking $c_\sigma=1.863\cdot10^{-3}$ in (b). We show the expectation value $\langle \sigma\rangle $ defined in \labelcref{eq:EoMphiGen} in MeV.  We also show the phase boundary for different RG-scales $k$ in the chiral case.}
	\label{fig:QMY_PD}
\end{figure*}

In summary, as discussed in \Cref{sec:PhysGap+Cutoff}, the standard choice is certainly not an optimal one. Moreover, the lack of similar cutoff scales with $m^2_{\phi,\textrm{gap}}\approx 0$ for $k\leq k_\chi$ also carries a significant destabilisation potential. This becomes more and more prominent with increasing field dependences if the latter are not accompanied with a better resolution of the momentum dependence.

As mentioned above, for $k_q = k_\phi$, the quark mass function $m^2_q(\rho)$ as well as the Yukawa coupling $h(\rho)$ diverge at $\rho = 0$. Technically, the reason for this destabilisation can be readily identified within the flow equation of the quark mass \labelcref{eq:flow_mq}, 
\begin{align}
\Delta=	\frac{m_q^2}{\rho} 
	\left(2 m_q^2 L_{(1,1)}^{(4)}\left(m_q^2,m_\pi^2;T,\mu_q\right)-k^2 l_1^{(B,4)}\left(m_\pi^2\right)\right)\,. 
\label{eq:subtract_mq}
\end{align}
A stable flow for the quark mass function in the limit $k\to 0$ is obtained if the difference $\Delta$ in \labelcref{eq:subtract_mq} is smaller than zero. 

However, at the value of the order parameter $\langle\sigma\rangle_\chi$ in the chiral limit (see \labelcref{eq:chiral_order}) the quark mass $m_q^2(\rho)$ can be seen to have a kink, similarly to the effective potential. Below $\langle\sigma\rangle$ we can obtain several solutions for $m_q^2$. Hence, we have several possible evolutions of $m_q^2$, depending on the size of $\Delta$ and the initial conditions.

\begin{enumerate}
	\item[(i)] $\Delta<0$ leads to the solution followed here. Then the quark mass function $m_q$ is linear in $\sigma$ for $\sigma$-values that are smaller than the expectation value in the chiral limit \labelcref{eq:chiral_order}: For $\sigma < \langle\sigma\rangle_\chi$ we find 
	\begin{align}
		m_q(\sigma < \langle\sigma\rangle_\chi) =m_q(\langle\sigma\rangle_\chi) \frac{\sigma}{\langle\sigma\rangle_\chi} \,, 
		\label{eq:mqSollinear}
	\end{align}
	see also \Cref{fig:compare_g_qmasses,fig:mq_h_high_mq} in \Cref{app:flow_quark_masses}.
	
\item[(ii)]  The extremal case with $\Delta<0$ leads us to 
\begin{align}
		m_q(\sigma < \langle\sigma\rangle_\chi) \equiv 0 \,. 
		\label{eq:mqSol0}
\end{align}
We have not been able to obtain this solution in our current setup for any $\gamma$. A more intricate fine-tuning is beyond the scope of the present work and we defer it to future investigations. 
	
\item[(iii)]  In turn, for $\Delta>0$ in the limit $k\to 0$, the quark mass function is $m_q^2$ is rising exponentially in the flat regime and finally leads to a diverging quark mass function for $\sigma<\langle\sigma\rangle_\chi$. 
\begin{align}
		1/ m_q(\sigma < \langle\sigma\rangle_\chi) =0 \,. 
		\label{eq:mqSolinfty}
\end{align}
This certainly also causes numerical instabilities in the solution of the flow which may even destabilise the physical regime with $\sigma>\langle\sigma\rangle_\chi$. 
Furthermore, this solution has a discontinuity at $\rho = 0$, where the fixed value of $m_q^2(\rho= 0) = 0$ jumps immediately to infinity.
Although we have investigated this solution, the numerical problem is getting worse at finite chemical potential, leading to highly fine-tuned phase diagrams.
We will not consider this numerical solution here. Note however, that it is a perfectly viable solution to the system for $\sigma<\langle\sigma\rangle_\chi$, but numerically very hard to access and control.
	
\item[(iv)]  Finally, in between the two solution branches with $\Delta>0$ with 	\labelcref{eq:mqSollinear,eq:mqSol0}, and $\Delta>0$ with \labelcref{eq:mqSolinfty}, there is the case of a constant quark mass, 
\begin{align}
		m_q(\sigma < \langle\sigma\rangle_\chi) =m_q(\langle\sigma\rangle_\chi) \,. 
		\label{eq:mqSolConst}
\end{align}
This is the solution discussed in \cite{Pawlowski:2014zaa} for the self-consistent LPA', where also $k$-dependent wave functions have been considered, see Figure 6 there and the respective discussion. In contrast to \labelcref{eq:mqSollinear} this solution can only be obtained in our case through extreme finetuning, see also \Cref{app:flow_quark_masses}.
\end{enumerate}
In summary, while all these solutions in the regime $\sigma < \langle\sigma\rangle_\chi$ can be triggered with appropriate choices of $k_\phi(k_q)$, this does not influence the solution for  $V_\textrm{eff}(\rho)$ and $h(\rho)$ with $\sigma > \langle\sigma\rangle_\chi$ and hence the physics. Note that the numerical instabilities occuring in \labelcref{eq:mqSolinfty} may also falsely influence this regime due to their exponential nature. Accordingly, we do not consider larger $\gamma$, including $\gamma=~1$, as this may trigger a false destabilising behaviour for  $\sigma >~\langle\sigma\rangle_\chi$.

In the remainder of this work we concentrate on the solution branch (i), which is the simplest and most stable solution to tune. We close this analysis with the side remark that while this holds true in the current functional renormalisation group set-up, the branch (i) is the most difficult one to tune in the Dyson-Schwinger equation (DSE) as there it only can be defined as the limit $k\to 0$ of solutions for $k\neq 0$: strictly at $k=0$ the mesonic loops in the DSE do not give any contribution to the potential, while the quark loop contributions are non-trivial. For $k\neq 0$ the mesonic loops are proportional to $V^{(n)}$ which tends towards zero for $k\to 0$, but the pion propagators are quasi-massless and tend towards infinity. Accordingly, the mesonic contributions can cancel the quark contributions for $k\to 0$. 

In \Cref{fig:gamma_scan} we show the effect of different $\gamma$ on the EoM value of $\sigma$ for a model with fixed Yukawa coupling versus a running $h_k(\rho)$ at $T=4\, \text{MeV}$ and both $\mu_q = 0$ and $\mu_q = 280\, \text{MeV}$. 
The maximal value for which we can obtain valid results using a running Yukawa coupling is $\mu_q$-dependent. This is because the Fermi-surface shifts with increasing $\mu_q$ and thus requires different momentum transfer scales for any value of $\mu_q$.
At vanishing chemical potential $\mu_q = 0$ valid results can be obtained for $\gamma \leq 0.94$, whereas $\mu_q = 280 \,$MeV already requires to choose $\gamma \leq 0.86$.

We note that numerically, decreasing $\gamma$ leads to easier manageable dynamics. This is due to the pion not becoming immediately quasi-massless, which also lessens the difficulties of convexity-restoring behavior as discussed in \cite{Ihssen:2023qaq}.

Furthermore, decreasing the value of $\gamma$ leads to a decrease in back-bending, as can be seen from the additional plots in \Cref{app:detail_PD}. In addition, the critical end point is moving towards smaller chemical potentials, and for sufficiently small $\gamma$, it is lost. Notably, this happens already for $\gamma>1/2$ and hence for our choice $\gamma_\mathrm{opt}=1/2$, the quark-meson model does not show a critical end point in the current approximation. 

We observe a sharp, but second-order transition line at high-densities in the chiral limit and a sharp crossover transition line with explicit symmetry breaking instead of the expected first-order behaviour. We emphasise that this does not suggest that full QCD has no CEP but rather it is a sign for the breakdown of the RG-consistency of the current LEFT at high densities and low temperatures. Note in this context that the convergent estimate for the critical end point in QCD in \cite{Fu:2019hdw, Gao:2020fbl, Gao:2020qsj, Gunkel:2021oya, Bernhardt:2022mnx} is at larger temperatures and smaller chemical potentials that are not directly affected by the above intricacies.

\subsection{Phase diagram of the QM model in the self-consistent LPA}
\label{sec:results:QMY}

Having discussed the necessary choice of relative cutoff scales $k_\phi(k_q)$, we are now in a position to investigate the phase diagram of the quark-meson model in scLPA, i.e.~with the inclusion of quark--antiquark--$n$-meson scatterings with an arbitrary number of mesons~$n$.

In addition to the field-space discretisation of $V_k(\rho)$ we apply the LDG discretisation formalism also to the running Yukawa coupling $h_k(\rho)$, (c.f. \Cref{sec:NumericalImpl}). As discussed above, the shift in cutoff scales of quarks and mesons is implemented with a fixed $\gamma_\mathrm{opt} = 0.5$ in the entire phase diagram, the optimal choice in the vacuum (c.f. \Cref{sec:InitialCondOptimisedk}).

The resulting phase diagram is shown in \Cref{fig:QMY_PD} and constitutes a main result of the present work. It is shown for both the chiral case (left) as well as for a physical value of the explicit symmetry breaking parameter $c_\sigma=~1.863\cdot10^{-3}\,\text{GeV}$ (right). In the chiral setup we obtain a critical temperature of $T_c=~144.8\,\textrm{MeV}$, to be contrasted with the case of initial vacuum physics but no running of the Yukawa coupling $T_c=148.3\,\textrm{MeV}$.

For the chiral case in the approximate vacuum, $(T,\mu_q) = (4,0)\,\text{MeV}$, we obtain a condensate and quark mass of
\begin{align}\nonumber 
	\langle\sigma\rangle_\chi = 88.2\,\text{MeV}\,,\qquad m_q = 343\,\text{MeV}\,,	\\[1ex] 
	m_\pi = 0\,\text{MeV}\,,\qquad 	m_\sigma = 233\,\text{MeV}\,.
\end{align}
With the explicit symmetry breaking parameter $c_\sigma$ we adjust the physical pion mass $m_\pi = 138\,\text{MeV}$ at $(T,\mu_q)=~(4,0)\,\text{MeV}$. The respective phase diagram is obtained from the data for $V(\rho)$ and $h(\rho)$ computed in the chiral limit. The condensate, quark, sigma and pion mass in the vacuum are given by 
\begin{align}
	\langle\sigma\rangle = 95.6\,\text{MeV}\,,\qquad  m_q = 363\,\text{MeV}\,,\notag \\[1ex] 
	m_\pi = 138\,\text{MeV}\,,\qquad m_\sigma = 538\,\text{MeV}
	\,.
\end{align}
Regarding the evolution of the quark mass, we see that within the chirally broken phase $h_k(\rho)$ flattens out below the EoM value of $\sigma$ for all densities, c.f.~\Cref{app:flow_quark_masses}. This entails that the minimal quark mass in any mesonic background is given by that in the chiral limit. Technically, this is also expected as the effective potential $V(\rho)=V_{k=0}(\rho)$ is necessarily flat (or linear in $\sigma$ due to $c_\sigma \neq 0)$ in the regime $\sigma<\langle\sigma\rangle_\chi$ where $\langle\sigma\rangle_\chi$ is the solution of the equation of motion \labelcref{eq:EoMsigma}. This also severely restricts the $\rho$-dependence of $h(\rho)=h_k(\rho)$ as well as both flows. 

The dynamical effects of the running Yukawa coupling can be extracted by a comparison to a setup with fixed Yukawa coupling, but an identical RG scale shift in order to separate the two effects. We compare our result to identical initial conditions in the simplified setup and also to a situation with vacuum physics adjusted to be identical to our flowing Yukawa coupling setup (see also \Cref{fig:transitions} and \Cref{fig:trans:mu0}).
Relative to the adjusted vacuum physics case, $T_c$ is marginally smaller when one uses a running Yukawa coupling, but identical when one chooses fixed initial conditions.

Furthermore, the inclusion of $h_k(\rho)$ does visibly diminish the back-bending at the high-density transition, but does not completely remove it. We remark in this context that this back-bending as well as the presence of a CEP is strongly regulator-dependent as well as parameter-dependent also in LPA, for a recent investigation see \cite{Otto:2022jzl}. This stresses the need for an even more sophisticated approximation than the self-consistent LPA in the regime of low temperatures, $T\lesssim 50-80$\,MeV and large chemical potentials, $\mu_q\gtrsim~200-300$\,MeV. A first step in this direction has been undertaken in \cite{Helmboldt:2014iya}, where the full momentum dependence of the propagators in the quark-meson model has been resolved at finite temperature and vanishing density. We hope to report on respective results in the phase structure of the quark-meson model at finite $T,\mu_q$ with DG methods soon.    

In order to fully appreciate the differences and similarities of the coupled dynamics of meson scattering and quark--antiquark meson scattering in self-consistent LPA and only meson scattering in LPA we provide a detailed collection of RG-time evolutions with and without the inclusion of higher order scatterings in \Cref{app:flow_quark_masses} and \Cref{app:flow_bos_masses}. Moreover, a collection of phase diagrams for $\gamma \in \{0.5,0.67,0.77,1.0\}$ can be found in \Cref{app:detail_PD}.

\section{Conclusions and Outlook}
\label{sec:sum}

In this work we investigated the phase structure of the quark-meson model in the self-consistent LPA. This approximation includes all scattering orders between two quarks and an arbitrary number of mesons included in a field-dependent Yukawa coupling $h(\rho)$, as well as the self-scattering of an arbitrary number of mesons included in the effective potential $V(\rho)$. On the technical side we have used an extended Discontinuous Galerkin method which allows for quantitative precision in field space. This method proved to be essential at high densities, where it also accommodates potentially relevant discontinuities in the flow. Indeed, higher scattering orders display non-analyticities which give large contributions to the RG-time evolution.

Crucially, this study necessitated a close examination of the relation between bosonic cutoff scales $k_\phi$ and fermionic cutoff scales $k_q$. Keeping these cutoff scales identical leads to a large difference of the physical quark and meson mass gaps, as can be seen in \Cref{fig:gamma_gaps}. This is due to the pions approaching the quasi-massless limit very quickly in the symmetry breaking regime. In this case one needs to accommodate large momentum transfers between $p^2\approx 0$ and $p^2 \approx k^2$, which cannot be obtained in a local potential approximation. Moreover, we have observed that the flow of the Yukawa coupling shows an unphysical divergence for identical cutoff scales $k_\phi=k_q$. We have evaluated the technical origin, see the discussion around \labelcref{eq:subtract_mq}, and it finally can be traced back to the missing momentum dependences of the self-consistent LPA.

As a solution, we adjusted the cutoff scales $k_q,k_\phi$ such that the physical gaps are kept close during the flow, see \Cref{sec:InitialCondOptimisedk}. The relative mass gap is controlled by a parameter $\gamma$: for $\gamma \rightarrow 0$ the quarks are integrated out first and one is led to an O(4)-type model. In turn, $\gamma \rightarrow \infty$ leads to the aforementioned large momentum transfers and one ends up within a simple approximation of an NJL-type model. 
Scanning the dependence of the vacuum physics on $\gamma$, we found a region of approximate stability in the quark mass with $0.4 \lesssim \gamma\lesssim 0.6$. We took $\gamma_\textrm{opt}=0.5$ as an approximate optimal choice, since it leads only to a small difference between the physical quark and meson mass gaps in the flow. This is explicitly shown in \Cref{fig:gammadep}.

At high densities a uniform choice of $\gamma$ is not fully sufficient and we see an increase in $m_q^2$ at higher $\mu_q$, which signals the need for a more sophisticated approximation. This approximation has to take into account the relevant momentum regime of the quarks moving with the Fermi surface, which necessitates an approximation with full momentum dependences of the propagators and potentially also the vertices. 

For the choice $\gamma=0.5$, that is optimised for small chemical potentials, we found that the inclusion of the running of the Yukawa coupling shifts both the high-temperature as well as the high-density transition to slightly lower values. Furthermore, both the adjustment of $\gamma$ as well as the inclusion of the higher order quark-meson scatterings lead to decrease of the back-bending behaviour commonly seen in LPA. 

We conclude from this study that even the self-consistent local potential approximation is not sufficient for quantitative precision at high densities and low temperatures within the quark-meson model and, by extension, also for a functional QCD setup with dynamical hadronisation. However, the resolution of this intricacy simply necessitates the inclusion of momentum dependences in propagators and vertices, which has been already achieved very successfully in the vacuum and at finite temperature. Moreover, a sufficiently advanced truncation can also accommodate large momentum transfers 
and the optimisation of the relative cutoff scales is not required anymore. Indeed, this has been already tested in two-flavour QCD computations in the vacuum in \cite{Cyrol:2017ewj}. There, the results did not depend on a rescaling of the relative cutoff scales of gluon and matter sector. We emphasise however, that the application of approximation schemes such as LPA' or generally low-order derivative expansions at low temperatures and high densities requires a careful analysis of relative cutoff scale dependences for the systematic error control and hence the quantitative or even qualitative reliability of the results.

\begin{acknowledgments}

We thank Christopher Busch and Bernd-Jochen Schaefer for discussions and for sharing results on the impact of relative cutoff scales in the QM model in LPA prior to publication \cite{BuschSchaefer2023}. This work is done within the fQCD collaboration \cite{fQCD} and is funded by the Deutsche Forschungsgemeinschaft (DFG, German Research Foundation) under Germany’s Excellence Strategy EXC 2181/1 - 390900948 (the Heidelberg STRUCTURES Excellence Cluster) and the Collaborative Research Centre SFB 1225 - 273811115 (ISOQUANT).
FI acknowledges support by the Studienstiftung des deutschen Volkes.
FRS acknowledges funding by the GSI Helmholtzzentrum f\"ur Schwerionenforschung.
NW acknowledges support by the Deutsche Forschungsgemeinschaft (DFG, German Research Foundation) – Project number 315477589 – TRR 211 and by the State of Hesse within the Research Cluster ELEMENTS (Project ID 500/10.006).

\end{acknowledgments}


\appendix

\begingroup
\allowdisplaybreaks

\section{$\mu_q$-dependence of relative cutoff scales}
\label{app:mu_dep_gamma}

We have investigated whether the quark-mass plateau found in \Cref{sec:InitialCondOptimisedk} from which we extracted $\gamma_\mathrm{opt} = 2$ can be extended to higher $\mu_q$. The situation at a high chemical potential, $\mu_q = 280\,\text{MeV}$ is shown in \Cref{fig:gamma_optimal_mu}, where we see that qualitatively the asymptotic regimes, i.e.~the NJL limit for $\gamma\rightarrow0$ and the O(4) limit for $\gamma\rightarrow\infty$, do not change. The relatively stable region around $\gamma_\mathrm{opt} = 2$ gets however a larger $\gamma$-dependence and we also find a higher quark mass at high densities, which we do not expect from physical QCD.

\begin{figure}[t]
	\includegraphics[width=1\linewidth]{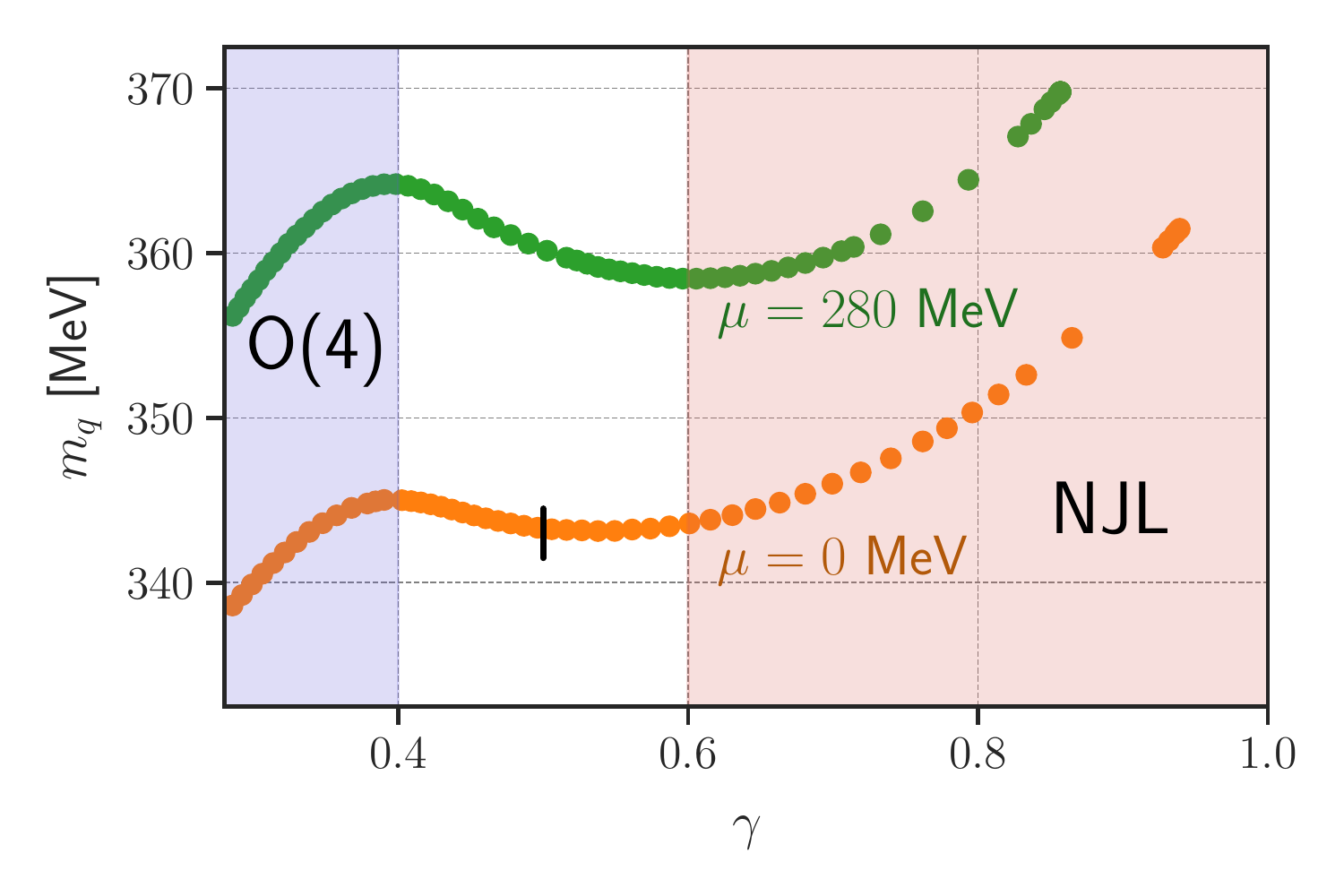}
	\caption{The quark mass as a function of the relative cutoff parameter $\gamma$ in the vacuum and at $\mu_q = 280\,\text{MeV}$. The previously relatively stable region for intermediate $\gamma$ gains a slope in the intermediate region at higher densities. Furthermore, the quark mass increases, signaling a breakdown of the model.}
	\label{fig:gamma_optimal_mu}
\end{figure}

The $\gamma$-dependence is still relatively mild (about $5\,\text{MeV}$ variation in the intermediate regime), but certainly larger, which can be explained due to the additional dynamics introduced by the Fermi-surface that also influence the relevant quark momentum scales - fluctuations of the quark should be either integrated out around the Fermi surface, as done in \cite{Braun:2020bhy}, or higher momentum dependences should be taken into account to include the relevant momentum transfers.
\begin{figure}[t]
	\centering
	\includegraphics[width=1\linewidth]{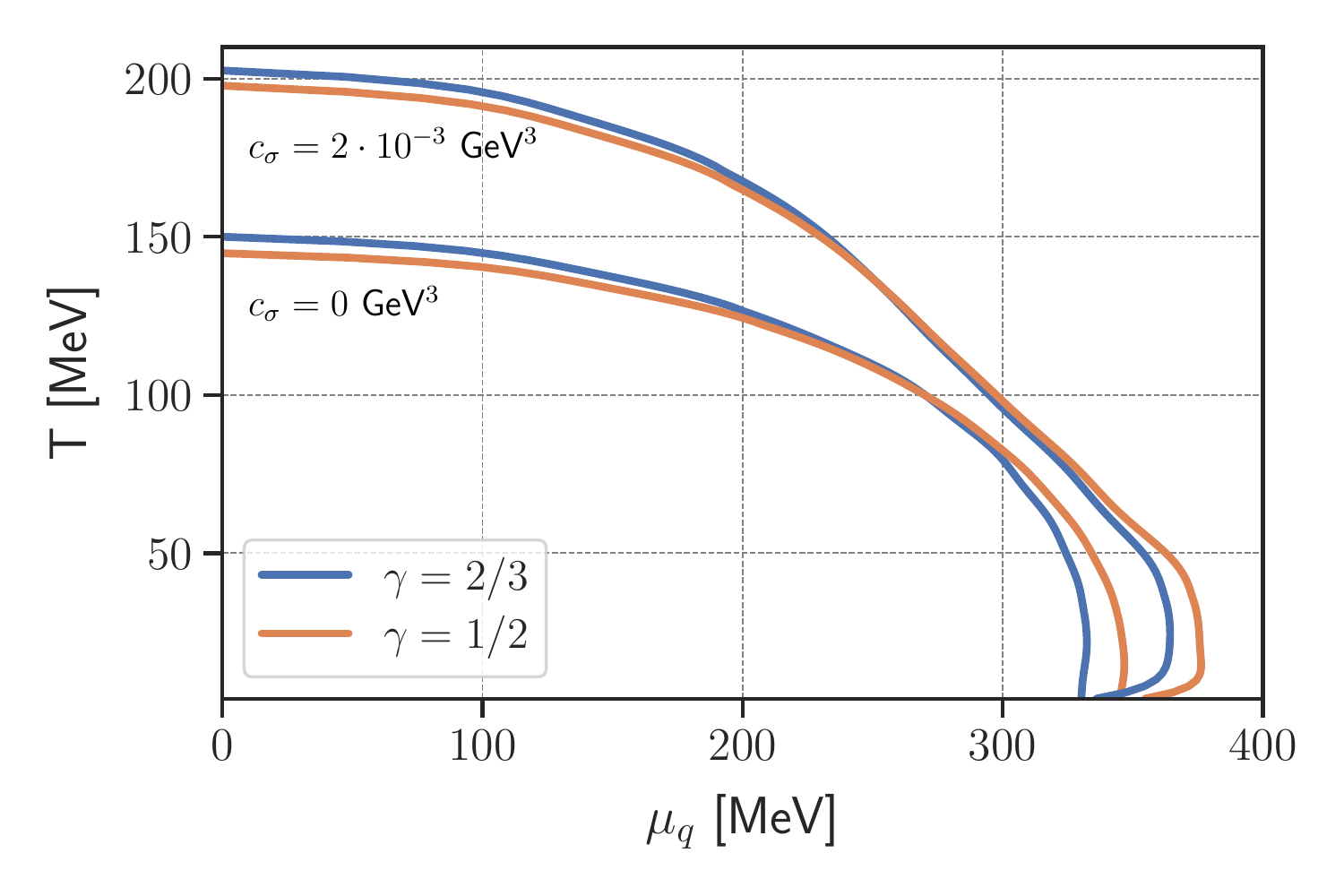}
	\caption{Comparison of the chiral phase boundaries for $\gamma={2}/{3}$ (\textit{blue}) and $\gamma={1}/{2}$ (\textit{orange}) for vanishing and large explicit symmetry breaking.}
	\label{fig:exsym}
\end{figure}
\begin{figure*}[ht]
	\centering
	\begin{subfigure}{0.5\textwidth}
		\includegraphics[width=1\textwidth]{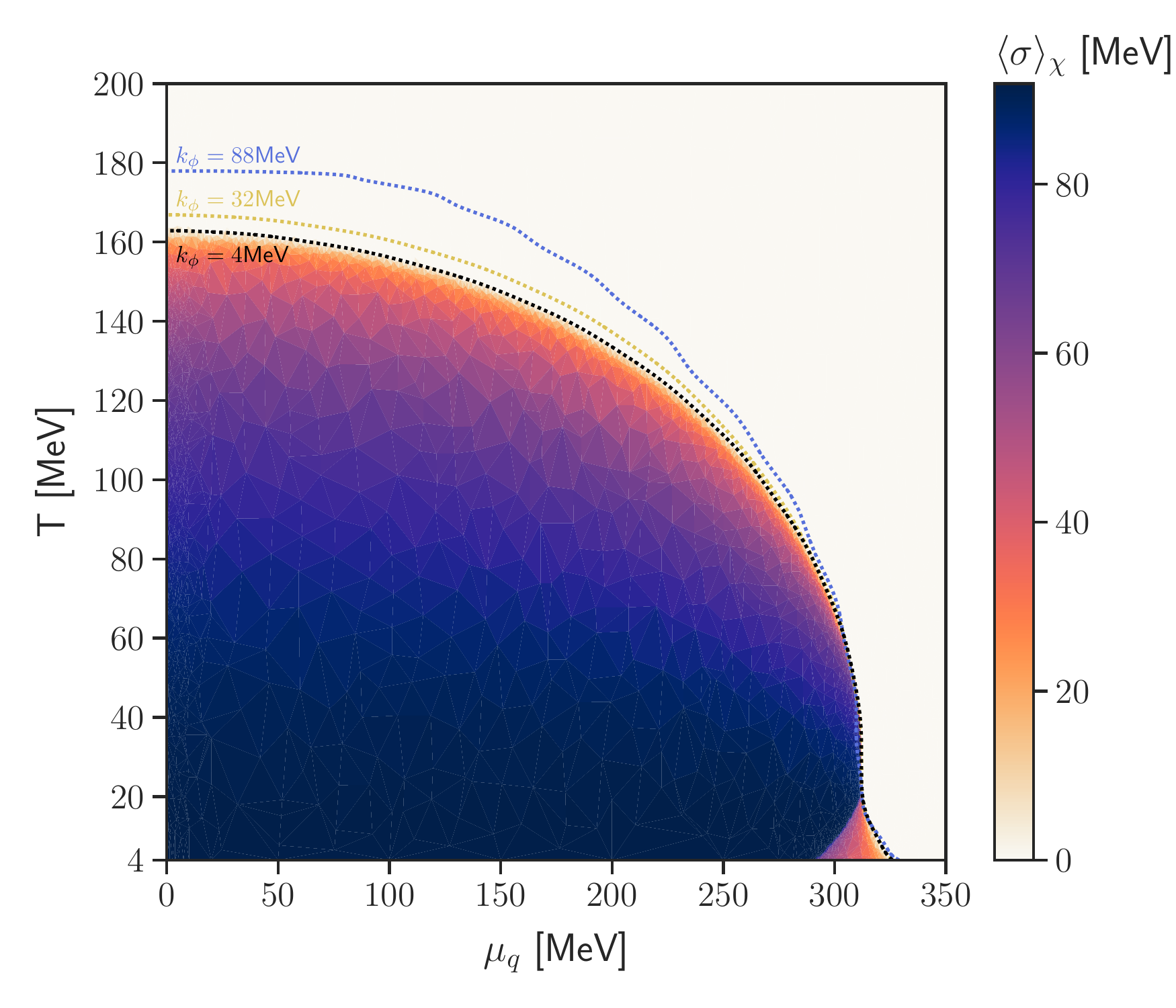}
		\caption{$c_\sigma=0$.}
	\end{subfigure}\hfill%
	\begin{subfigure}{0.5\textwidth}
		\includegraphics[width=1\textwidth]{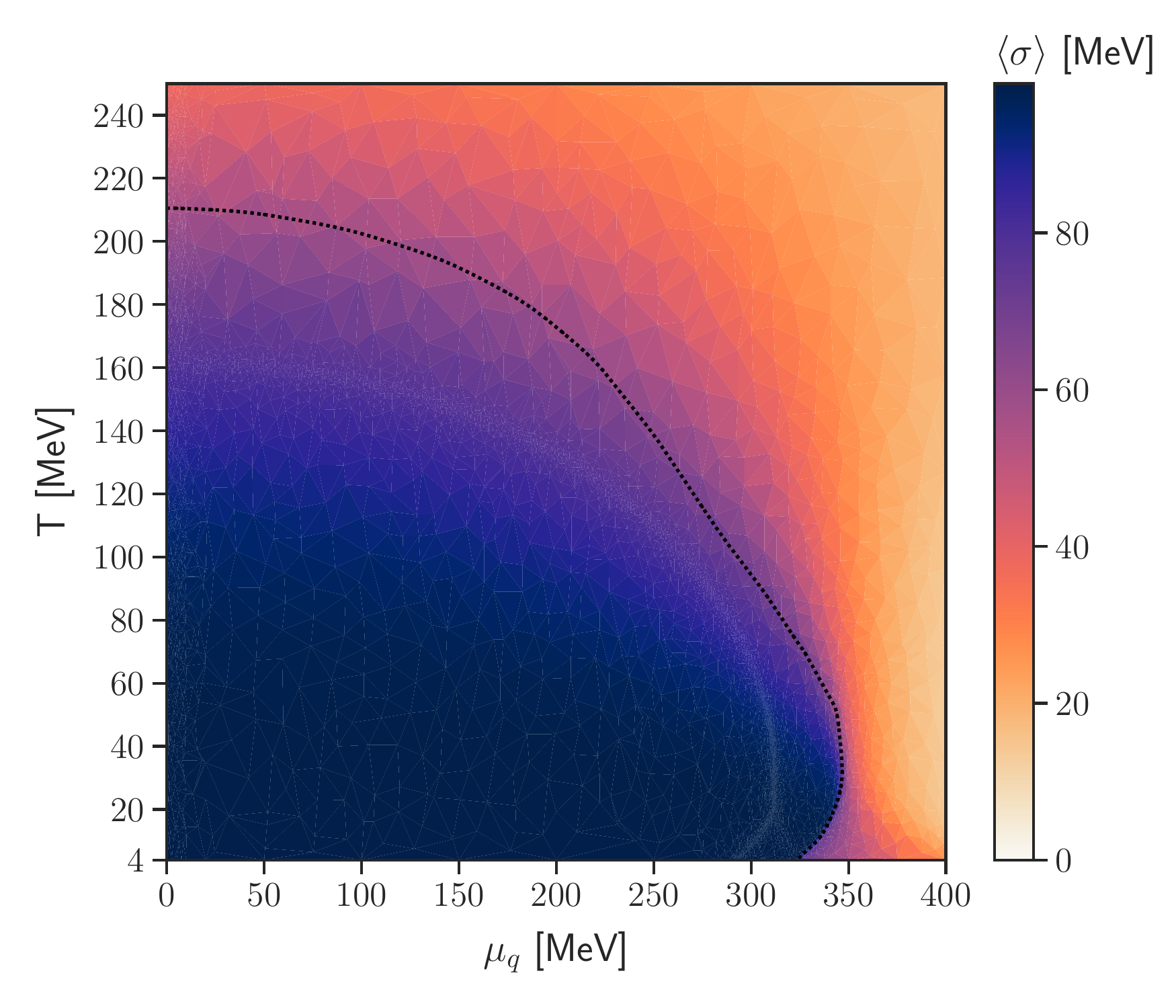}
		\caption{$c_\sigma=1.89\cdot10^{-3}\,\text{GeV}^3$.}
	\end{subfigure}
	\caption{Phase diagram of the quark-meson model without higher quark-meson scatterings (LPA) in the chiral limit (a) and with explicit chiral symmetry breaking with $c_\sigma=1.89\cdot10^{-3}\,\text{GeV}^3$ (b). We show a heat map (blue: large $\langle \sigma\rangle$, orange: small $\langle \sigma\rangle$) of the condensate expectation value $\langle \sigma\rangle $ defined in \labelcref{eq:EoMphiGen} in MeV.  We also depict the phase boundary for different RG-scales $k$ in the chiral case.}
	\label{fig:QM_PD}
\end{figure*}
%

\section{Explicit symmetry breaking and $\gamma$-dependence}%
\label{app:exsym}

As mentioned in \Cref{sec:results:QMY}, a change in $\gamma$ also changes the phase diagram of the quark-meson model. We showcase this explicitly in \Cref{fig:exsym} for a chiral system as well as for a larger explicit symmetry breaking $c_\sigma = 2\cdot10^{-3}\,\textrm{GeV}^{-3}$. A decrease in $\gamma$ visibly pushes out the high-$\mu_q$ phase boundary and in turn decreases back-bending in the chiral setup. At the same time, a decrease in $\gamma$ lowers $T_c$ in both cases.

\section{Fixed Yukawa coupling}
\label{app:QM}

\begin{figure*}[t]
	\centering
	\begin{minipage}[t][1.2cm]{.48\textwidth}
		\centering
		\includegraphics[width=1\linewidth]{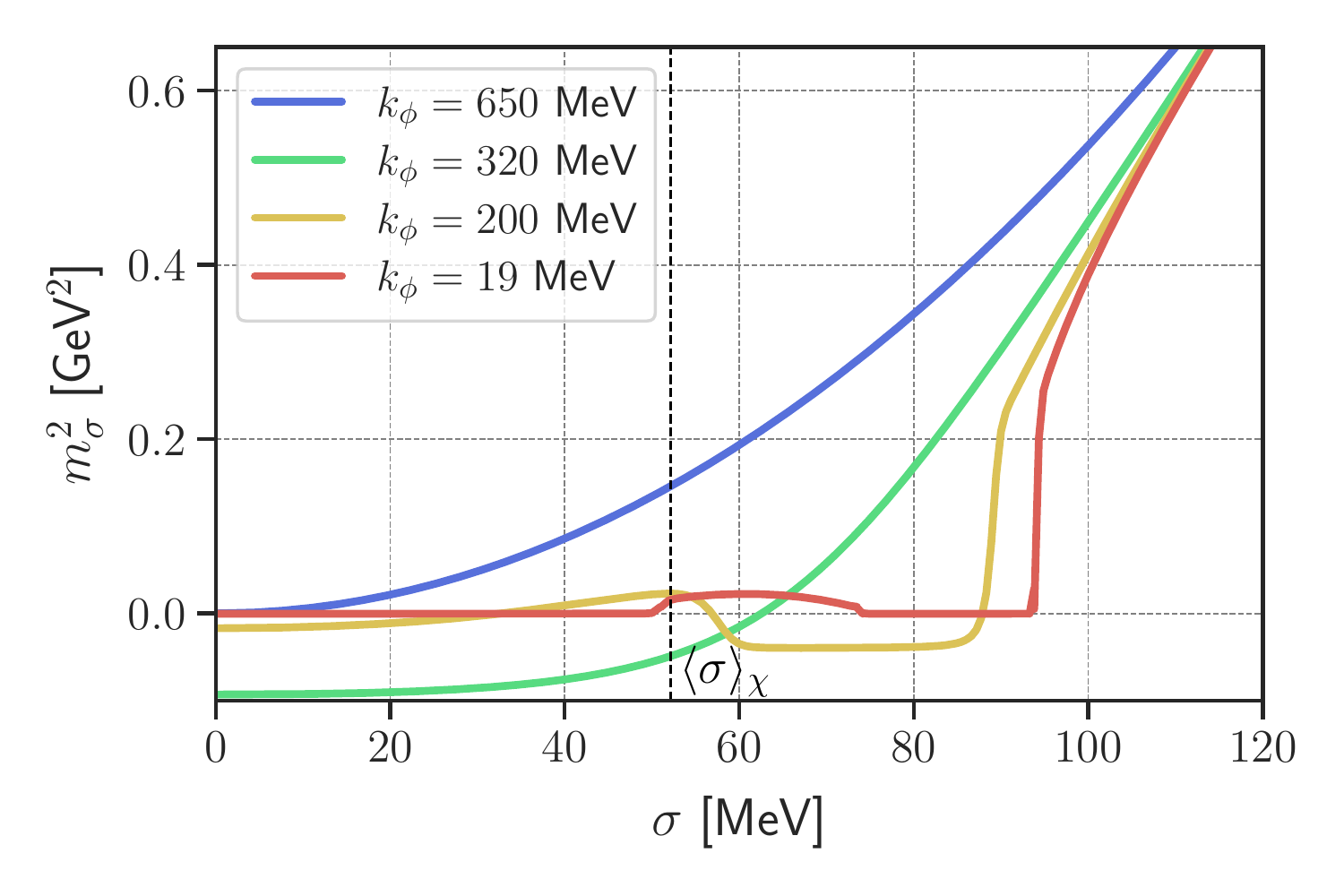}%
		\caption{Build-up of a discontinuity in the derivative term $\partial_\rho m_\pi^2$, i.e.~a shock-like structure in $m_\sigma^2$ at $(T, \mu_q) =~(4,300)$ MeV.}%
		\label{fig:sigma_shock}
	\end{minipage}%
	\begin{minipage}[t][1.2cm]{.04\textwidth}
		\hfill
	\end{minipage}%
	\begin{minipage}[t][1.2cm]{.48\textwidth}
		\centering
		\includegraphics[width=0.95\linewidth]{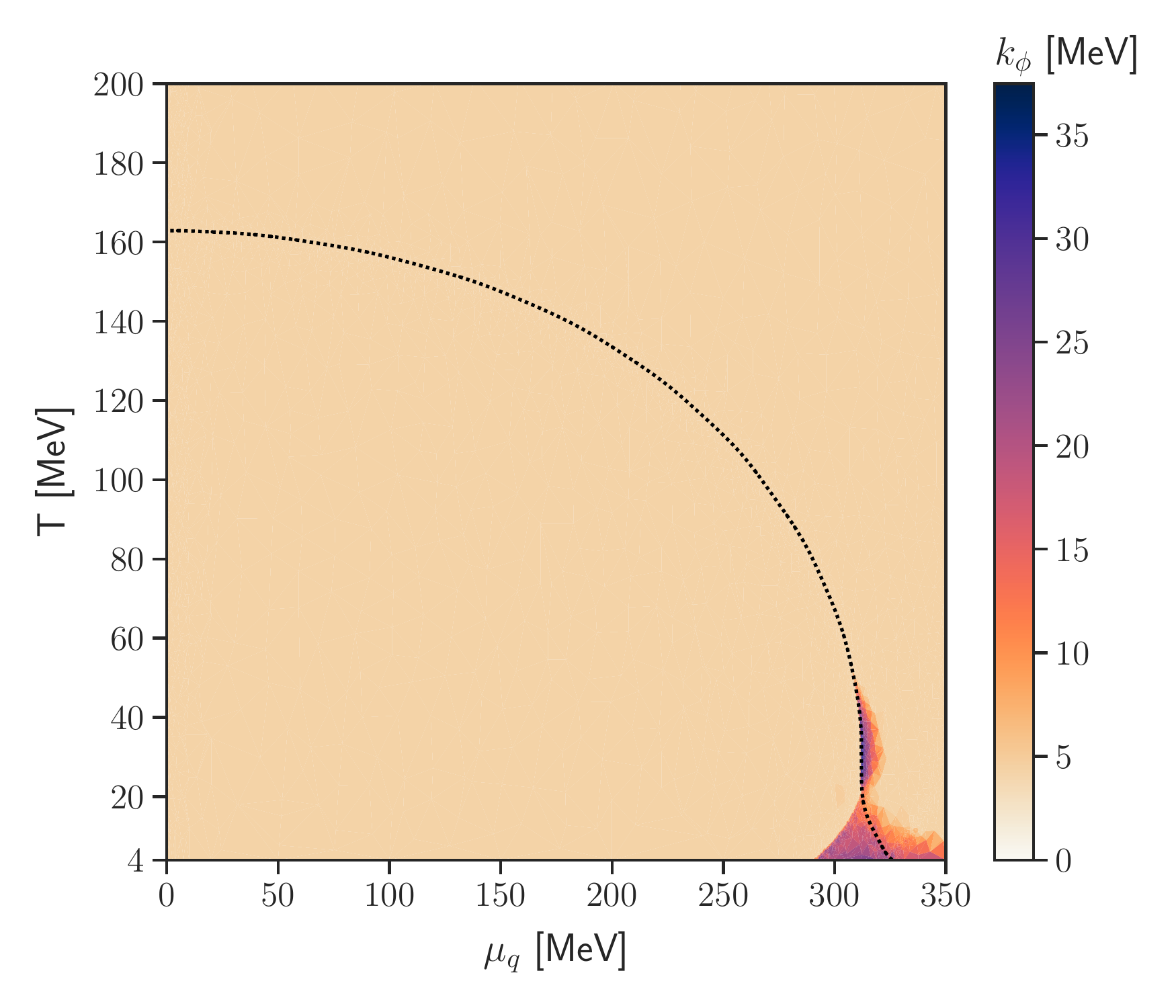}%
		\caption{The convergence diagram of the quark-meson model without higher quark-meson scatterings.}%
		\label{fig:QM_LPA_convergence}
	\end{minipage}
\end{figure*}

In this Appendix we compute the phase diagram of the quark-meson model in LPA with a fixed Yukawa coupling $h_k(\rho)=h$. In particular, the comparison with the results in the present systematically improved approximation provides a non-trivial reliability estimate and hence non-trivial support for the LPA results in the literature at sufficiently large temperature. 

Using this approximation, the flow equations are equivalent to the setting in \cite{Schaefer:2006ds, Grossi:2021ksl, Tripolt:2013jra}, but evaluated using the elaborate numerical set-up outlined in \Cref{sec:LDG}.

We put special emphasis on the evaluation of the phase transition regime for large quark chemical potential $\mu_q$ and low temperature~$T$. In this regime, the current results are the first ones obtained a numerical method that has the capacity of capturing the intricate dynamics of competing order effects and potential shock development. The possibility of shock development in proximity to first order phase transition has previously been pointed out in \cite{Grossi:2019urj, Grossi:2021ksl, Ihssen:2022xkr}. Importantly, the current setup is the first with the ability to confirm or exclude the presence of shocks for a given set of model parameters. A thorough scan for shock development in sets of different initial conditions is deferred to future work.

We use the initial conditions from \cite{Grossi:2021ksl} which are tuned to physical parameters for $\gamma=1$, 
\begin{align}
	\Lambda = 650\,\text{MeV}\,,\qquad
	\lambda_\Lambda = 71.6\,,\qquad
	h = 7.2 \,.
	\label{eq:InitialParametersQM}
\end{align}
The corresponding phase diagram with $\gamma=1$ is shown in \Cref{fig:QM_PD} both in the chiral case as well as for a small explicit quark mass set through $c_\sigma=1.89\cdot10^{-3}\,\text{GeV}$. This choice of explicit symmetry breaking $c_\sigma$ has been made by tuning to a physical pion mass $m_\pi^2 = 138\,\text{MeV}$. We are able to integrate the flows down to $k\approx4\,\text{MeV}$, an explicit convergence diagram of the results is shown in \Cref{fig:QM_LPA_convergence}, see also \Cref{app:detail_PD} for other observables.

In the high-density region we find that below $T \approx 20 \text{MeV}$ the transition line splits into a first order phase transition which exhibits back-bending, i.e.~moves to smaller $\mu_q$ at decreasing temperature, and a second order line. The second order transition bends in the opposite direction towards larger $\mu_q$. This has been seen before e.g.~in \cite{Tripolt:2013jra}. It is strongly regulator-dependent though, see \cite{Otto:2022jzl}. In the case of explicit chiral symmetry breaking with $c_\sigma = 1.89\cdot10^{-3}\,\text{GeV}^3$, we also see that the second-order transition turn into a crossover. For a more detailed view, we refer to \Cref{app:PD_lines}, where transition lines along fixed $\mu_q$ and $T$ are plotted separately. 

We find that the second order transition is correlated with shock-like structures in the sigma mass $m_\sigma^2(\rho)$. We show such a structure in \Cref{fig:sigma_shock} at $(\mu_q,T)=(300,4)$ MeV, within the region between the two transition lines. After a build-up and travel towards larger $\langle\sigma\rangle$ it freezes out. Then, its tail is pushed down again, so that some residual symmetry breaking remains which is however smaller compared with the case without the shock.

We comment further, that these discontinuities get washed out in the case of $\gamma=0.5$, as can be seen in the mass plots of \Cref{app:flow_bos_masses}.

With the initial conditions \labelcref{eq:InitialParametersQM} we obtain in the chiral limit and approximate vacuum, $(\mu_q, T) = (0,4)\,\text{MeV}$,  
\begin{align}
	\langle\sigma\rangle = 91.7\,\text{MeV}\,,\qquad m_q = 330\,\text{MeV}\,,\notag	\\[1ex] 
	m_\sigma = 176\,\text{MeV}\,.
\end{align}
For vanishing chemical potential the chiral phase transition temperature in the chiral limit is given by 
\begin{align}
	T^\chi_c = 162.9 \,.
\end{align}

When choosing an explicit symmetry breaking of $c_\sigma = 1.89\cdot10^{-3}\,\text{GeV}^3$ we get from the same data at $(\mu_q, T) = (0,4)\,\text{MeV}$, condensate, quark mass and the pion mass as
\begin{align}
	\langle\sigma\rangle = 99.0\,\text{MeV}\,,\qquad  m_q = 357\,\text{MeV}\,,\notag \\[1ex] 
	m_\pi = 138\,\text{MeV}\,,\qquad m_\sigma = 606\,\text{MeV}
	\,.
\end{align}
%
	
\section{Numerical details}
\label{app:numdet}

The LDG method described in \Cref{sec:LDG} has been implemented in the C++ finite elements library \texttt{deal.II} \cite{dealII94}.
All phase diagrams have been computed using the \texttt{adaptive} python library \cite{Nijholt2019}, which efficiently computes the diagram while ensuring that all features are sufficiently detailed.

The discretisation in field space is done using Legendre polynomials up to order $P = 4$ in every cell. 
Accordingly, we perform all integral calculations by employing a Gauss-Legendre quadrature of order $P+1$, which integrates polynomials of order $P$ exactly. Increasing this order was not found to be of any difference to the convergence or results of the simulations.

We specify the vertices of the $\rho$-grid as
\begin{align}
	&[0:\num{1e-5}:\num{5e-4}] \cup [\num{5e-4}:\num{5e-5}:\num{3e-3}] \notag\\[1ex]
	&\cup [\num{3e-3}:\num{1e-5}:\num{4.2e-3}] \cup [\num{4.2e-3}:\num{5e-5}:\num{5e-3}] \notag\\[1ex]
	&\cup [\num{5e-3}:\num{1e-4}:\num{1e-2}] \cup [\num{1e-2}:\num{1e-3}:\num{1.5e-2}]\,,
\end{align}
where we used a python-like slice notation, i.e.~
\begin{equation*}
	[a:b:c] = \{ a + nb : n \in \mathbb{N}_0, \, a + nb \leq c\}\,.
\end{equation*}
We have therefore 291 cells and 1455 degrees of freedom per simulation for $m_q^2(\rho)$ and $m_\pi^2(\rho)$ each.

We also have used an uniform value of $\alpha = 2.5\times10^{-8}$, as introduced in \Cref{sec:LE-QCD}, for the flows of both $m_\pi^2$ and~$h(\rho)$.

Regarding time-stepping, at any time-step an assembly of the residual (i.e.~the evaluation of \labelcref{eq:residual}) involves first the computation of all projections $g_\eta,\,w_{\eta,\xi}$. These projections are in the current case linear in $u$ and can be thus performed through matrix multiplication, i.e.
\begin{equation}
	g_\eta = \frac{\partial g_\eta}{\partial u} u\quad\text{and}\quad w_{\eta\xi} = \frac{\partial w_{\eta\xi}}{\partial u} u\,,
\end{equation}
which is much faster than the explicit construction of $g_\eta$ and $w_{\eta\xi}$. We therefore store the matrices $\frac{\partial g_\eta}{\partial u}$ and $\frac{\partial w_{\eta\xi}}{\partial u}$ which only have to be computed once.

Similarly, the evaluation of the Jacobian involves computation of all nested Jacobians for all equations. As \texttt{SUNDIALS} uses a lazy Jacobian update, this costly calculation can be done relatively few times during the time integration. However, Jacobians and their inverses are evaluated exactly, where the inversion is done using \texttt{UMFPACK}~\cite{UMFPACK} and the derivatives are performed utilizing automatic differentiation via the C++ \texttt{autodiff} library \cite{autodiff}.

\section{Regulators and threshold Functions}
\label{app:thfkt}

In the present work we use the 3-dimensional flat or Litim regulator, see  \cite{Litim:2000ci}. We perform shifting of the RG scales of bosons and fermions relative to each other. This accounts to choosing two RG scales
\begin{align}
	k_\phi = \Lambda e^{\gamma t}\,,\qquad
	k_q = \Lambda e^{t}\,,
\end{align}
which are both parameterized in terms of the same RG-time $t$ but shifted using $\gamma > 0$.
Furthermore, by noting that
\begin{equation}
	\partial_t = \gamma\, k_\phi \partial_{k_\phi} = k_q \partial_{k_q}\,
\end{equation}
we express all results in terms of the same RG-time. The flat bosonic regulator is given by,   
\begin{align}\label{eq:regphi} 
	\hspace{-.2cm} R_\phi =&\, \mathbf{p}^2 \,r_\phi(x_\phi)\,,\quad r_\phi(x_\phi) = \,\left(\frac{1}{x_\phi}-1\right)\theta(1-x_\phi)\,,
\end{align}
and the fermionic one reads 
\begin{align}
	R_q=&\,  \slashed{\mathbf{p}}  \, r_q(x_q)\,, \quad \hspace{.2cm}  r_q(x_q) =  \left(\frac{1}{\sqrt{x_q}}-1\right)\theta(1-x_q)\,, 
	\label{eq:regq}
\end{align}
with 
\begin{align}
x_\phi=\mathbf{p}^2/k_\phi^2\,,\qquad x_q=\mathbf{p}^2/k_q^2 \,.
\end{align}
In the following we use the scalar part of the propagators to calculate the momentum loops,
\begin{align}
	G_{\phi,n} &= \frac{k_\phi^2}{\omega_n^2 + k_\phi^2 \, x\,\left[1+r_\phi(x)\right] + m_\phi^2} \,,\notag \\[1ex]
	G_{q,n} &= \frac{k_q^2}{(\nu_n+i\mu_q)^2 +k_q^2 \, x\,\left[1+r_q(x)\right]^2 +  m_q^2} \,.
\end{align}
The Matsubara frequencies read $\omega_n = 2 n \pi T $, with $n \in \mathbb{Z}$ for the Bosons and $\nu_n = (2n +1) \pi T $ for the Fermions.

Using the definition of the regulators, we can evaluate the momentum-loop integration and Matsubara summation for a given temperature $T$ and quark chemical potential $\mu_q$. We obtain the finite temperature threshold-functions for the 3d flat regulator, which we state for completeness in the following. Firstly, the threshold functions $l_0^{(B/F,d)}$ are given by simple bosonic/fermionic loops in $d$ dimensions, 
\begin{align}
	l_0^{(B,d)}
	&= \frac{T}{2k_\phi} \sum_{n \in \mathbb{Z}} \int_{x_\phi} x_\phi^{\frac{d-1}{2}} \partial_t r_\phi(x_\phi) G_{\phi,n}(m_\phi^2)\notag\\[1ex]
	&= \gamma\,\frac{k_\phi}{d-1} \frac{\coth \left(\frac{\epsilon_\phi}{2 T}\right)}{\epsilon_\phi}\,,
\end{align}
and
\begin{align}\notag
	l_0^{(F,d)}
	&= \frac{T}{k_q} \sum_{n \in \mathbb{Z}} \int_{x_q} x_q^{\frac{d-1}{2}} \partial_t r_q(x_q) G_{q,n}(m_q^2)\\[1ex]
	&= \frac{k_q}{2(d-1)} \frac{\tanh \left(\frac{\epsilon_q - \mu_q }{2 T}\right) + \tanh \left(\frac{\epsilon_q + \mu_q}{2 T}\right)}{ \epsilon_q} \,.
\end{align}
with the dissipation relation
\begin{align}
	\epsilon_\xi =\sqrt{k_\xi^2+{ m_\xi}^2}\,.
\end{align}
for the respective field $\xi \in \{\phi,q\}$. Note that we have used hyperbolic functions instead of the usual Bose-Einstein and Fermi-Dirac distributions in our notation. Since libraries make hyperbolic functions available up to very high numerical precision, using them makes for a more stable numerical implementation.

The functions $l_n^{(B/F,d)}$ correspond to diagrams with loops containing $n+1$ bosonic/fermionic propagator terms. They are obtained by taking a derivative with respect to $m^2$, 
\begin{align}\label{eq:ln}
	k^2 \, \partial_{m^2} l_n^{(B/F,d)} (m^2) = -(n + \delta_{n0}) \, l_{n+1}^{(B/F,d)} (m^2)\,.
\end{align} 
Furthermore, we consider mixed loops with a regulator insertion given by,
\begin{align} \nonumber
		L_{(1,1)}^{(d)}(m_q^2, m_\phi^2) =& \,\frac{T}{2k_\phi}\! \sum_{n\in\mathbb{Z}}\!\int\! dx_\phi \, x_\phi^{\frac{d-1}{2}} 
		\partial_t r_\phi(x_\phi)  G_{\phi,n}^2 G_{q,n} \\[1ex]
		&+\,\frac{T}{2k_q} \!\sum_{n\in\mathbb{Z}}\!\int dx_q \, x_q^{\frac{d-1}{2}} 2\left[ 1+r_q(x_q)\right]\nonumber\\[0.5ex]
		&\hspace{1.5cm}\times\partial_t r_q(x_q)G_{\phi,n} G_{q,n}^2\,.
		\label{eq:L11}
\end{align} 
This expression is not easily calculated analytically, as the regulator contributions in the propagators are now mismatched with the cutoff. This leads to some non-trivial integrals, which we solve in our simulations simply by evaluating \labelcref{eq:L11} numerically, i.e.~by performing the Matsubara sum up to order $n=20$ and using a Gauss-Legendre quadrature of order $20$ to perform the $p^2$-integral. We have checked that this yields sufficient precision, as we have observed no differences in the results up to numerical precision when doubling the order of both the quadrature and Matsubara sum.

\section{Additional plots}
\label{app:AdditionalPlots}

In the following, we present further detailed plots of the data used in the present work. This should facilitate the access to the details of our findings. 

In \Cref{app:PD_lines} we show the phase transition lines across fixed $\mu_q$ and $T$ for different $\gamma$ and compare the situation with and without flowing Yukawa coupling.

\Cref{app:flow_quark_masses} and \Cref{app:flow_bos_masses} showcase respectively the flows of quark and boson masses for different parts of the phase diagram.

Finally, in \Cref{app:detail_PD} further phase diagram plots are shown for different $\gamma$, with and without flowing Yukawa coupling.

\clearpage
\onecolumngrid
\subsection{Phase transition plots}%
\label{app:PD_lines}%
\begin{figure}[h]%
	\centering
	\begin{tabular}{c c}	
			\begin{subfigure}[b]{0.45\textwidth}
				\includegraphics[width=0.91\textwidth]{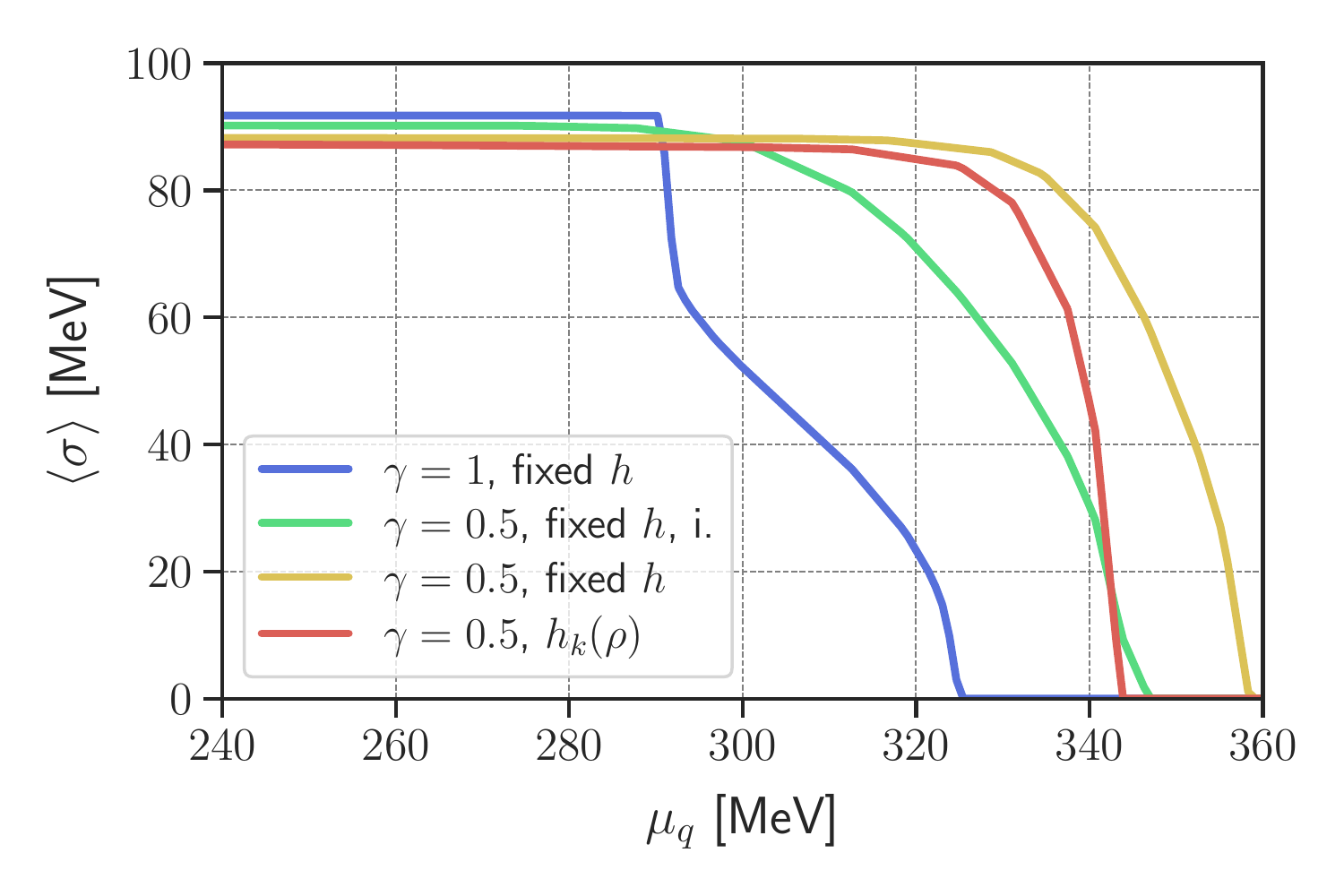}
				\includegraphics[width=0.91\textwidth]{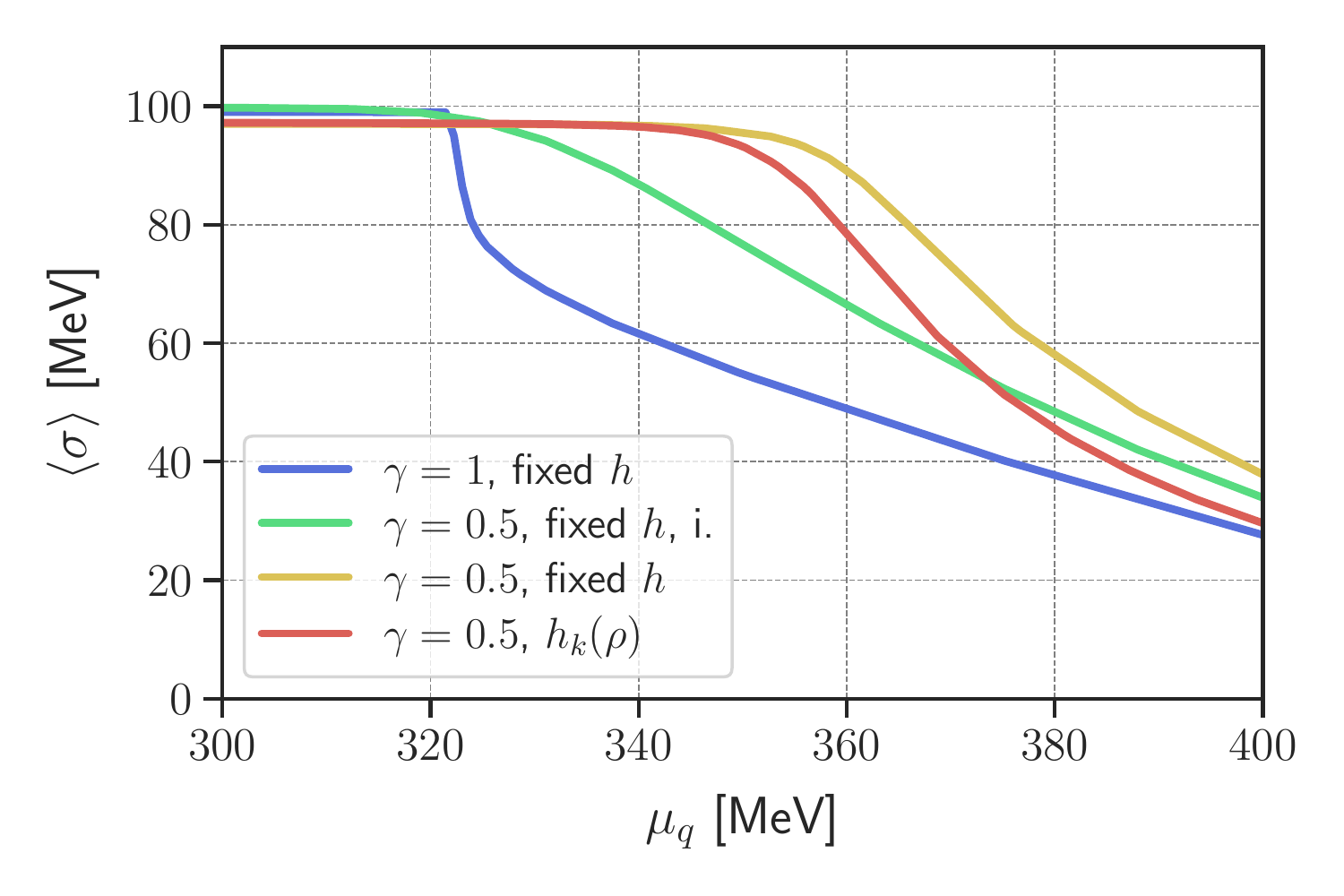}
				\caption{$\text{T}=4\,\text{MeV}$}
				\label{fig:trans:T0}
			\end{subfigure} &
			\begin{subfigure}[b]{0.45\textwidth}
				\includegraphics[width=0.91\textwidth]{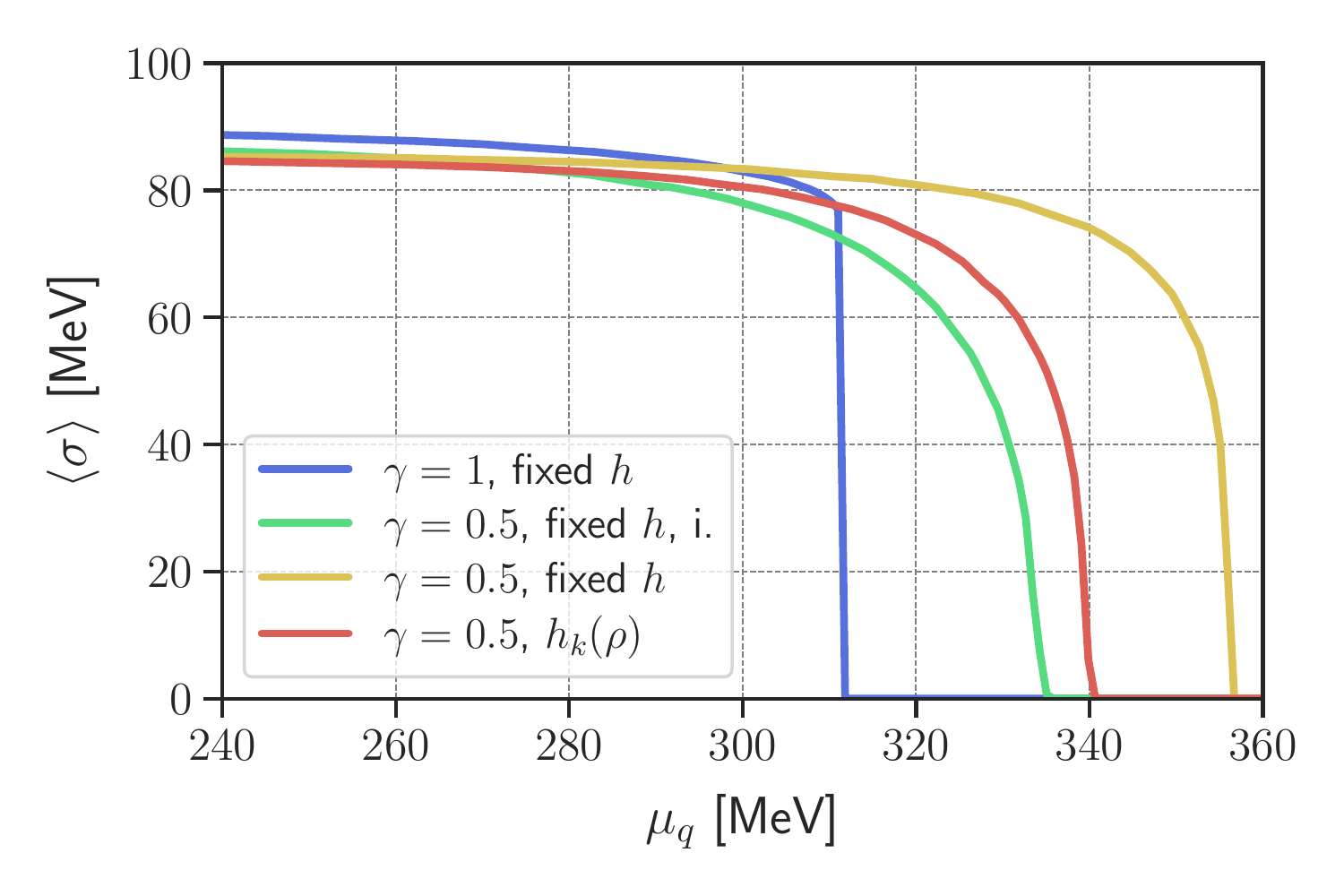}
				\includegraphics[width=0.91\textwidth]{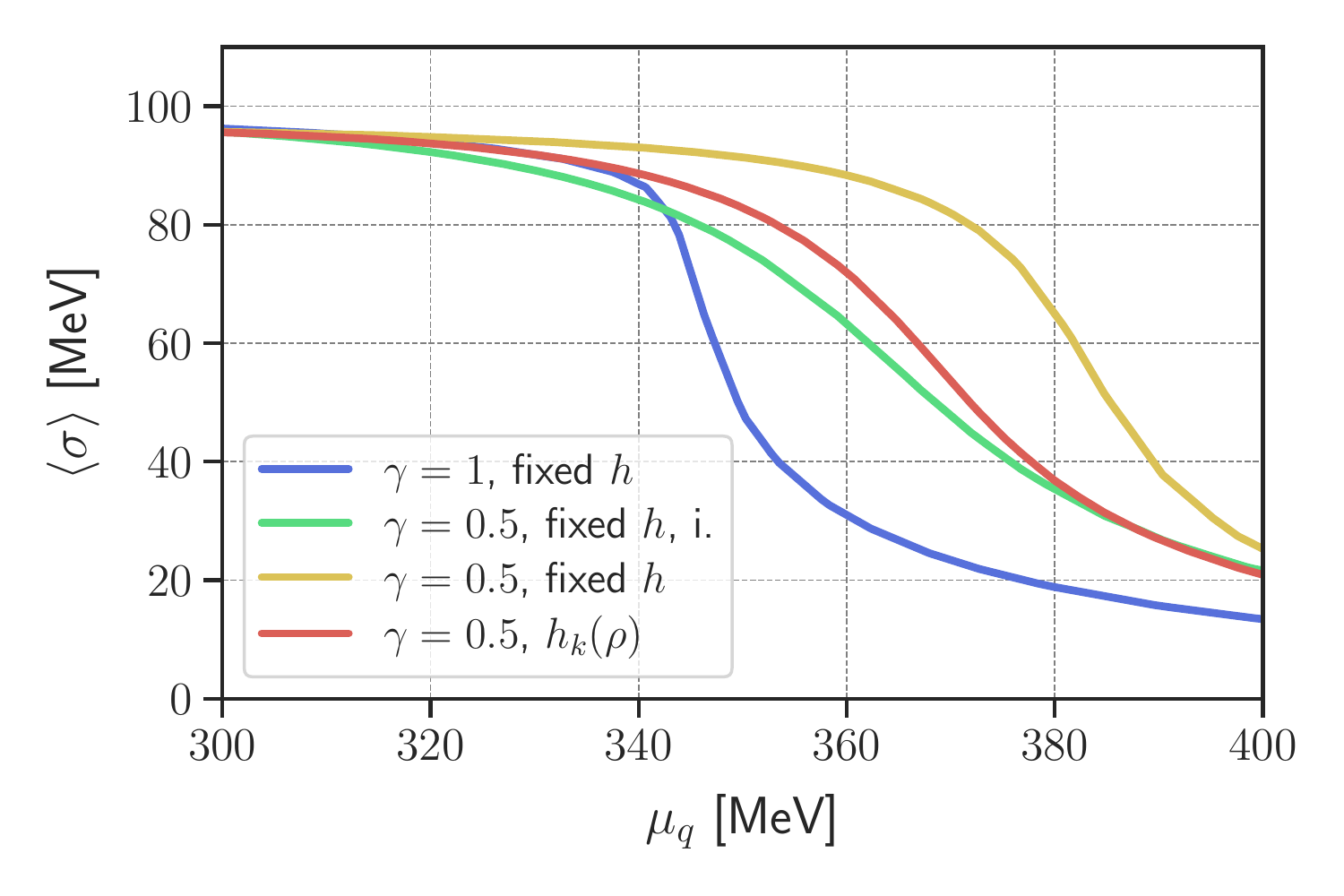}
				\caption{$\text{T}=40\,\text{MeV}$}
				\label{fig:trans:T1}
			\end{subfigure}
	\end{tabular}
	\caption{Chemichal potential dependence of the condensate $\langle \sigma \rangle$ defined in \labelcref{eq:EoMphiGen} for temperatures $T=4,40$\,MeV and different $\gamma$. The upper row is taken in the chiral limit with $c_\sigma=0$, whereas the lower ones show the case with explicit symmetry breaking, where we have tuned every $c_\sigma$ on a pion mass of $m_\pi^2=138\,\textrm{MeV}$. We show results obtained with  $\gamma=1$ and $\gamma=0.5$ with (scLPA) and without (LPA) running Yukawa coupling. The \texttt{i.} refers to simulations where the initial condition has not been tuned to physics but identified with the one used for scLPA. Furthermore, $c_{\textrm{LPA},\sigma}(\gamma=0.5)=1.848\cdot10^{-3}\,\textrm{GeV}^3$ and all other $c_\sigma$ are set as noted in \Cref{sec:results:QMY} and \Cref{app:QM}.}
	\label{fig:transitions}
		
		\begin{subfigure}[b]{0.48\textwidth}
			\includegraphics[width=1\linewidth]{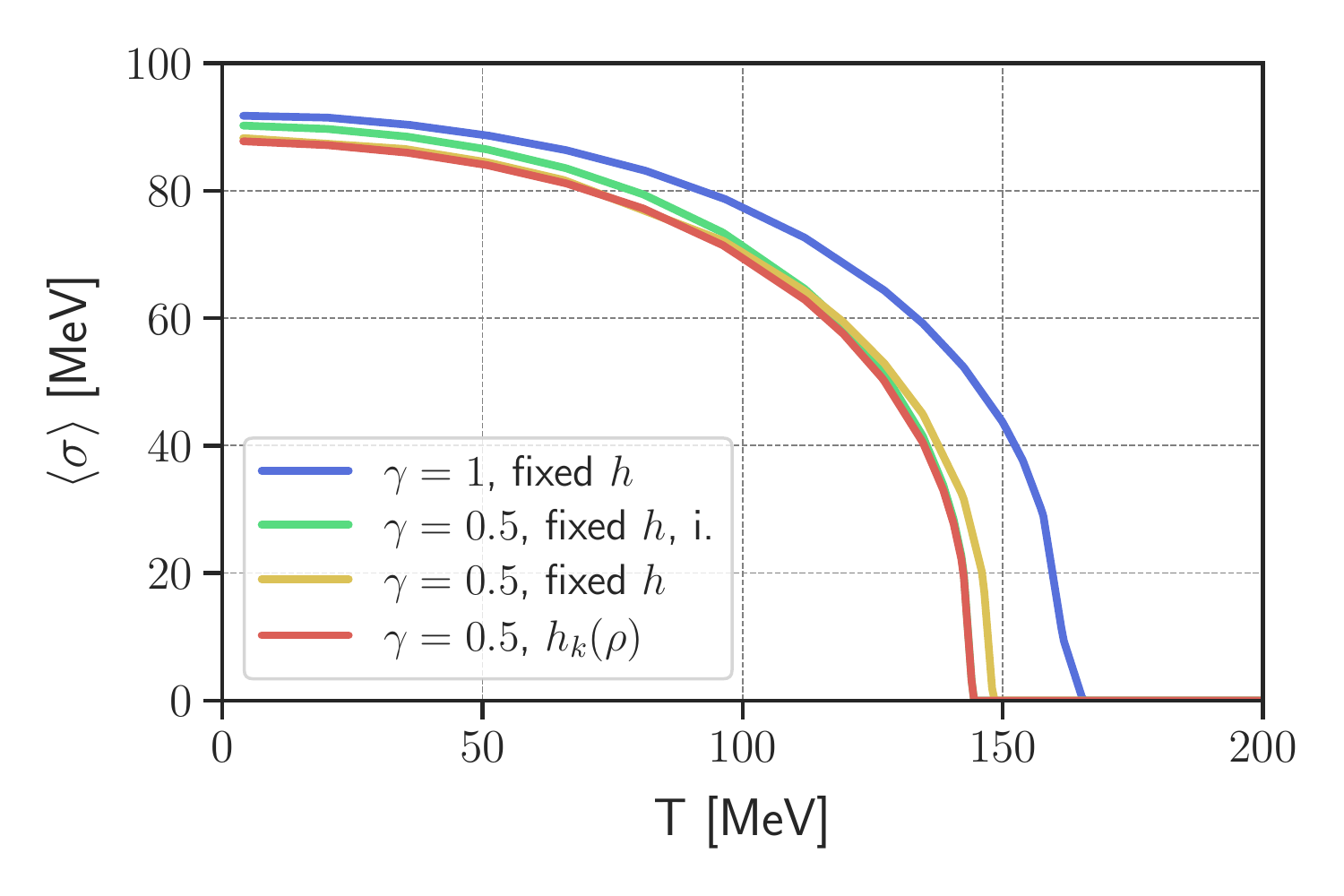}
			\caption{Chiral phase transition with $c_\sigma=0$ (chiral limit).\\\hfill}	
		\end{subfigure}
		\begin{subfigure}[b]{0.48\textwidth}
			\includegraphics[width=1\linewidth]{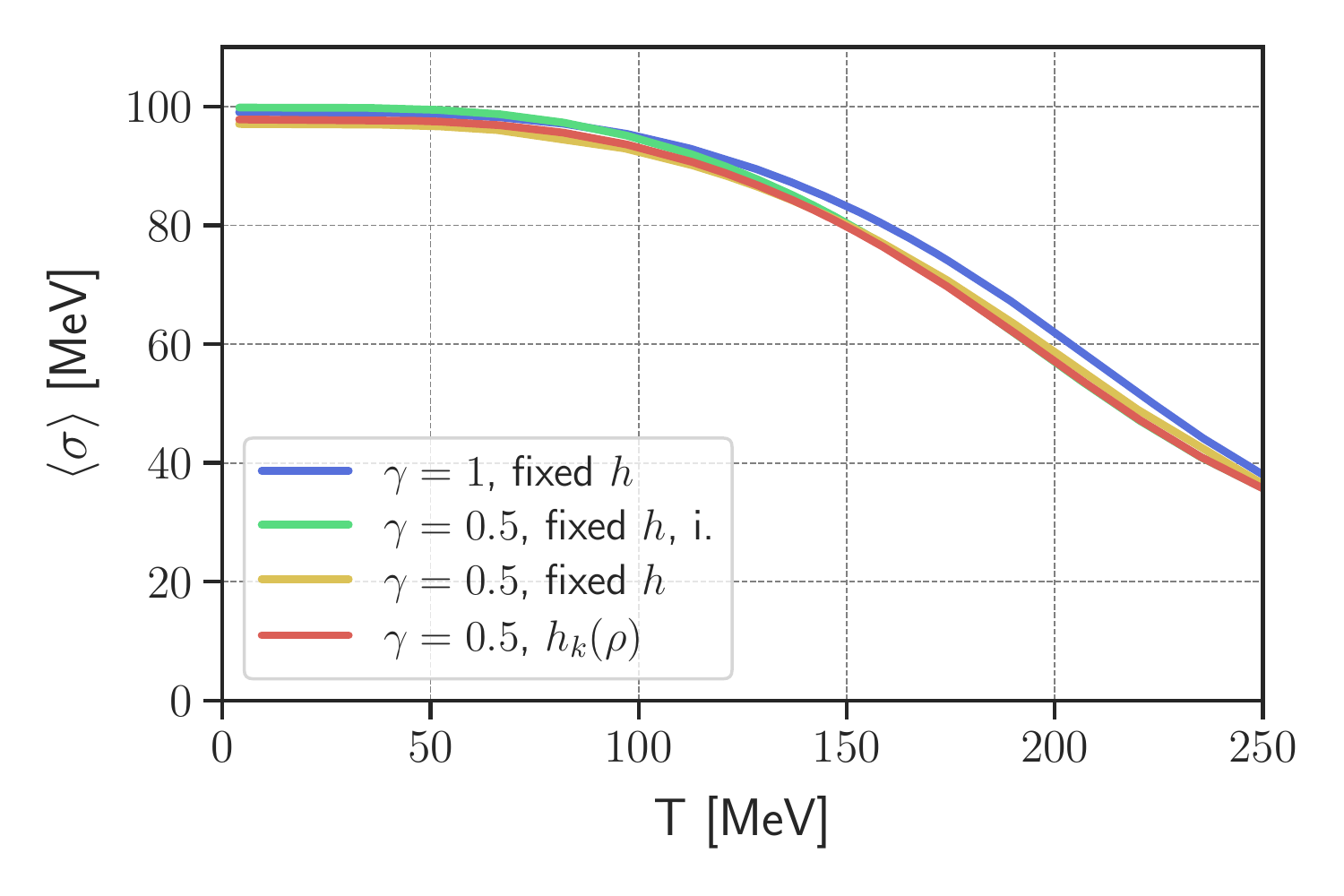}
			\caption{Chiral phase transition with explicit chiral symmetry breaking as in \Cref{fig:transitions}.}	
		\end{subfigure}
	\caption{Temperature dependence of the condensate $\langle \sigma \rangle$ defined in \labelcref{eq:EoMphiGen} for vanishing chemical potential, $\mu_q=0$. We show results in the quark-meson model at $\gamma=1$ without running Yukawa coupling and $\gamma=0.5$ with (scLPA) and without (LPA) running Yukawa coupling. The \texttt{i.} refers to simulations where the initial condition has not been tuned to physics but identified with the one used for scLPA.}
	\label{fig:trans:mu0}
	\vspace{-2cm}
\end{figure}

\clearpage
\subsection{RG-time evolution of quark masses}%
\label{app:flow_quark_masses}%
\begin{figure*}[h]
	\begin{subfigure}[b]{1\textwidth}
		\includegraphics[width=0.49\linewidth]{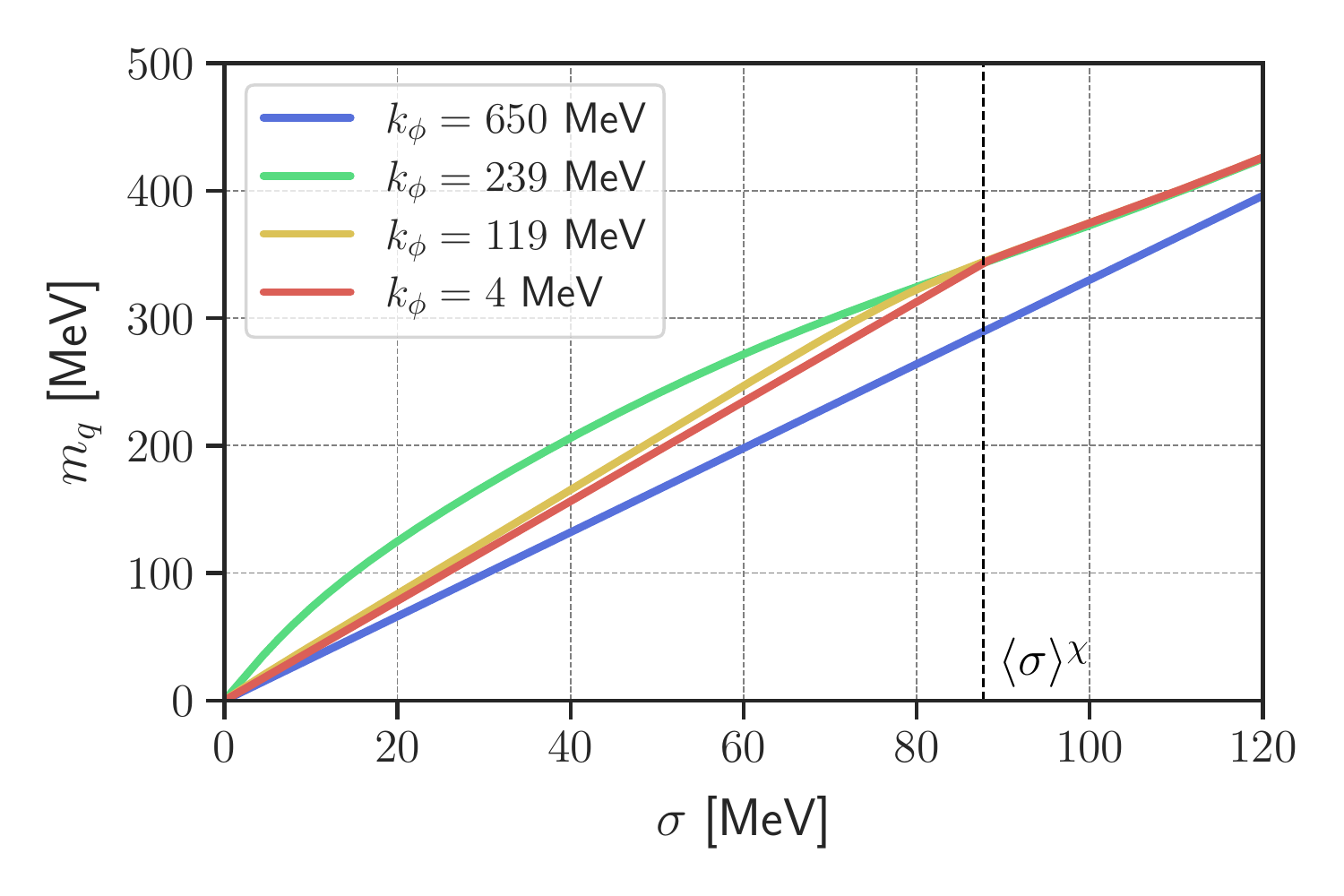}
		\includegraphics[width=0.49\linewidth]{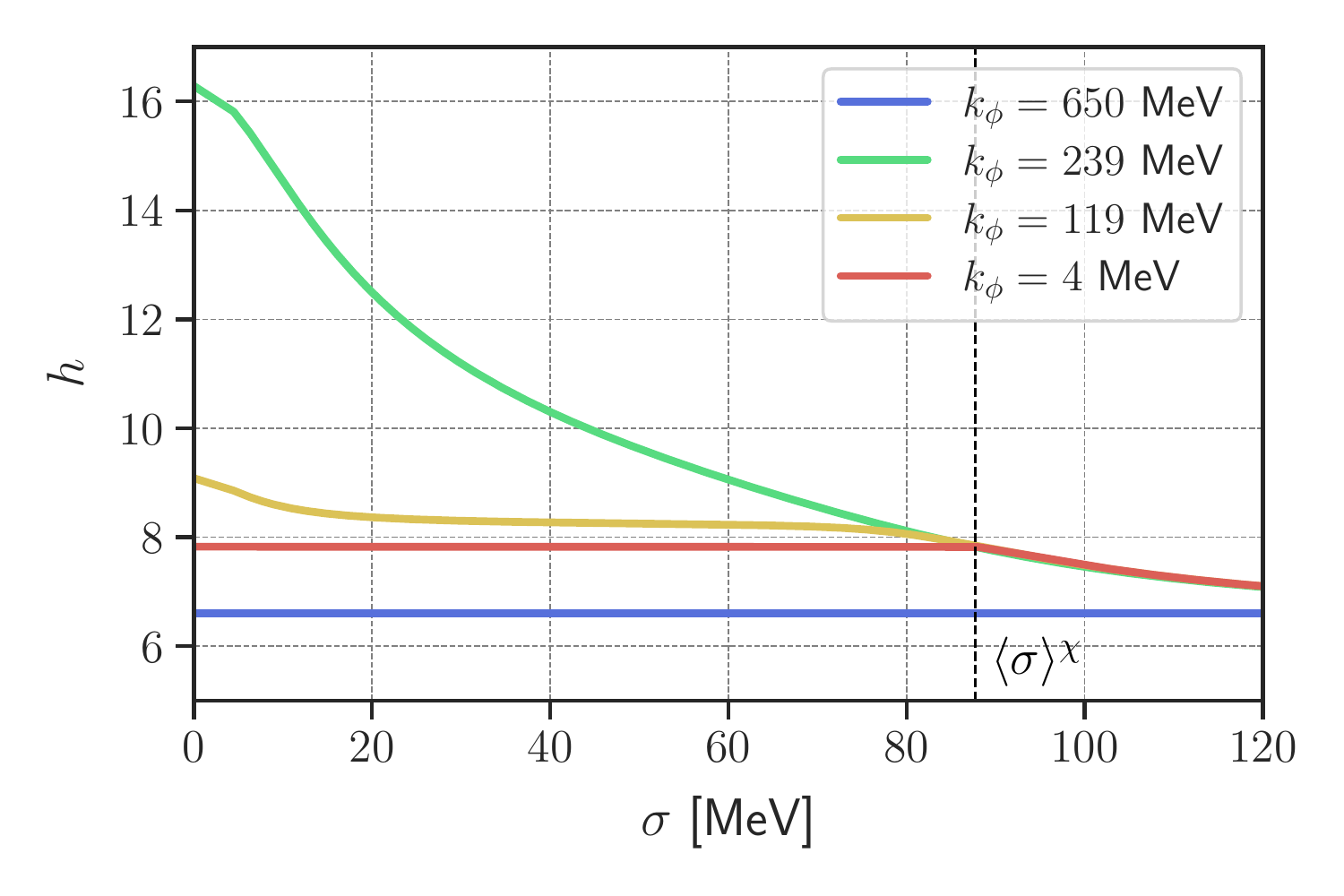}
		\caption{$\gamma=0.5$.}
		\label{fig:mq_h_low_mq}
	\end{subfigure}
	\begin{subfigure}[b]{1\textwidth}
		\includegraphics[width=0.49\linewidth]{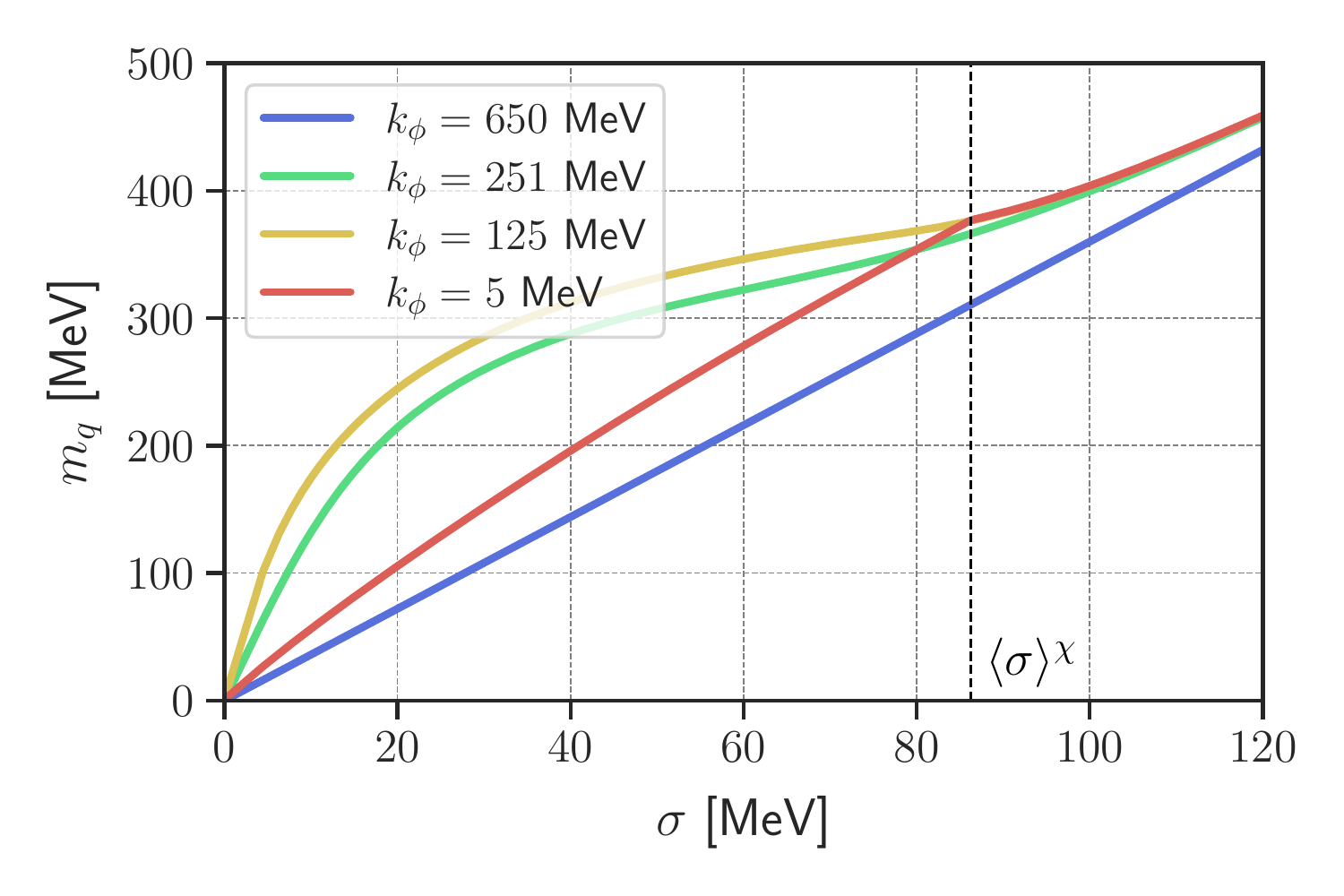}
		\includegraphics[width=0.49\linewidth]{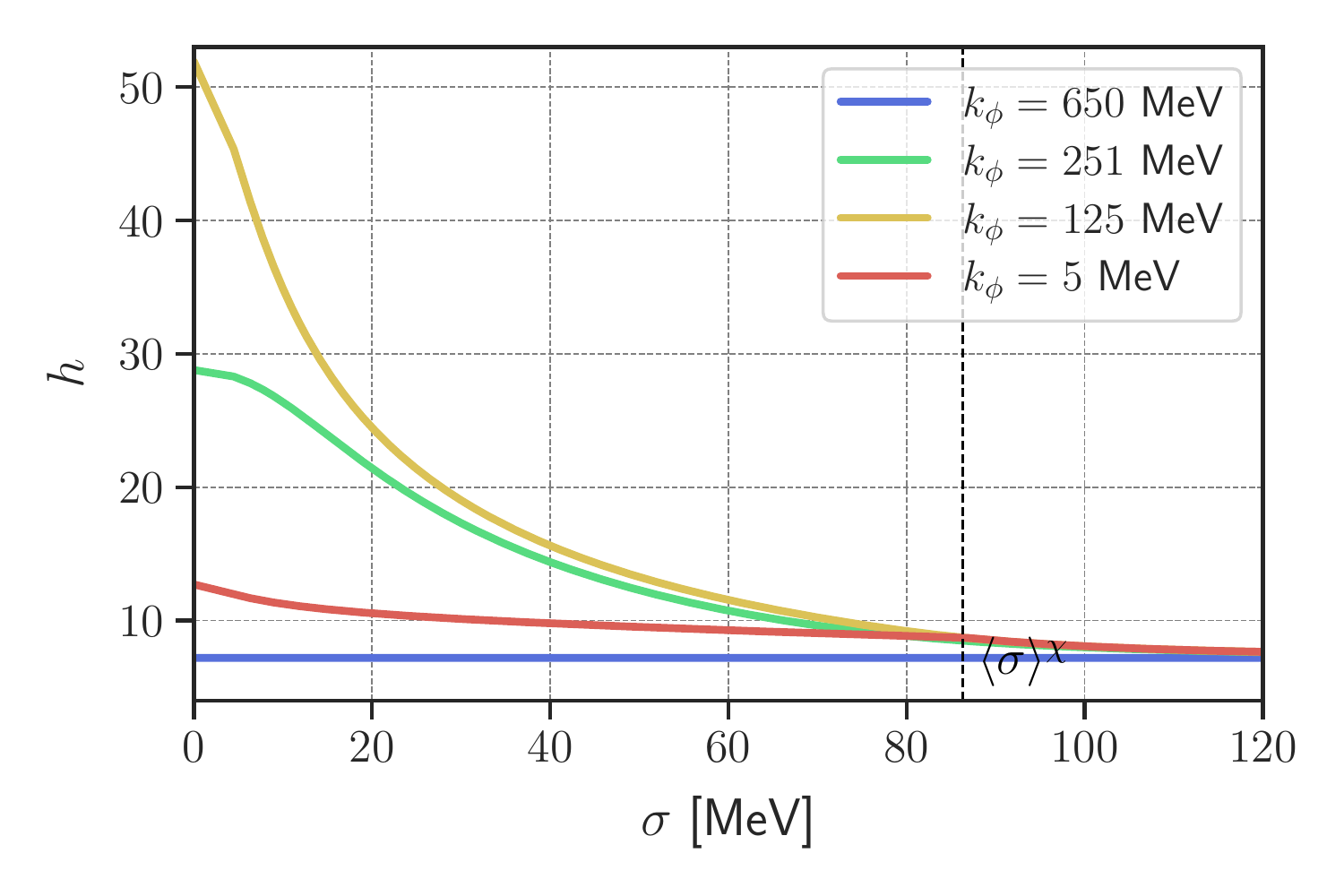}
		\caption{$\gamma=0.91$.}
		\label{fig:mq_h_low_mq_diff_g}
	\end{subfigure}
	\caption{RG-time evolution of the field-dependent quark mass $m_q^2(\rho)$ and Yukawa coupling $h(\rho)$ in scLPA at $(\mu_q,T)=~(0,4)~\textrm{MeV}$ for our optimal choice of $\gamma_\mathrm{opt} = 0.5$ and a $\gamma=0.91$ close to $1$,  quark and boson cutoff-scales very close to each other. 	It can be seen that the solution first moves towards a solution with $m_q^2 = \mathrm{const}$, i.e.~$h\propto \frac{1}{\sigma}$ but settles instead into the $h=\mathrm{const}$ solution, as discussed in \Cref{sec:results:relative_cutoffs}. Although this suggests that one could tune $\gamma\rightarrow1$ together with the inital conditions to give a constant quark mass, we have not found such a set of parameters. Instead, the quark mass shows in that case the divergence as discussed in \Cref{sec:results:relative_cutoffs}.}
	\label{fig:compare_g_qmasses}
\end{figure*}

\begin{figure*}[h]
	\begin{subfigure}[b]{1\textwidth}
		\includegraphics[width=0.49\linewidth]{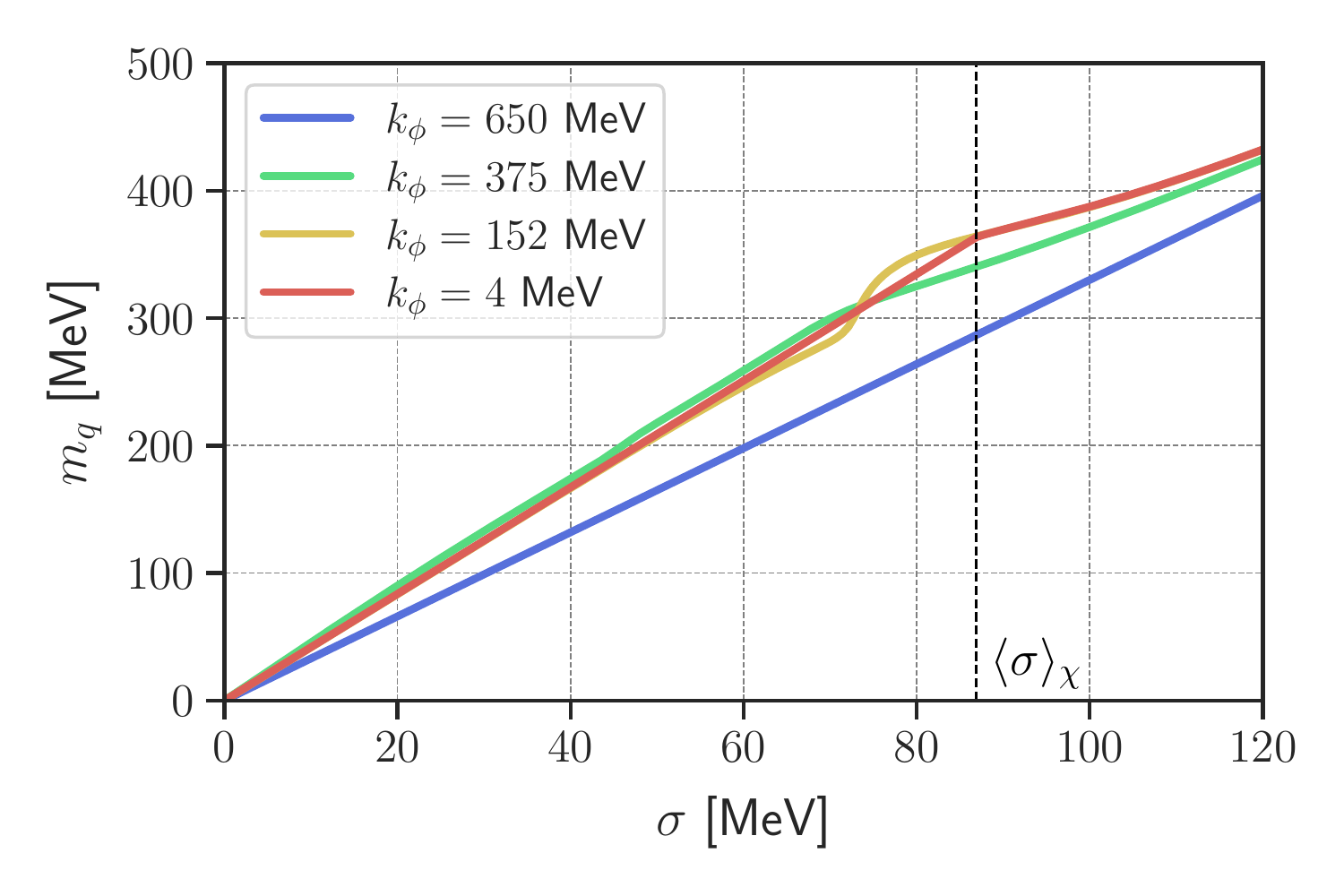}
		\includegraphics[width=0.49\linewidth]{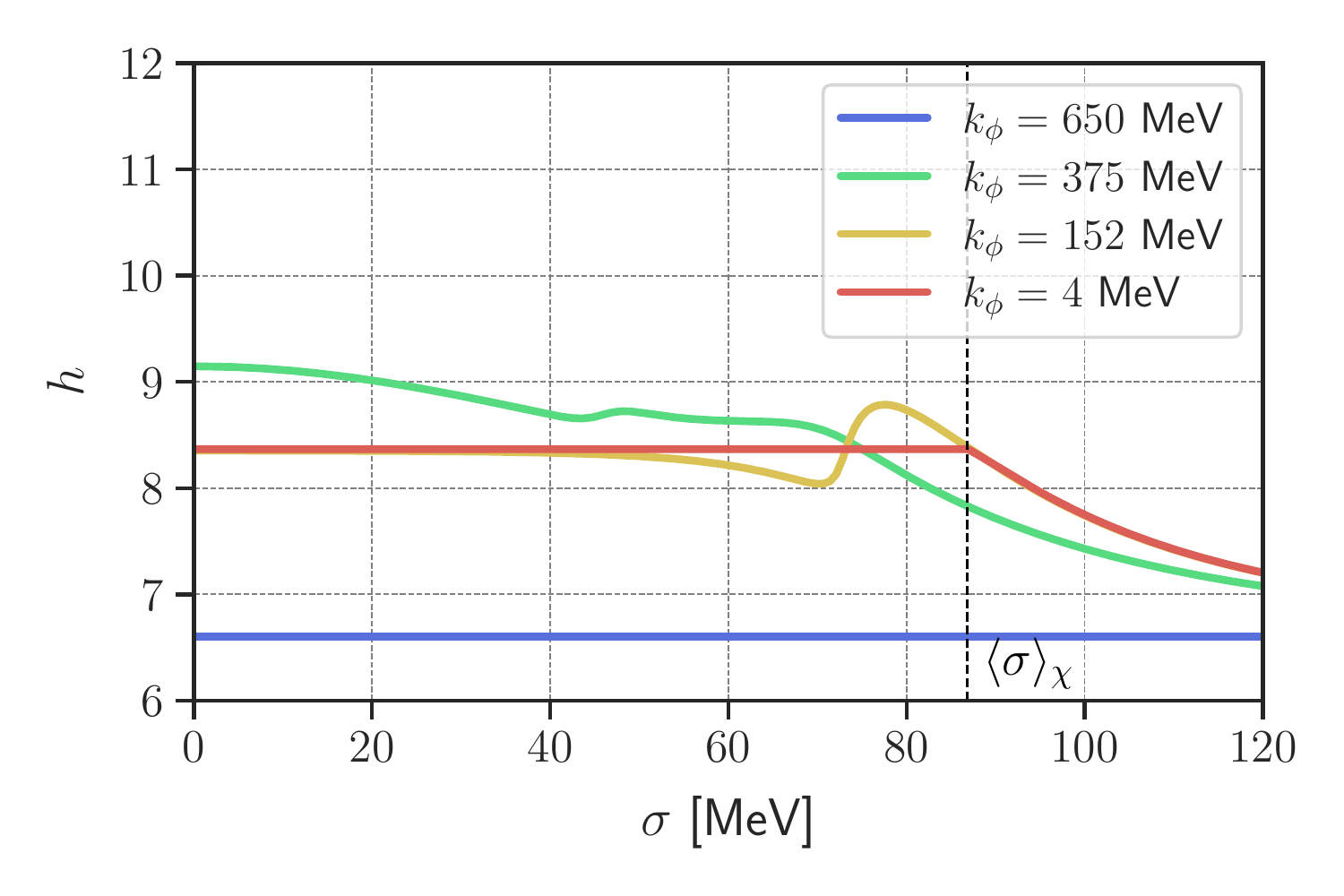}
		\caption{$(\mu_q,T)=~(300,4)~\textrm{MeV}$}
	\end{subfigure}
	\begin{subfigure}[b]{1\textwidth}
		\includegraphics[width=0.49\linewidth]{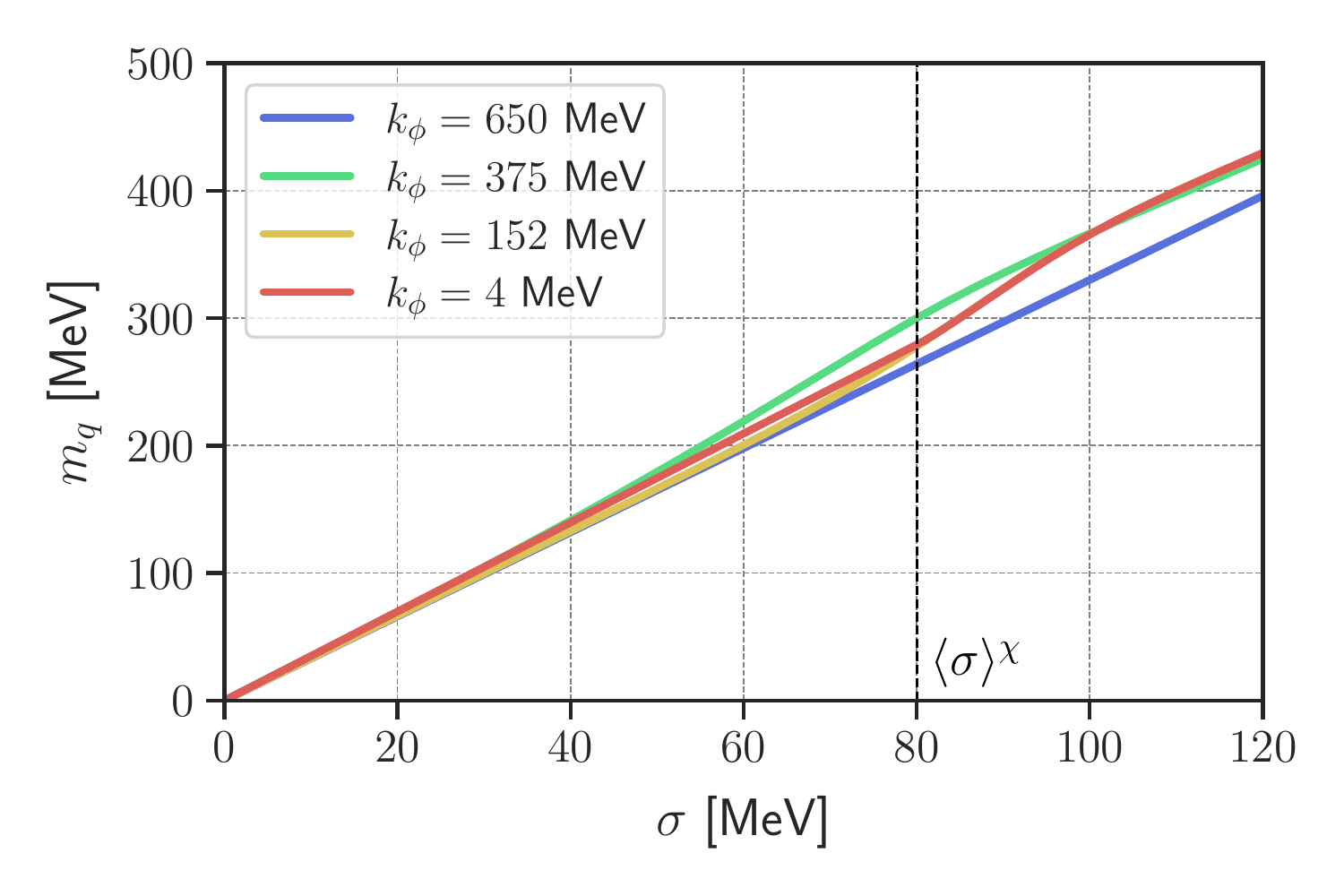}
		\includegraphics[width=0.49\linewidth]{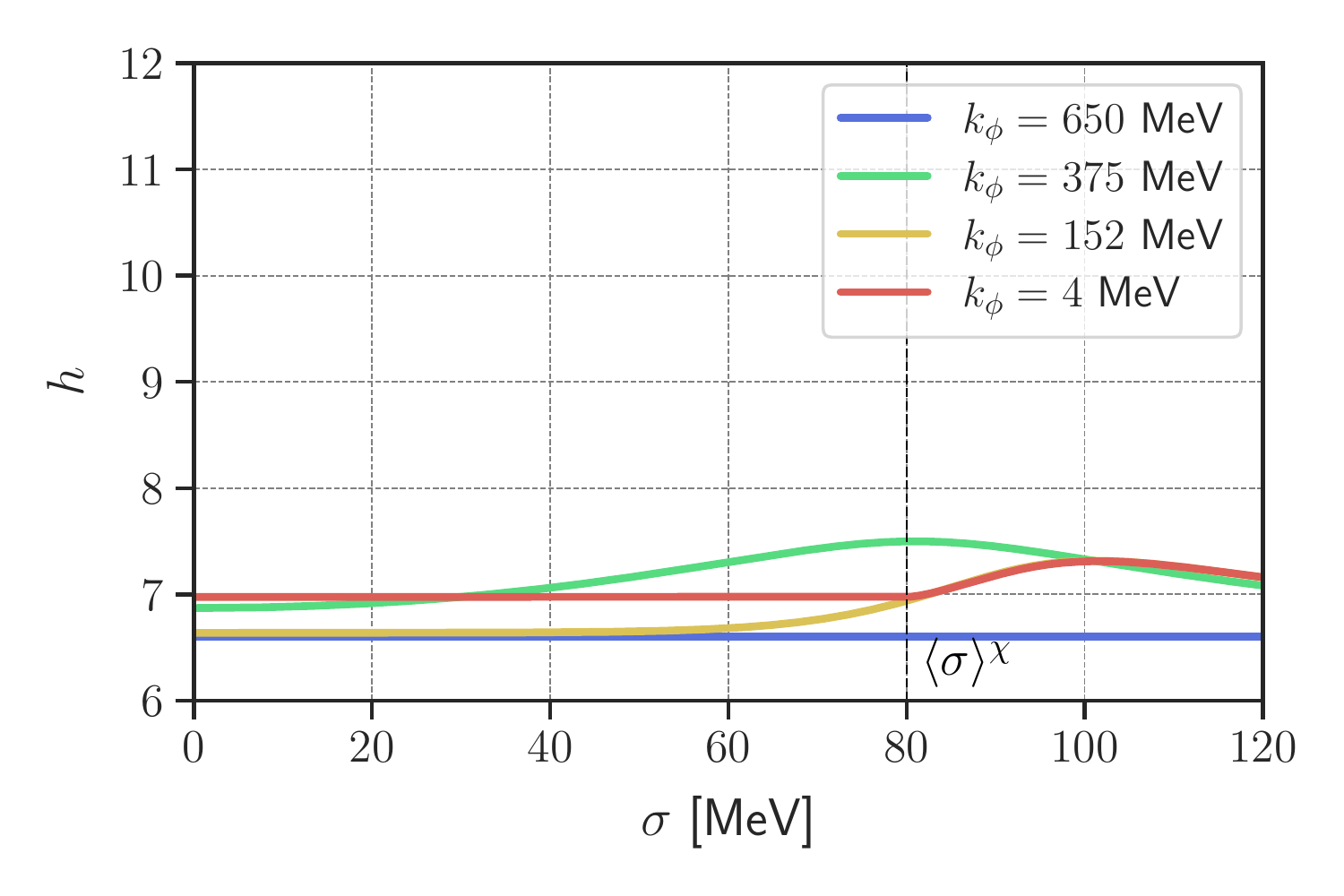}
		\caption{$(\mu_q,T)=~(300,40)~\textrm{MeV}$}
	\end{subfigure}
	\caption{RG-evolution of the quark mass in scLPA. We show here the evolution at low and intermediate temperatures $T=4,40\,\textrm{MeV}$ close to the phase boundary at $\mu_q = 300\,\textrm{MeV}$. The shown results are obtained using $\gamma_\mathrm{opt}=1/2$. In particular, the dynamics at low $\sigma$ are less pronounced compared to \Cref{fig:compare_g_qmasses}.}
	\label{fig:mq_h_high_mq}
\end{figure*}

\clearpage
\subsection{RG-time evolutions of bosonic masses}%
\label{app:flow_bos_masses}%
\begin{figure}[h]
	\centering
	\vspace{-0.6cm}%
	\begin{tabular}{ c  c }
		\begin{subfigure}[b]{0.5\linewidth}
			\centering
			\includegraphics[width=0.8\linewidth]{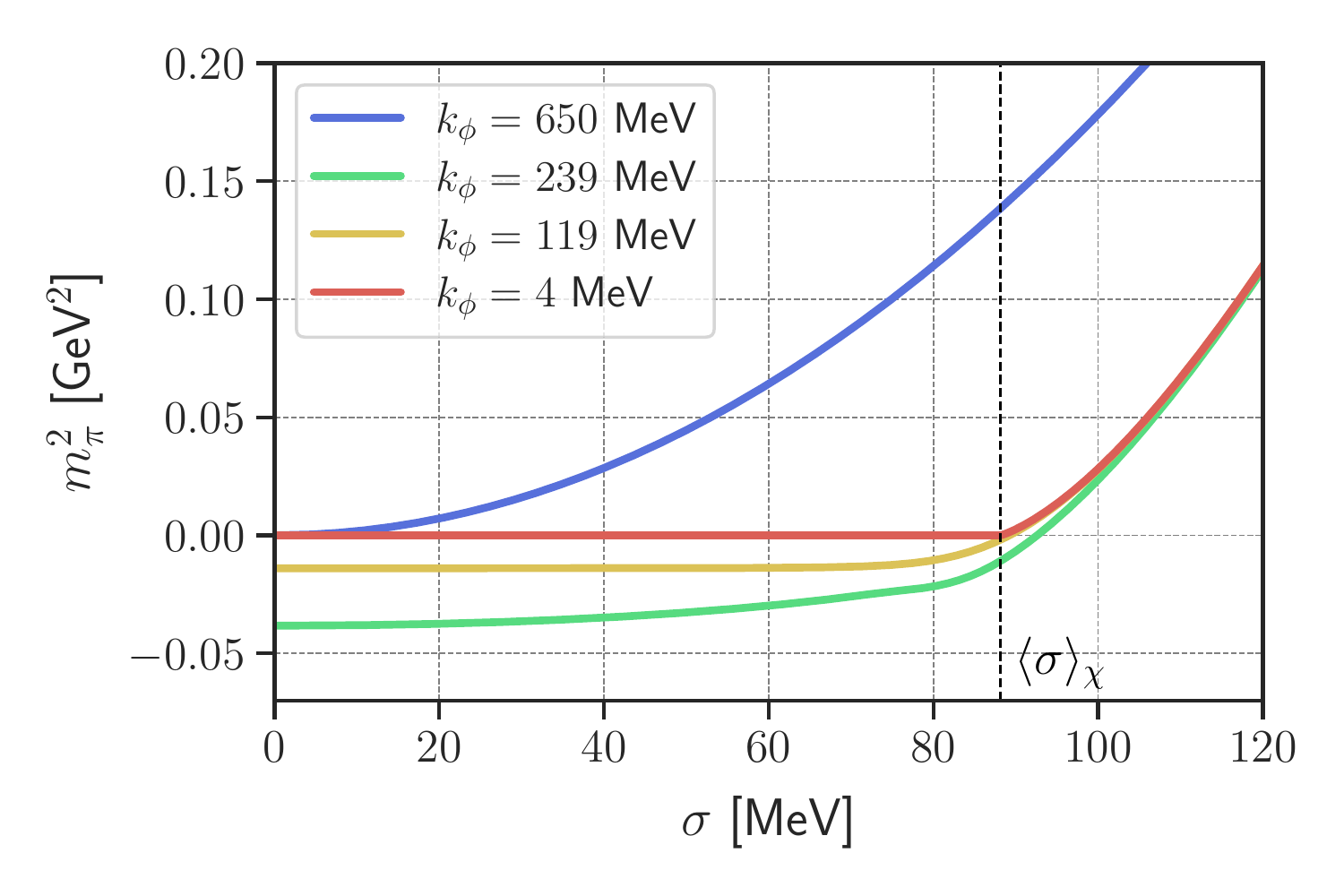}
			\includegraphics[width=0.8\linewidth]{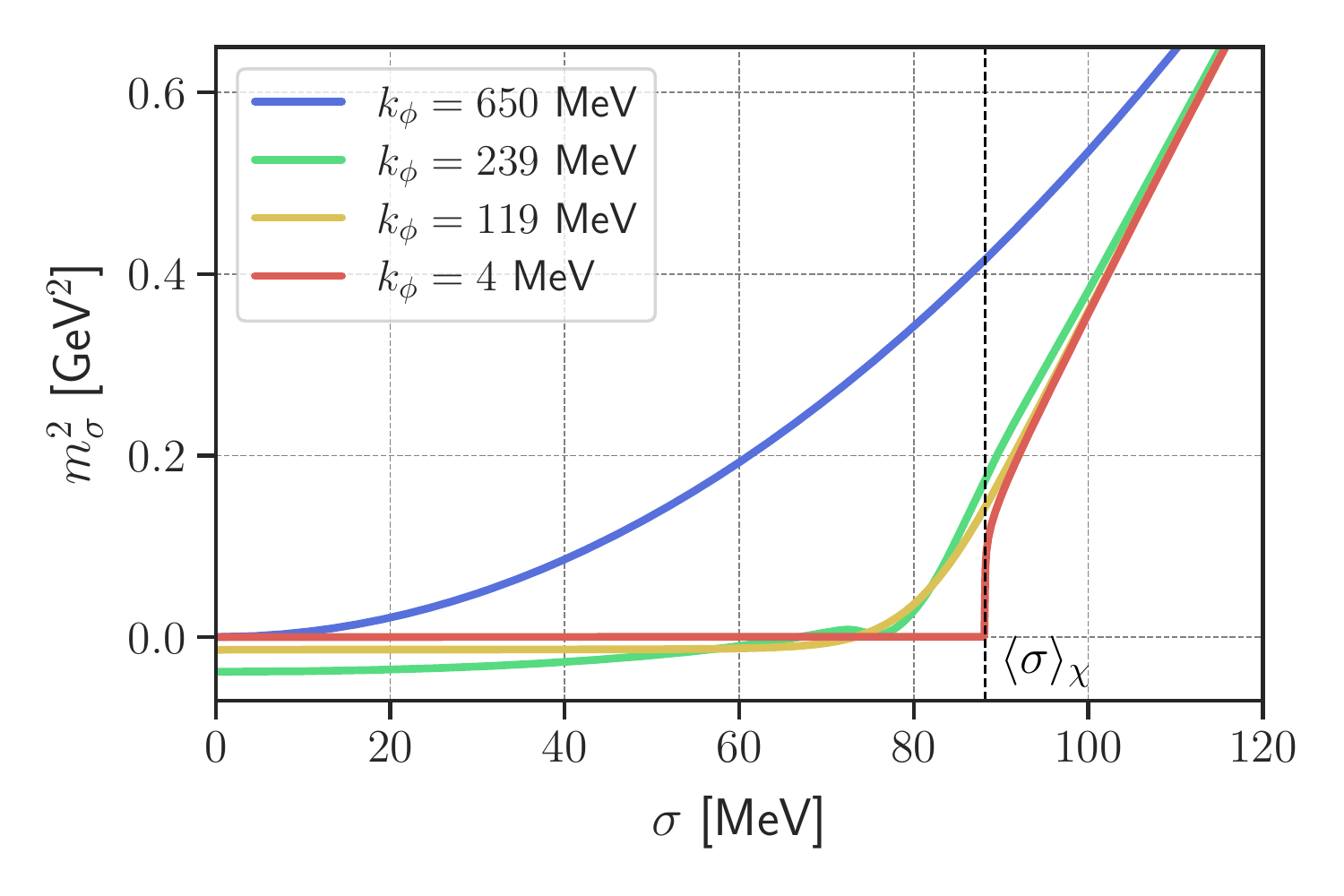}
			\caption{Fixed $h$ and $(\mu_q,T)=(300,4)\,\text{MeV}$.}
		\end{subfigure}
		&
		\begin{subfigure}[b]{0.5\linewidth}
			\centering
			\includegraphics[width=0.8\linewidth]{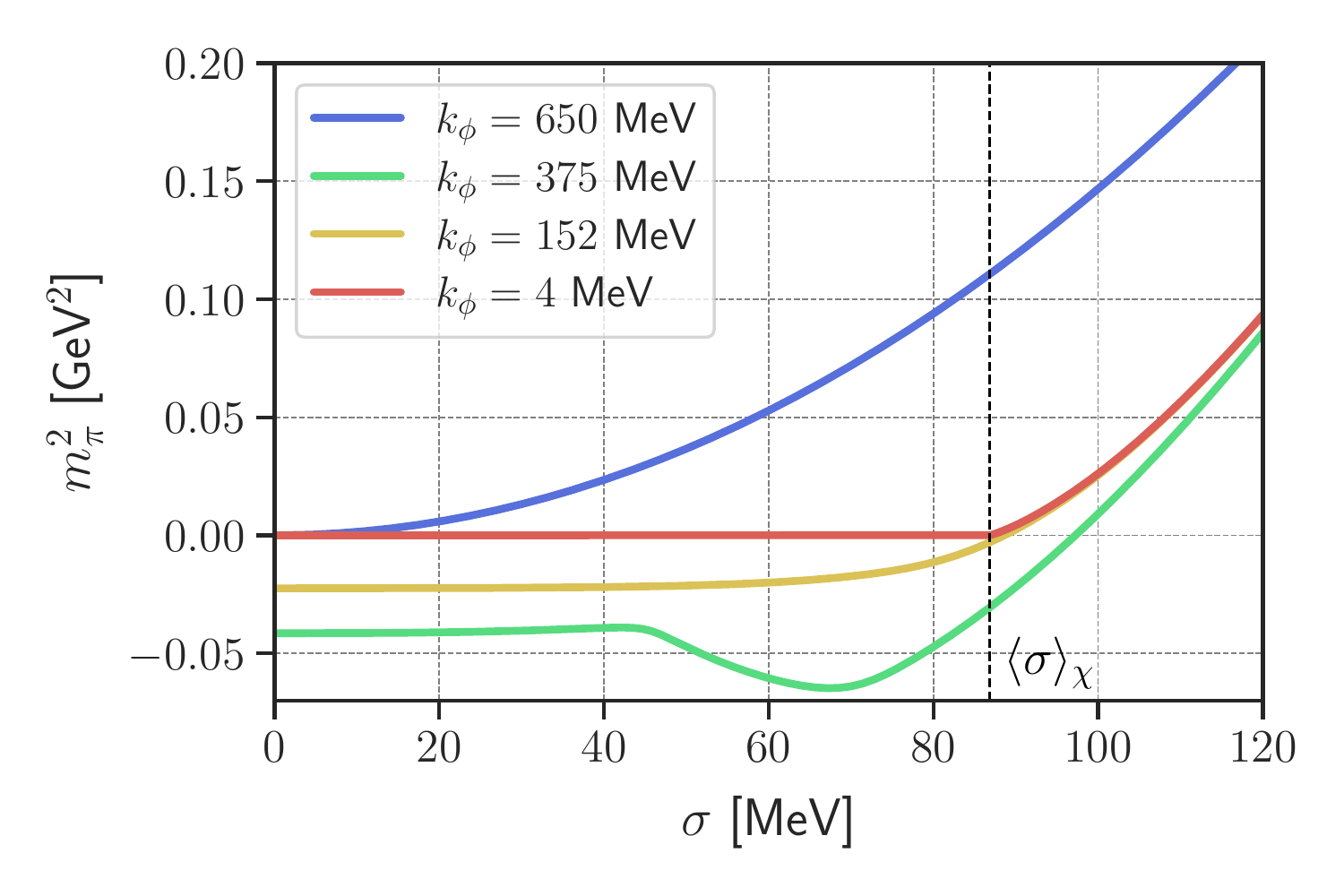}
			\includegraphics[width=0.8\linewidth]{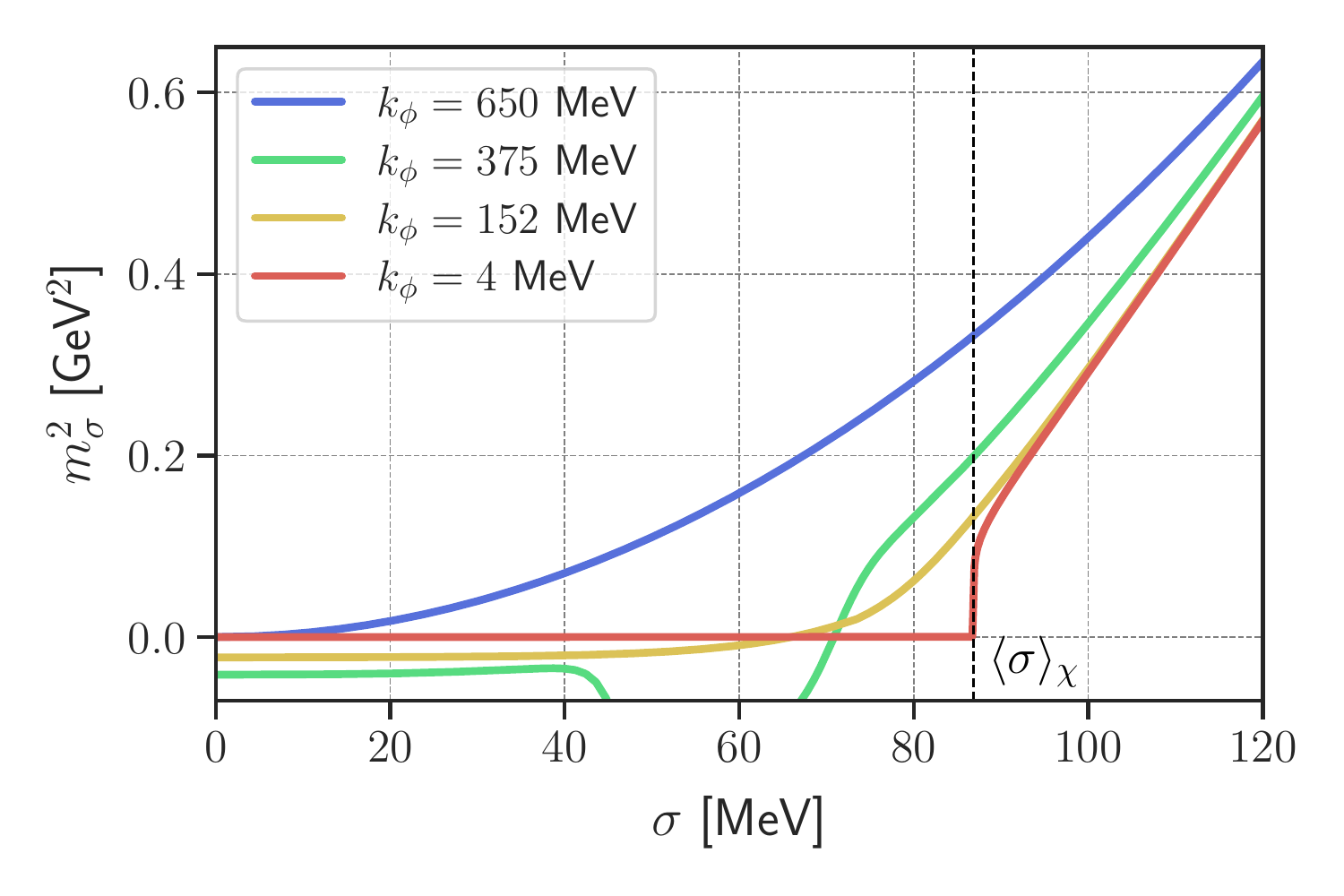}
			\caption{Full $h_k(\rho)$ and $(\mu_q,T)=(300,4)\,\text{MeV}$.}
		\end{subfigure}\\
		\begin{subfigure}[b]{0.5\linewidth}
			\centering
			\includegraphics[width=0.8\linewidth]{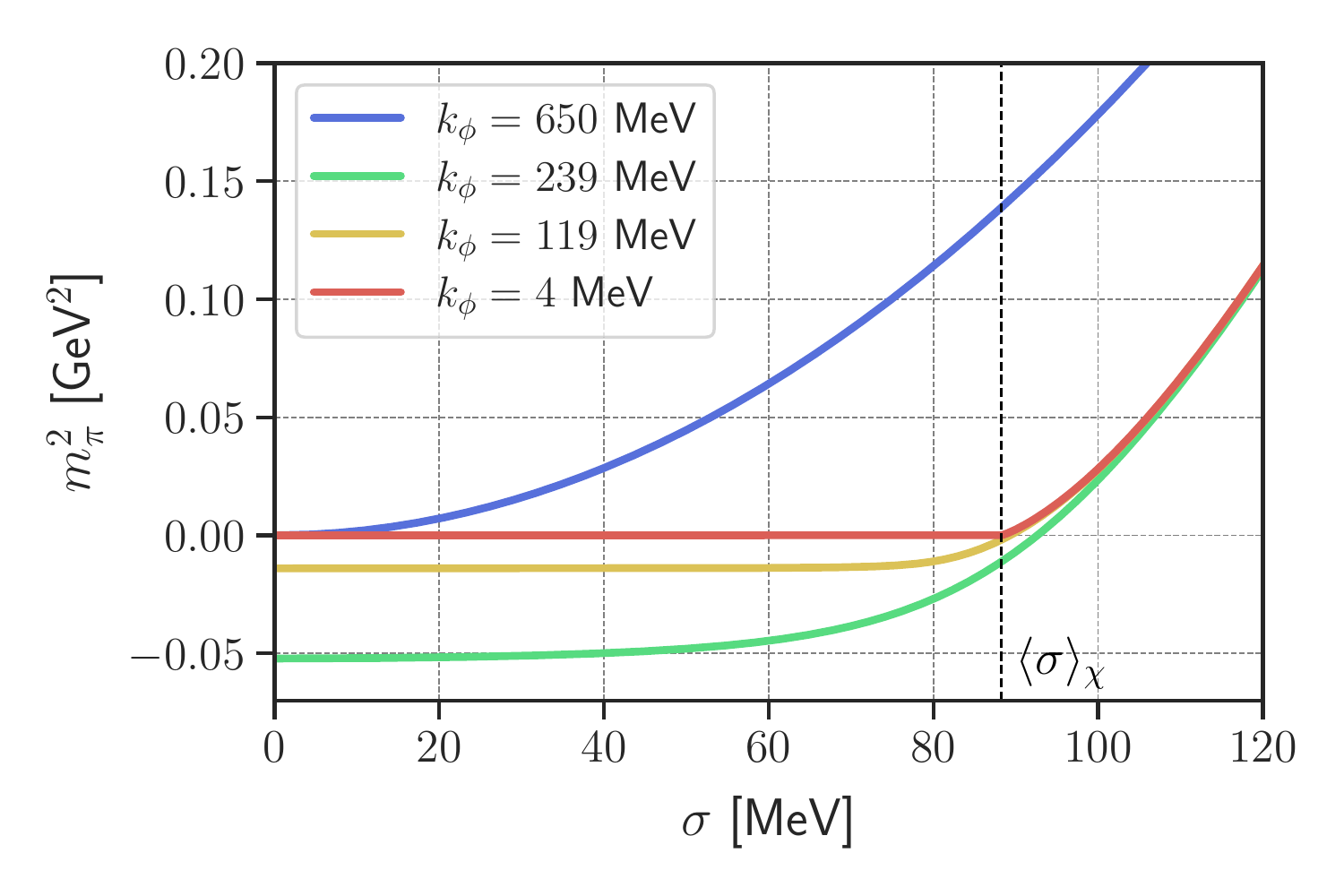}
			\includegraphics[width=0.8\linewidth]{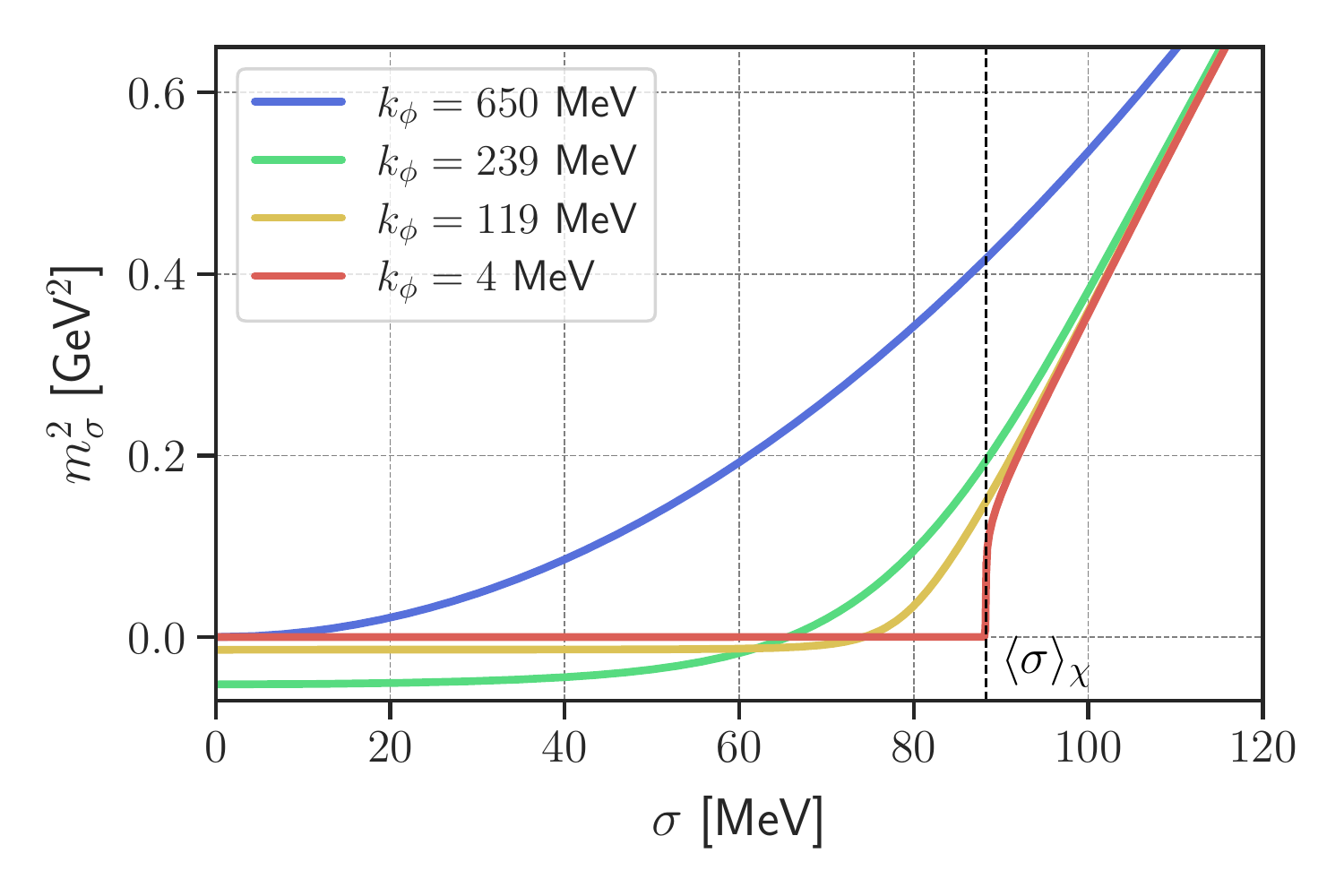}
			\caption{Fixed $h$ and $(\mu_q,T)=(0,4)\,\text{MeV}$.}
		\end{subfigure}
		&
		\begin{subfigure}[b]{0.5\linewidth}
			\centering
			\includegraphics[width=0.8\linewidth]{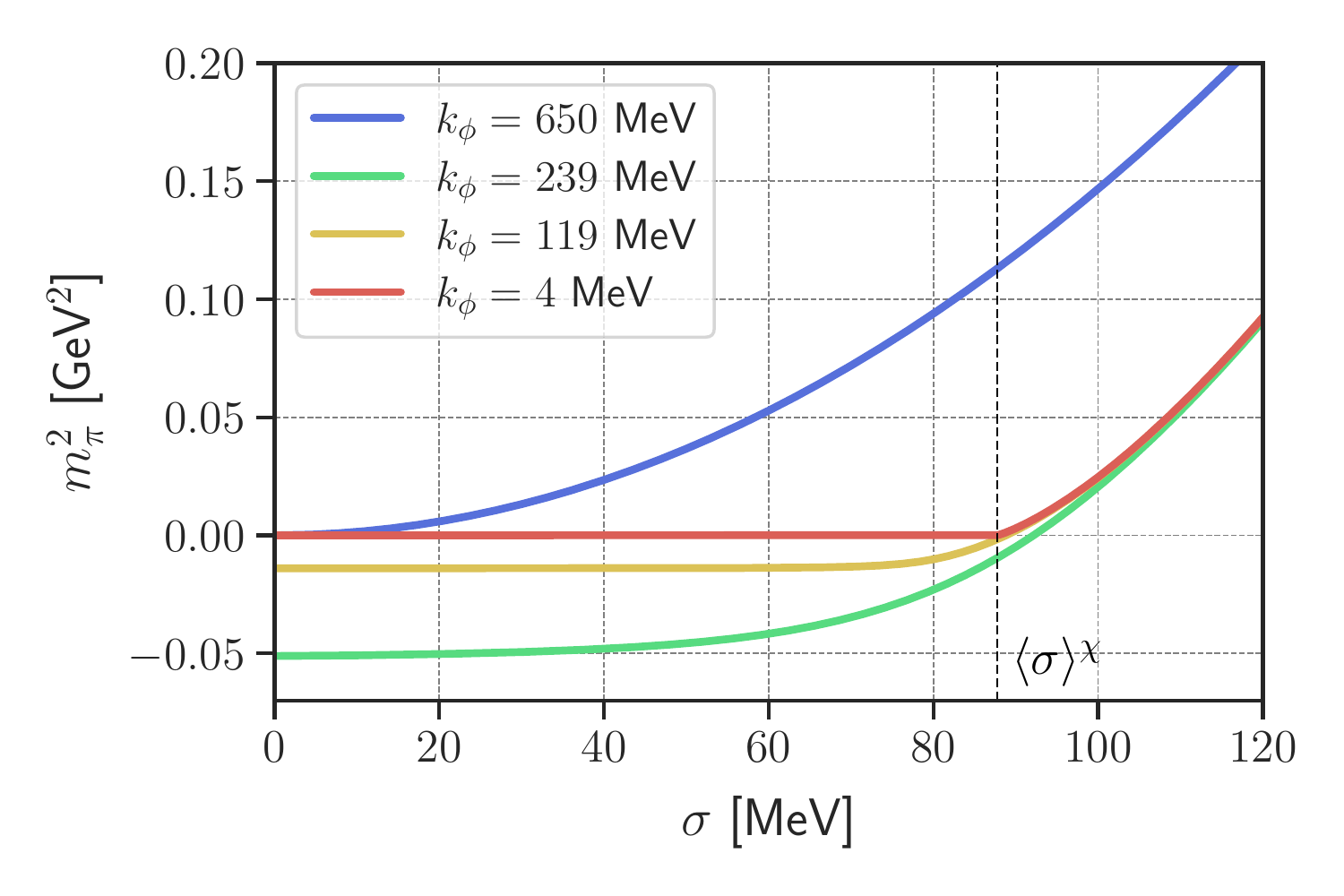}
			\includegraphics[width=0.8\linewidth]{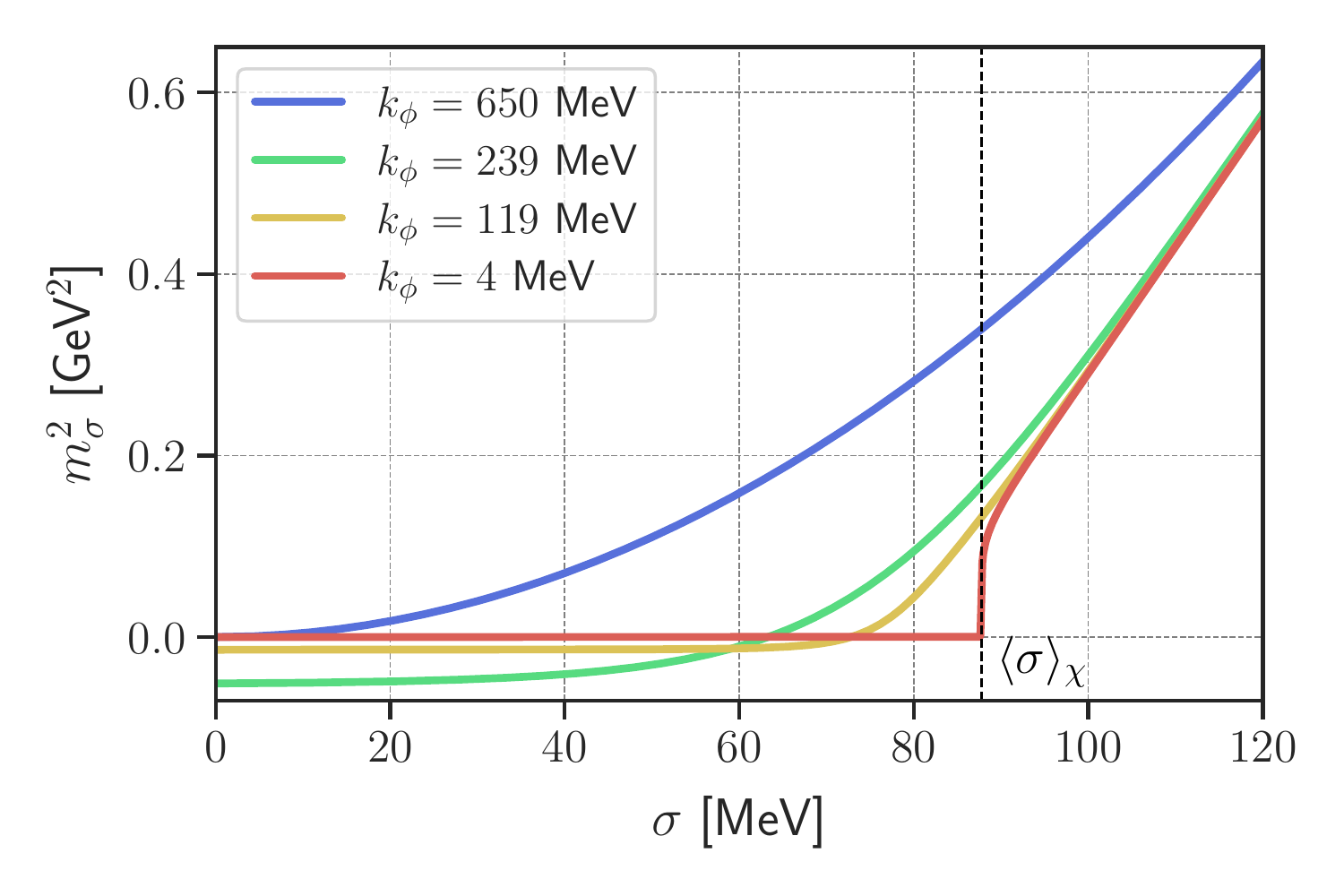}
			\caption{Full $h_k(\rho)$ and $(\mu_q,T)=(0,4)\,\text{MeV}$.}
		\end{subfigure}
	\end{tabular}
	\caption{Pion and $\sigma$ masses $m^2_\pi$ and $m^2_\sigma$ as a function of $\sigma$ for $T=4$\,MeV (approximate vacuum) and $\mu_q=0,200$\,MeV for $\gamma=0.5$ without higher order quark-meson scatterings (LPA, left) and with (scLPA, right). For $T=35, 40$\,MeV see \Cref{fig:Masses3540}. }
	\vspace{-1cm}
	\label{fig:Masses4}
\end{figure}
\begin{figure}[h]
	\begin{tabular}{ c  c }
		\begin{subfigure}[b]{0.5\linewidth}
			\centering
			\includegraphics[width=0.8\linewidth]{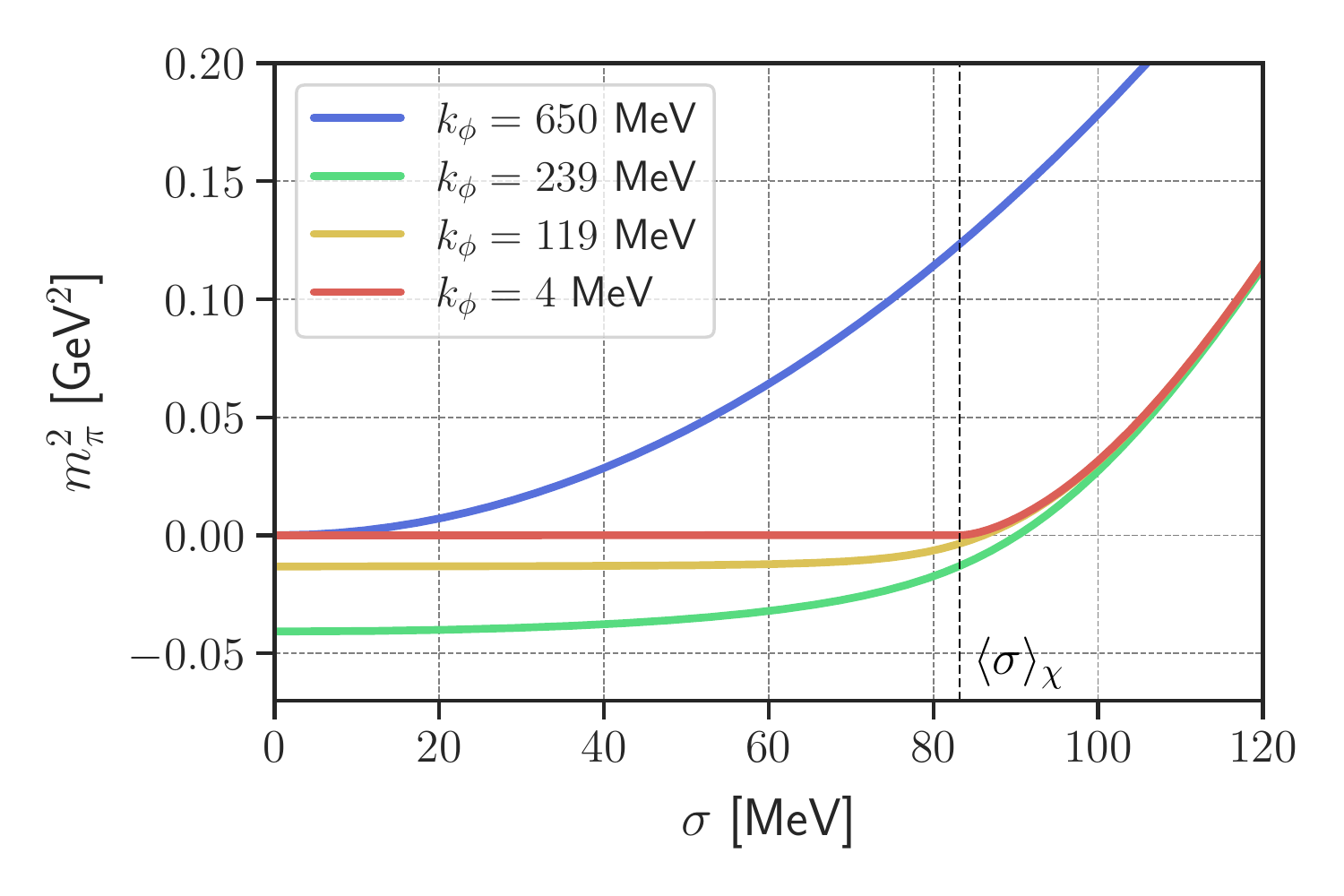}
			\includegraphics[width=0.8\linewidth]{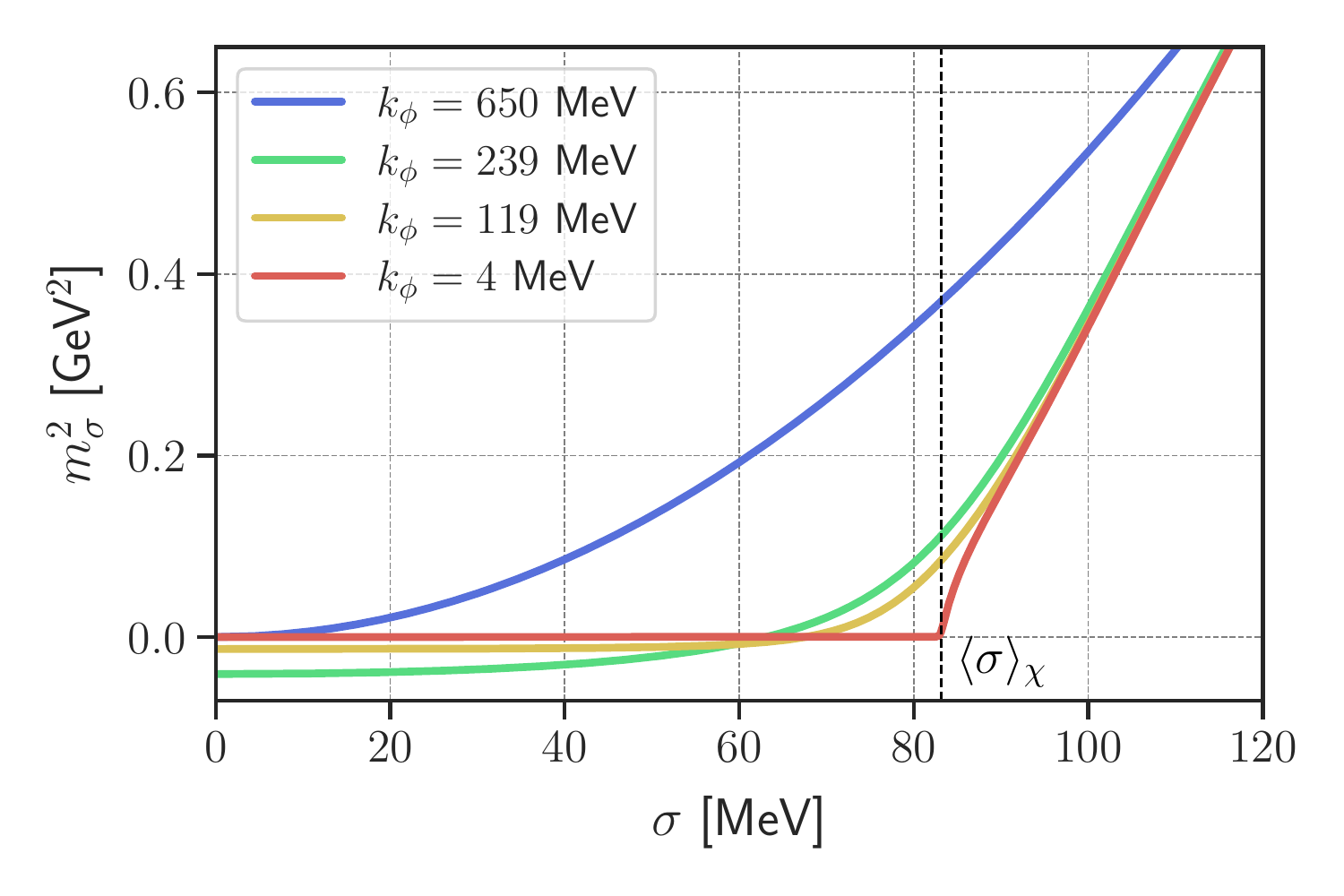}
			\caption{Fixed $h$ and $(\mu_q,T)=(300,40)\,\text{MeV}$.}
		\end{subfigure}
		&
		\begin{subfigure}[b]{0.5\linewidth}
			\centering
			\includegraphics[width=0.8\linewidth]{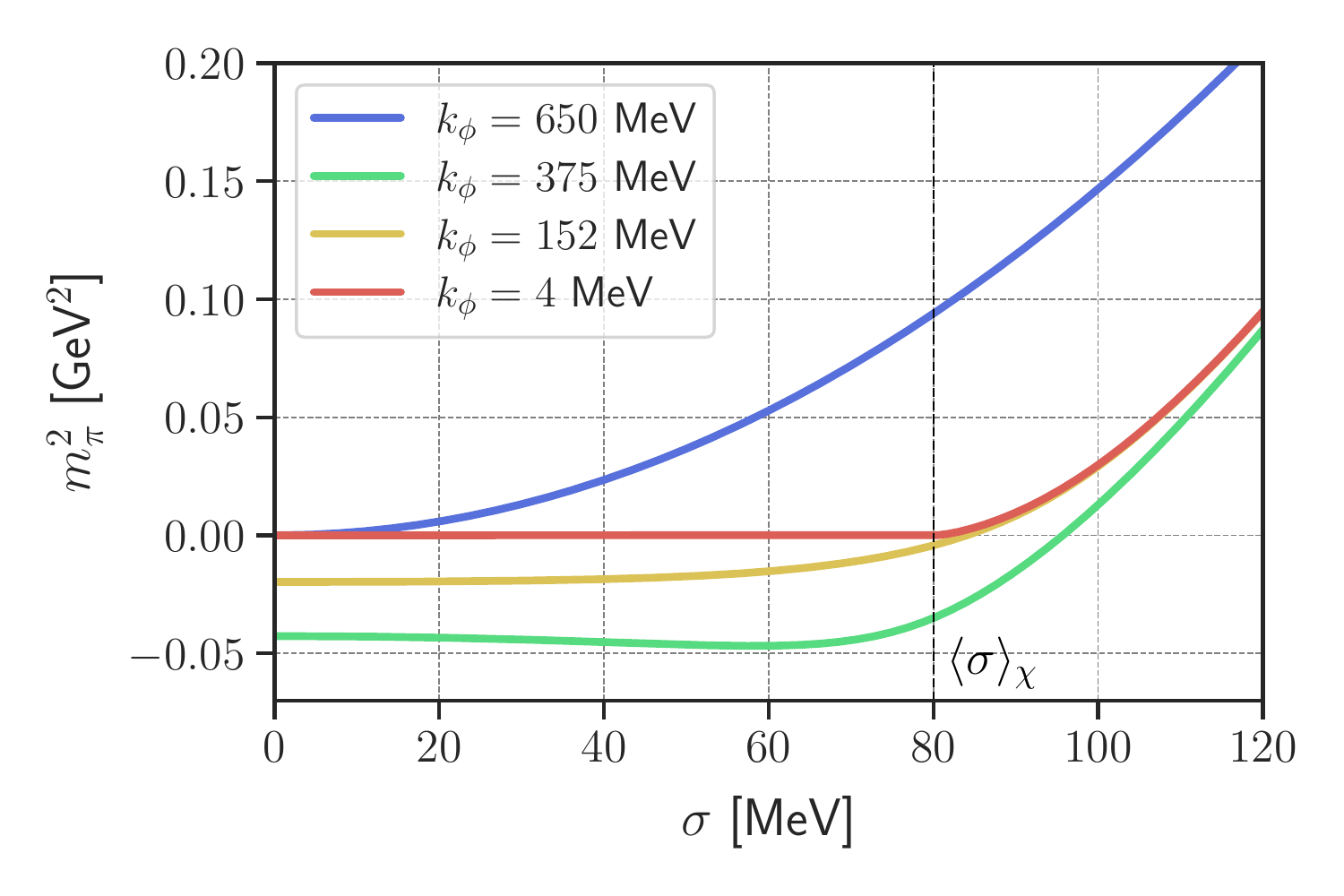}
			\includegraphics[width=0.8\linewidth]{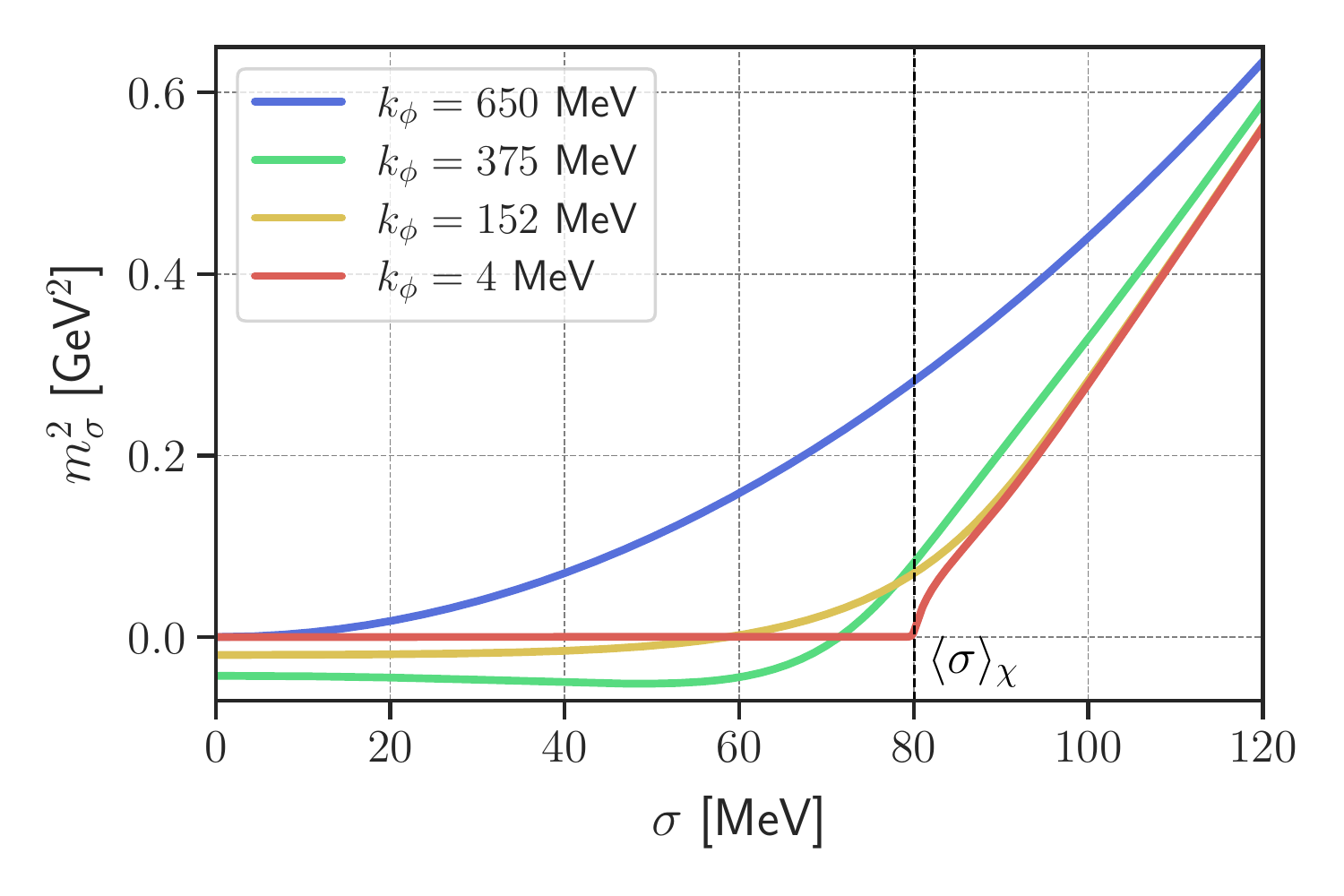}
			\caption{Full $h_k(\rho)$ and $(\mu_q,T)=(300,40)\,\text{MeV}$.}
		\end{subfigure}
		\\
		\begin{subfigure}[b]{0.5\linewidth}
			\centering
			\includegraphics[width=0.8\linewidth]{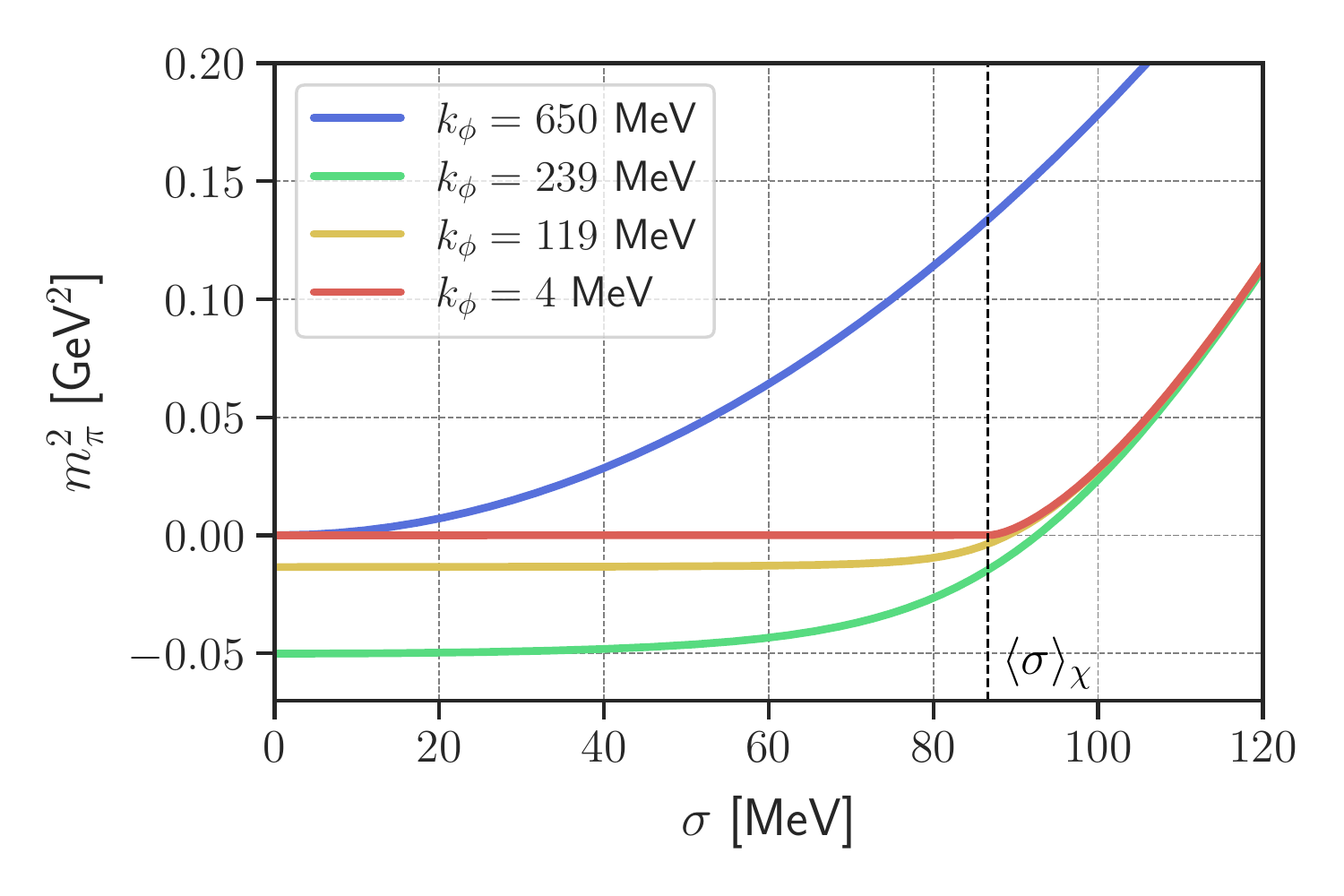}
			\includegraphics[width=0.8\linewidth]{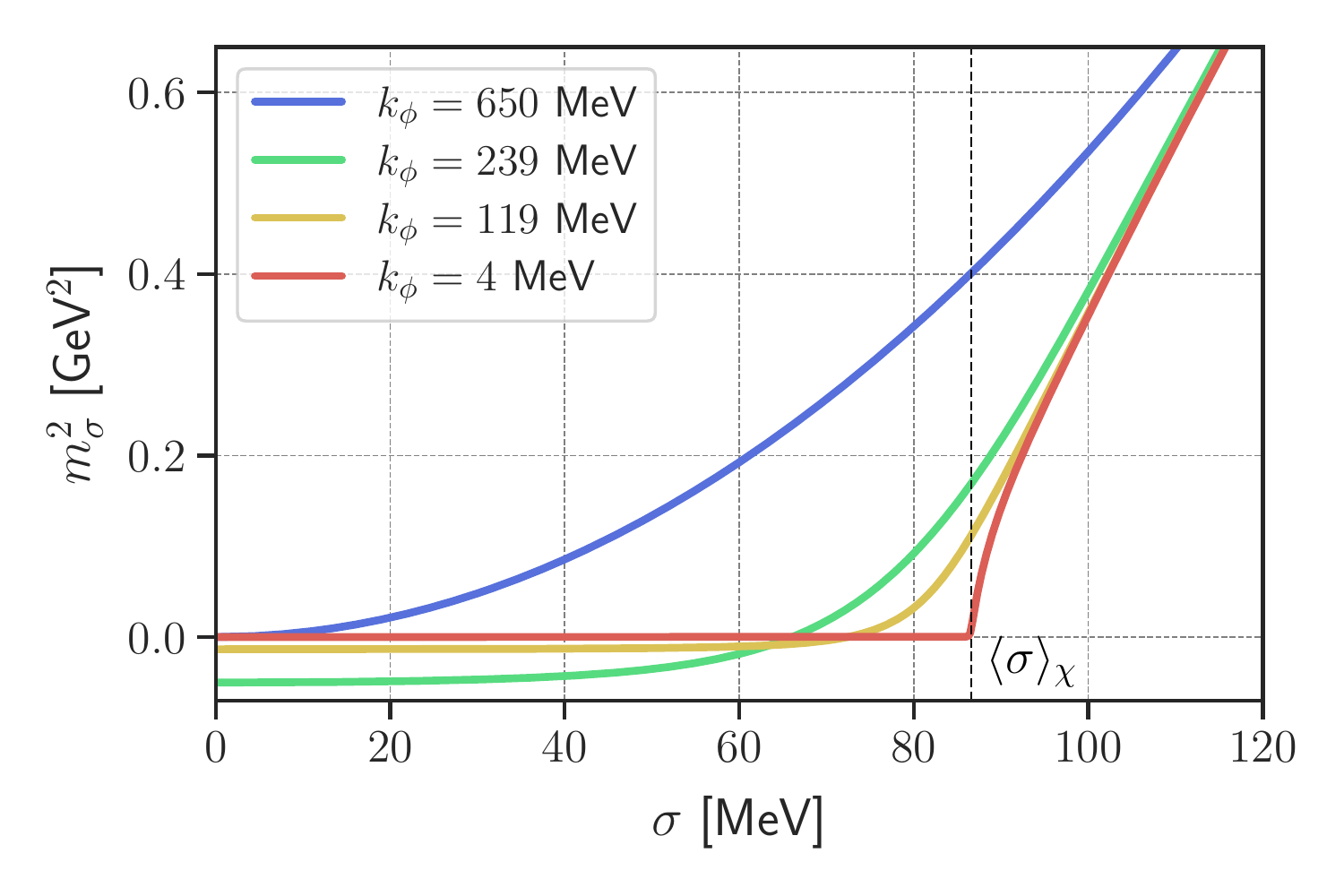}
			\caption{Fixed $h$ and $(\mu_q,T)=(0,35)\,\text{MeV}$.}
		\end{subfigure}
		&
		\begin{subfigure}[b]{0.5\linewidth}
			\centering
			\includegraphics[width=0.8\linewidth]{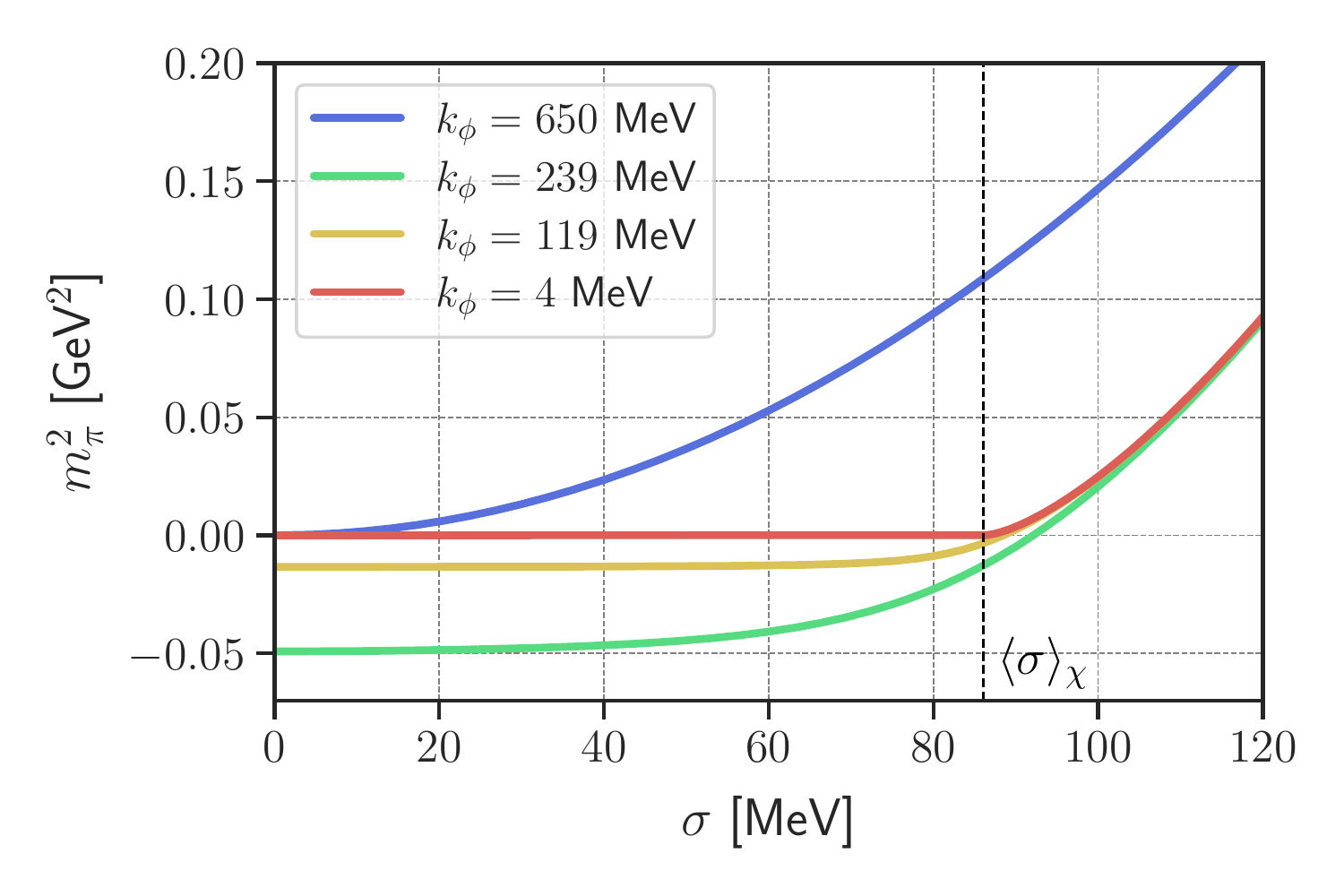}
			\includegraphics[width=0.8\linewidth]{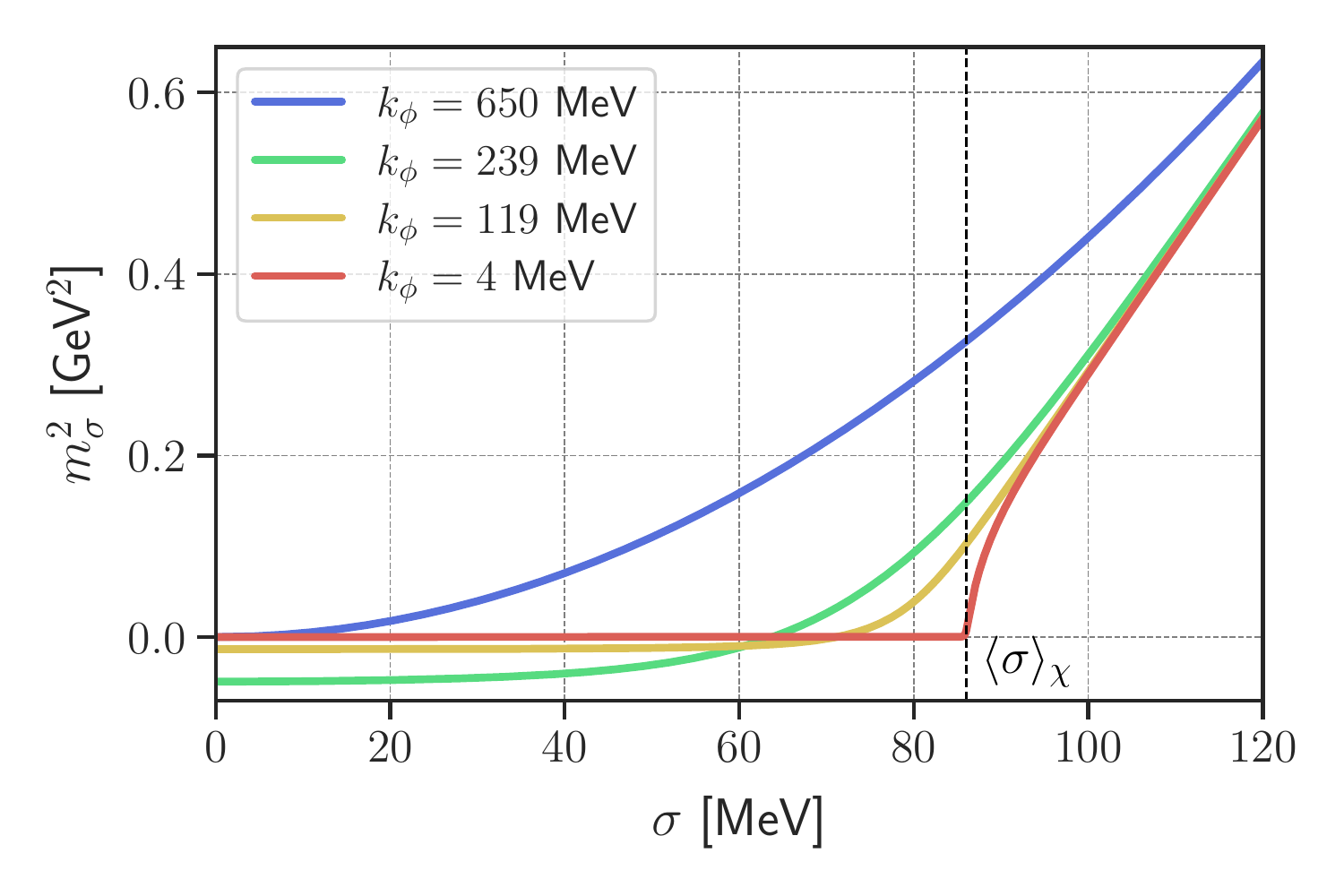}
			\caption{Full $h_k(\rho)$ and $(\mu_q,T)=(0,35)\,\text{MeV}$.}
		\end{subfigure}
	\end{tabular}
	\caption{Pion and $\sigma$ masses $m^2_\pi$ and $m^2_\sigma$ as a function of $\sigma$ for $T=35, 40$\,MeV and $\mu_q=0,200$\,MeV for $\gamma=0.5$ without higher order quark-meson scatterings (LPA, left) and with (scLPA, right). For $T=4$\,MeV (approximate vacuum) see \Cref{fig:Masses4}.}
	\label{fig:Masses3540}
\end{figure}
%

\clearpage
\subsection{Additional phase diagrams}
\label{app:detail_PD}

\begin{figure}[h]
	\centering
	\begin{subfigure}{0.5\linewidth}
		\centering
		\includegraphics[width=1\linewidth]{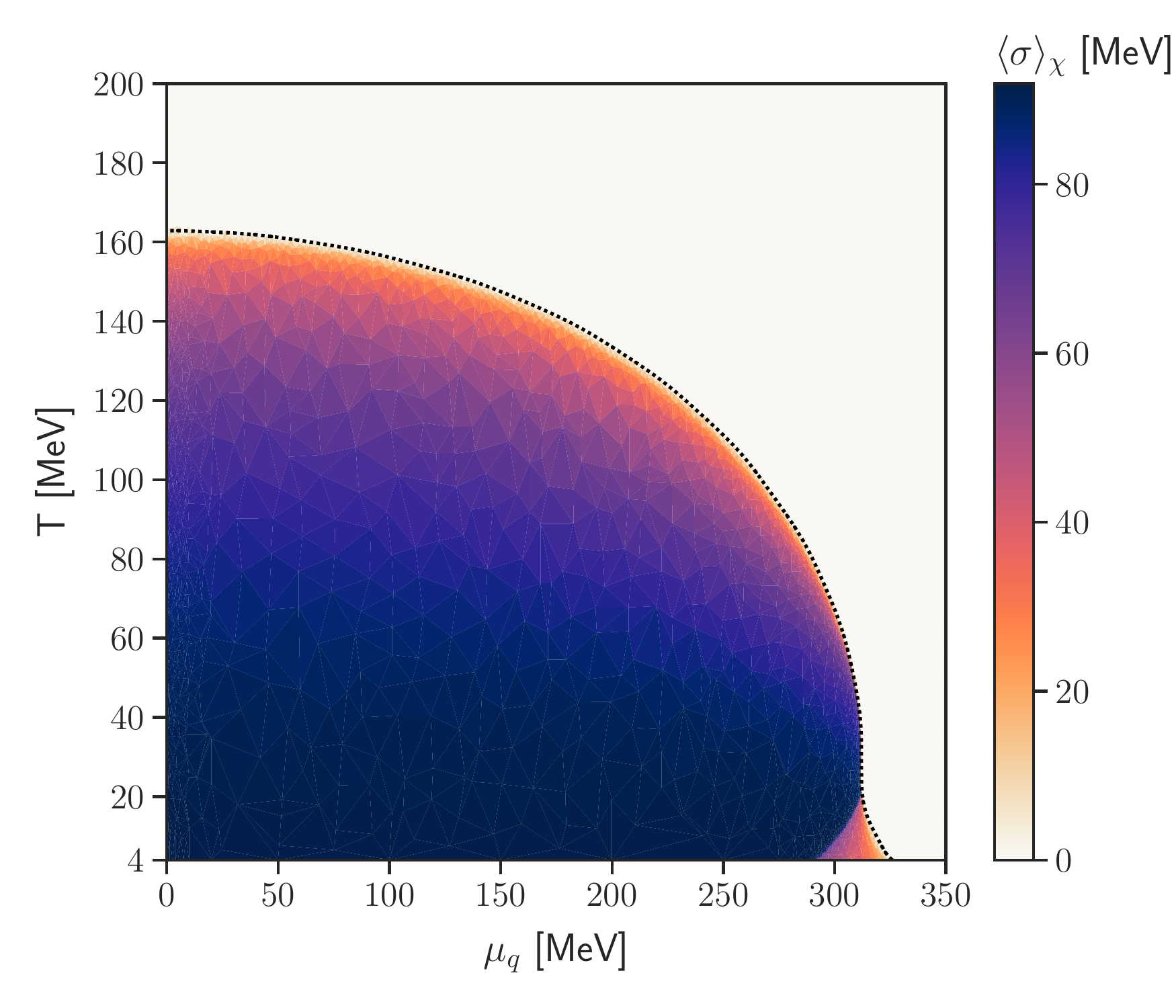}
	\end{subfigure}%
	\begin{subfigure}{0.5\linewidth}
	\centering
	\includegraphics[width=1\linewidth]{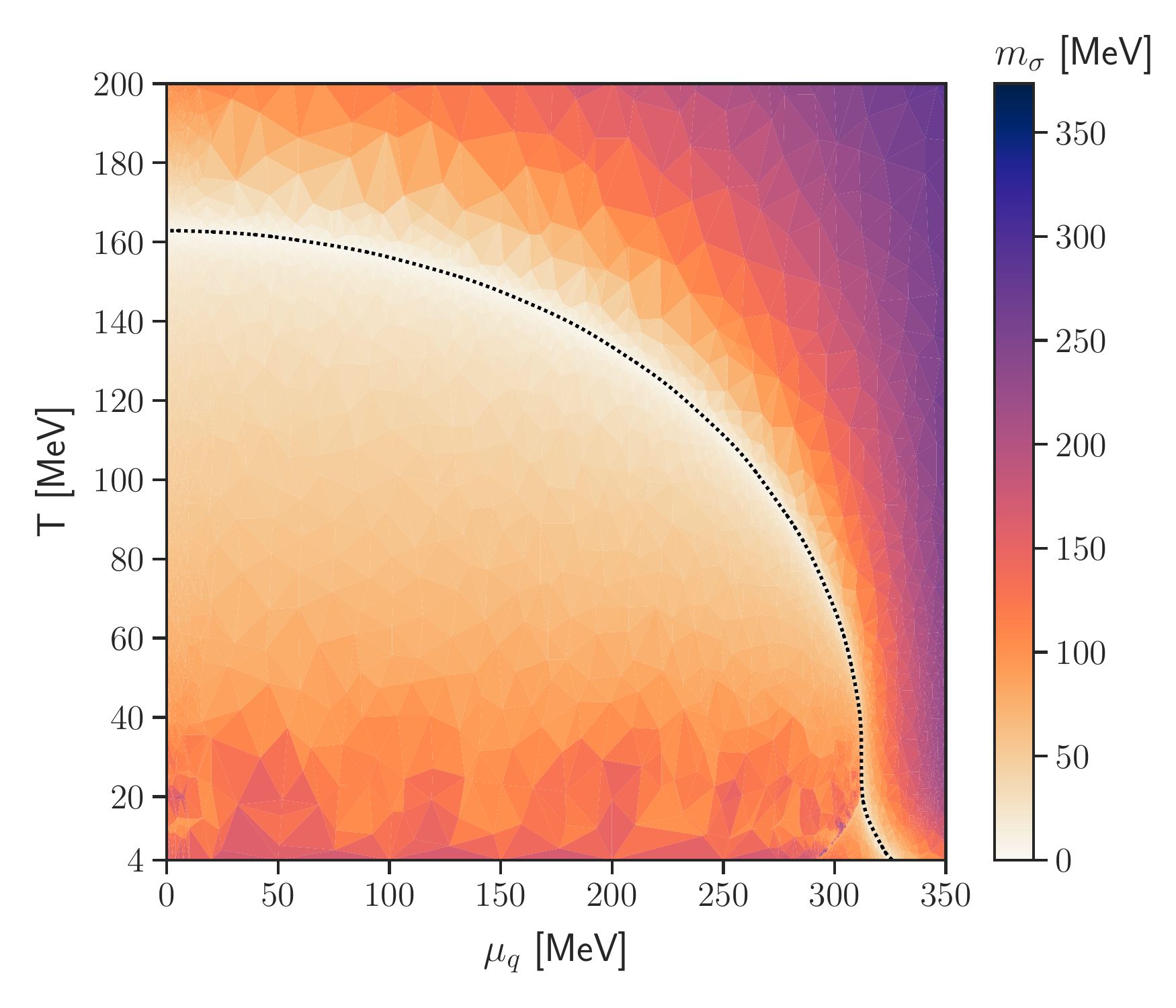}
	\end{subfigure}
	\caption{Order parameter $\langle\sigma\rangle$ and sigma mass $m_\sigma$ for the quark-meson model at $\gamma = 1$ and fixed Yukawa coupling (LPA). Note the large back-bending region with an additional second order phase transition at low temperatures and high chemical potential. Of all phase diagrams, in this case the feature is the most pronounced.}
	\label{fig:QM_PD_1}
\end{figure}
\begin{figure}[h]
	\centering
	\begin{subfigure}{0.5\linewidth}
		\centering
		\includegraphics[width=1\linewidth]{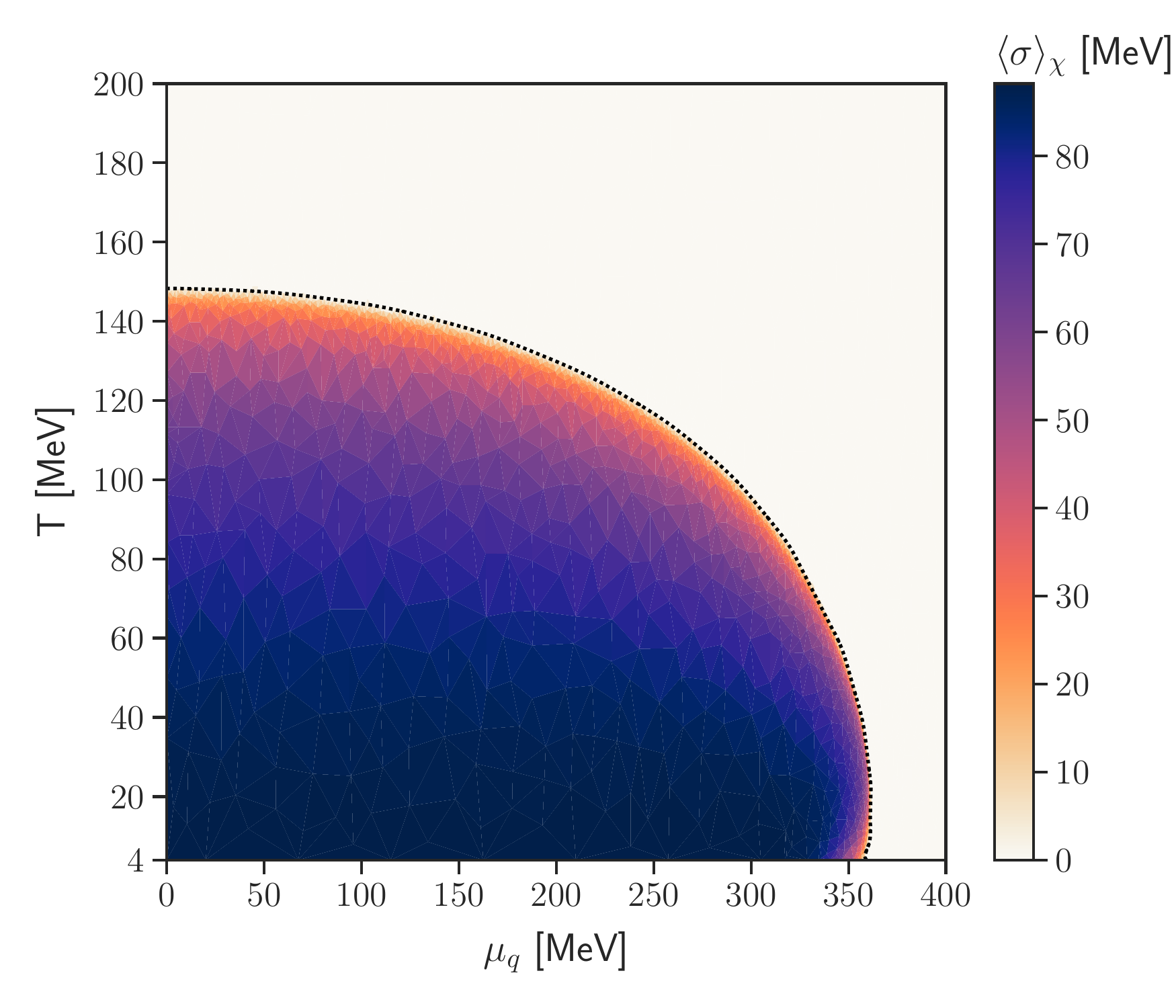}
	\end{subfigure}%
	\begin{subfigure}{0.5\linewidth}
		\centering
		\includegraphics[width=1\linewidth]{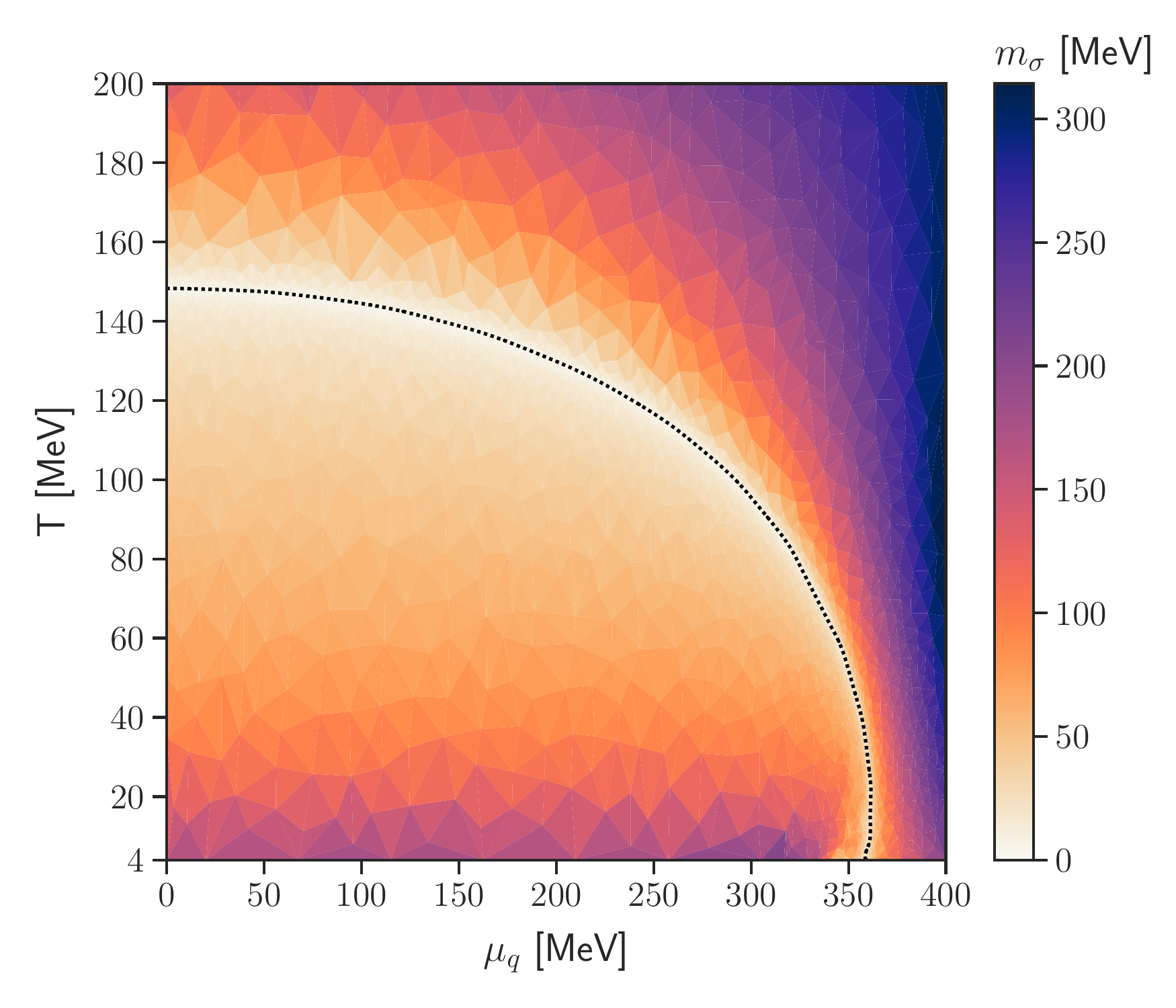}
	\end{subfigure}
	\caption{Order parameter $\langle\sigma\rangle$ and sigma mass $m_\sigma$ for the the quark-meson model at $\gamma = 0.5$ and fixed Yukawa coupling (LPA). The CEP is no longer present, as one can see in the sigma mass - it is equal to zero all along the phase boundary, signalling the second-order nature of the transition.}
	\label{fig:QM_PD_2}
\end{figure}
\begin{figure}[h]
	\begin{subfigure}{0.5\linewidth}
		\centering
		\includegraphics[width=1\linewidth]{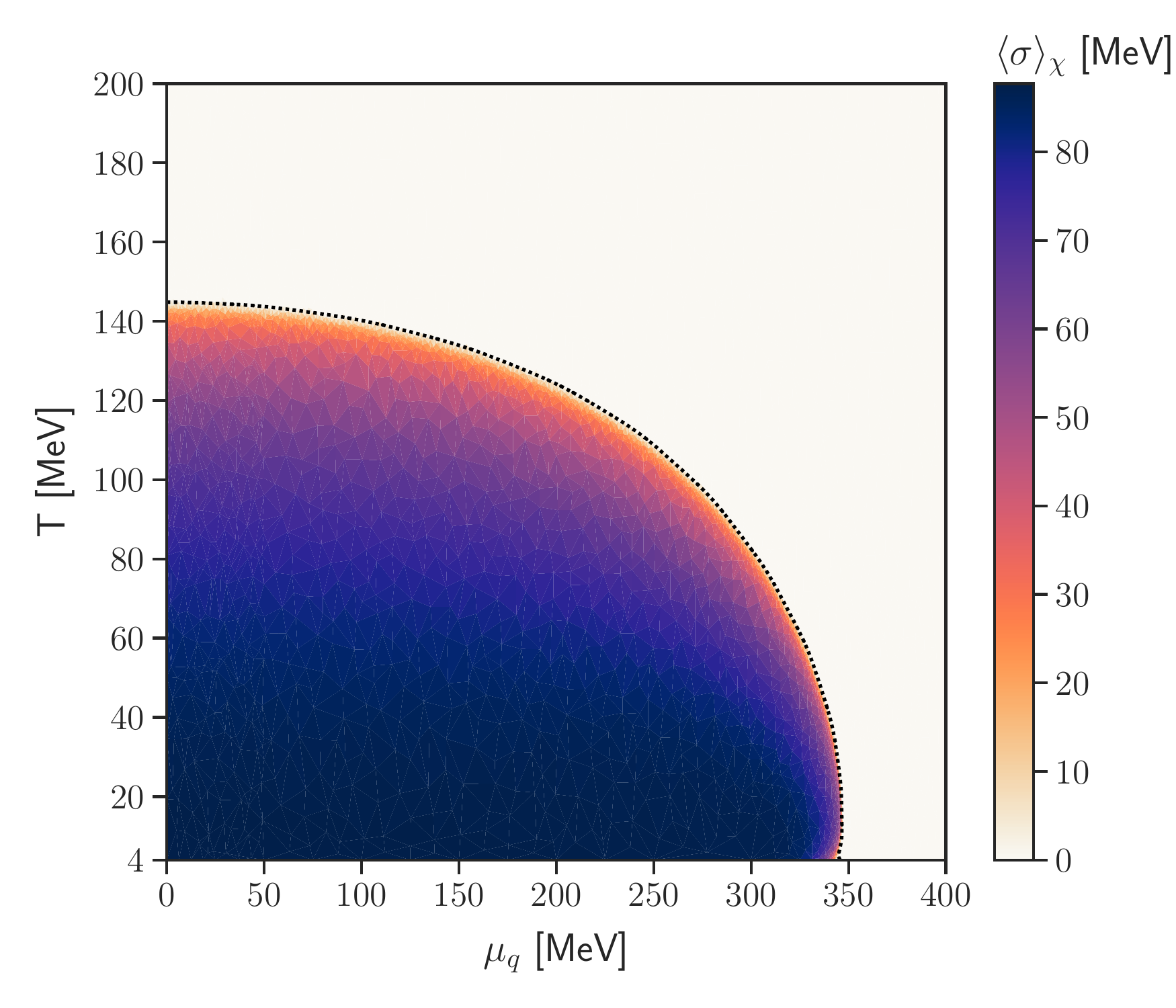}
	\end{subfigure}%
	\begin{subfigure}{0.5\linewidth}
		\centering
		\includegraphics[width=1\linewidth]{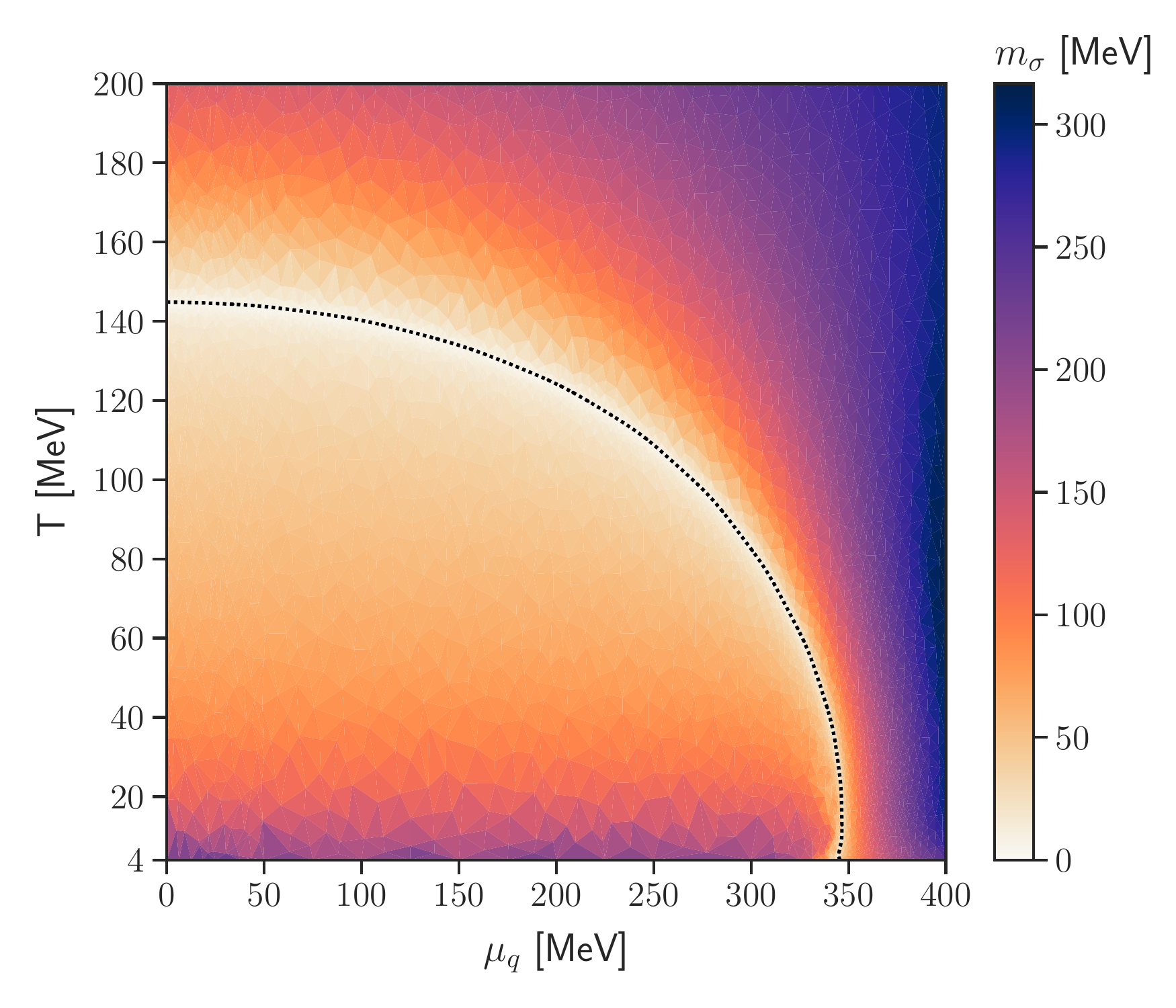}
	\end{subfigure}
	\begin{subfigure}{0.5\linewidth}
		\centering
		\includegraphics[width=1\linewidth]{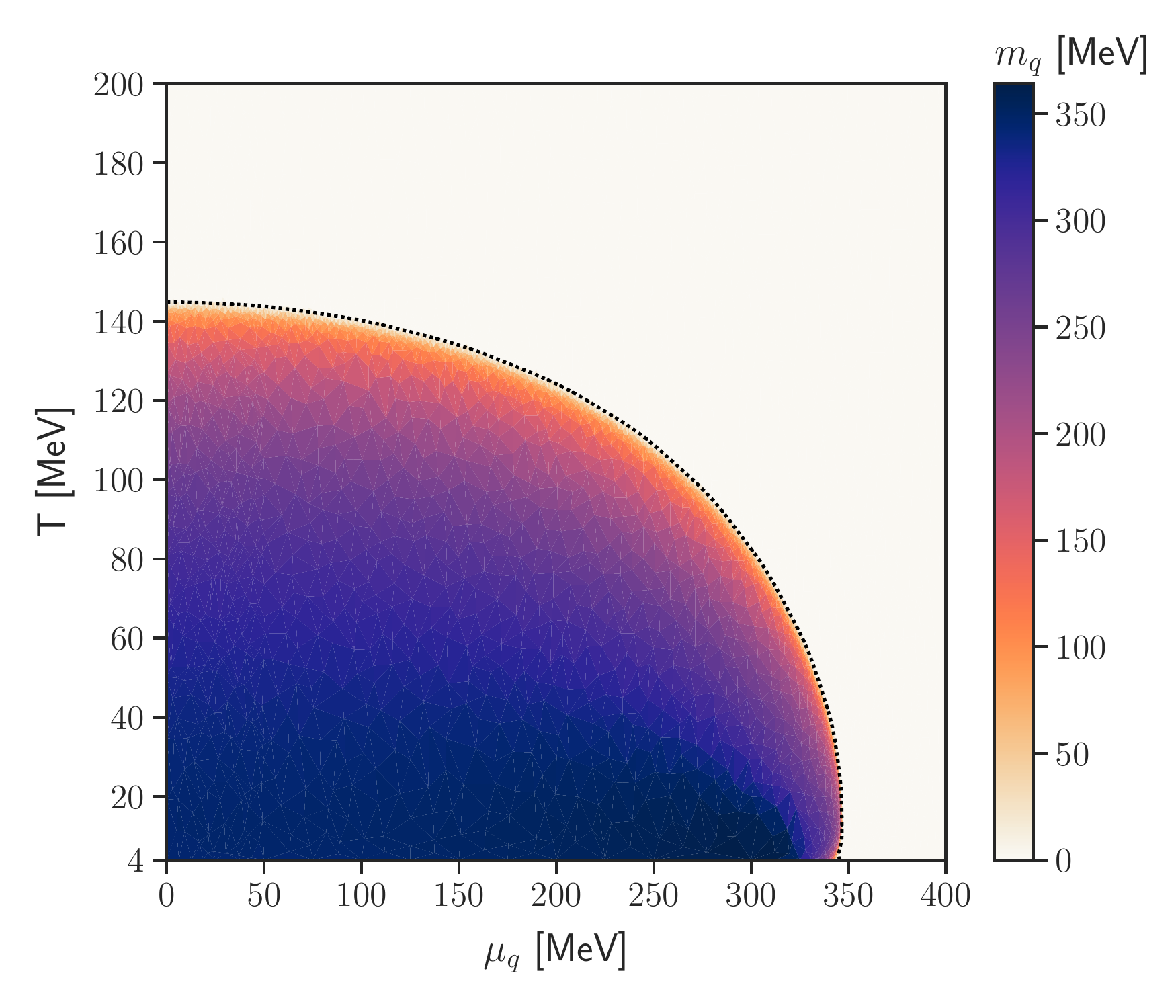}
	\end{subfigure}%
	\begin{subfigure}{0.5\linewidth}
		\centering
		\includegraphics[width=1\linewidth]{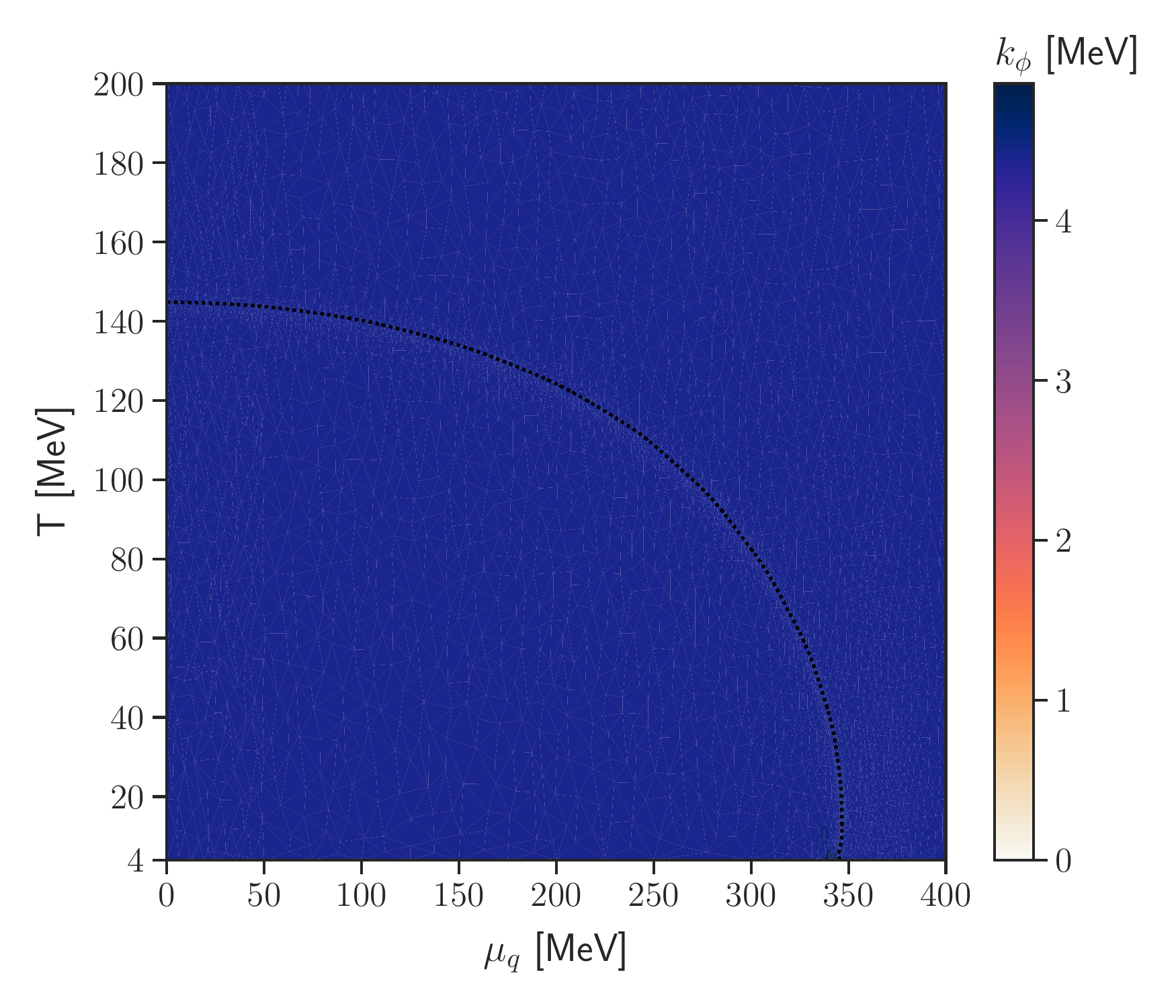}
	\end{subfigure}
	\begin{subfigure}{0.5\linewidth}
		\centering
		\includegraphics[width=1\linewidth]{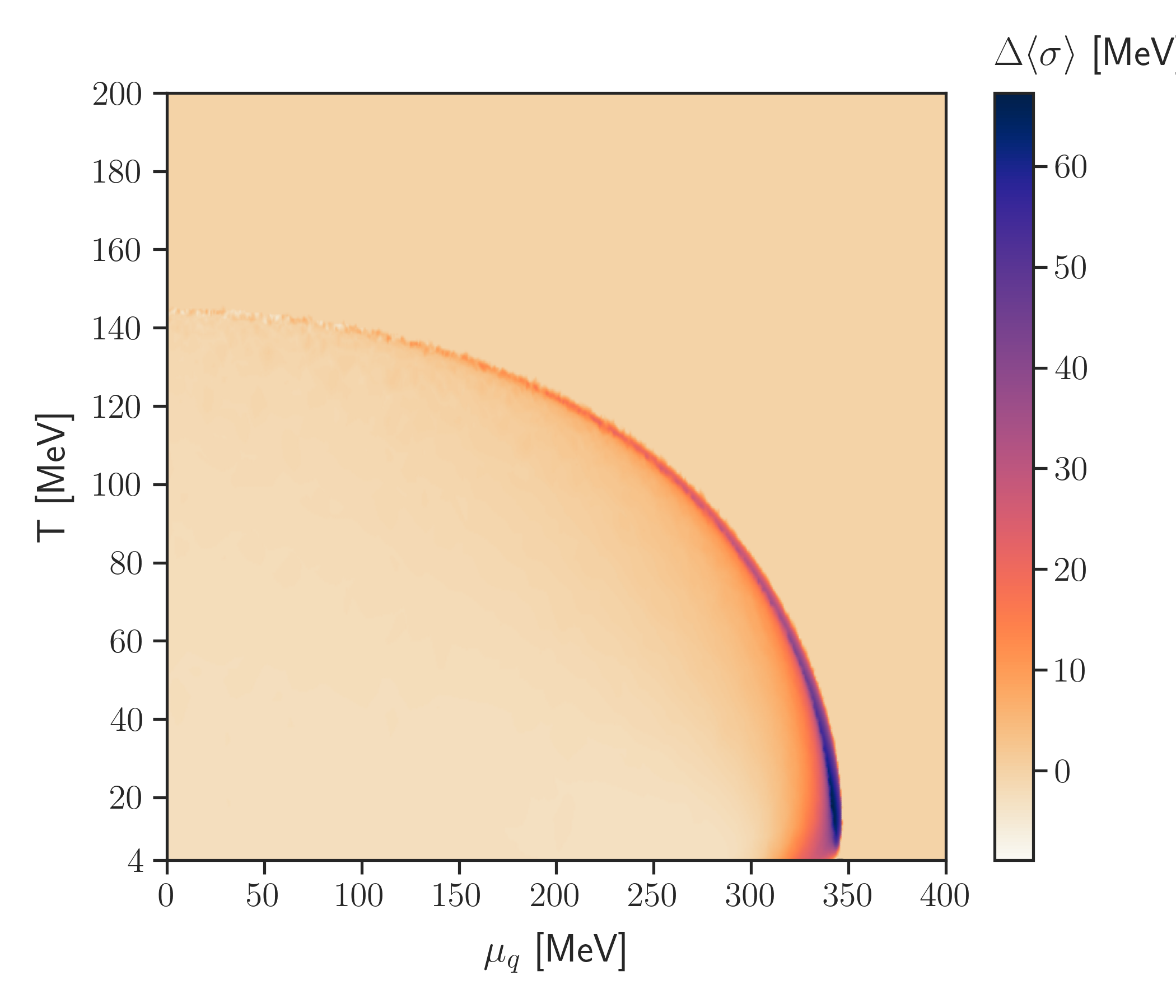}
	\end{subfigure}%
	\begin{subfigure}{0.5\linewidth}
		\centering
		\includegraphics[width=1\linewidth]{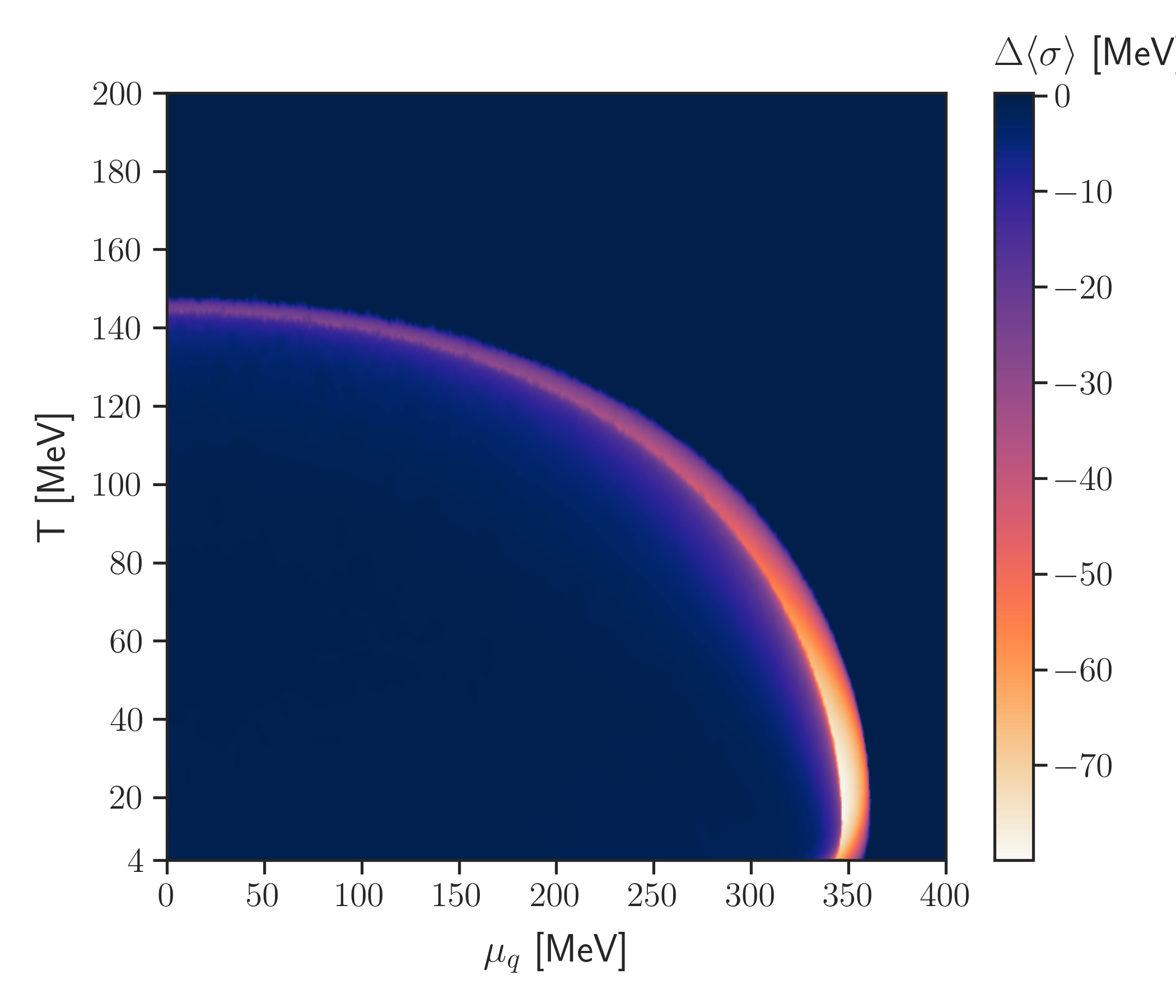}
	\end{subfigure}
	\caption{Order parameter $\langle\sigma\rangle$, sigma mass $m_\sigma$ and quark mass $m_q$ for the quark-meson model at $\gamma = 0.5$ and running $h_k(\rho)$ (scLPA), i.e.~the setup considered in \Cref{sec:results:QMY}. In the middle right plot we show a convergence diagram in $k_\phi$. At the bottom, the two difference plots show $\Delta(\sigma)=\langle\sigma\rangle_{\chi,\textrm{scLPA}} - \langle\sigma\rangle_{\chi,\textrm{LPA}}$ for identical initial conditions (left) and identical vacuum physics (right).}
	\label{fig:QMY_PD_2}
\end{figure}
\begin{figure}[h]
	\centering
	\begin{subfigure}{0.5\linewidth}
		\centering
		\includegraphics[width=1\linewidth]{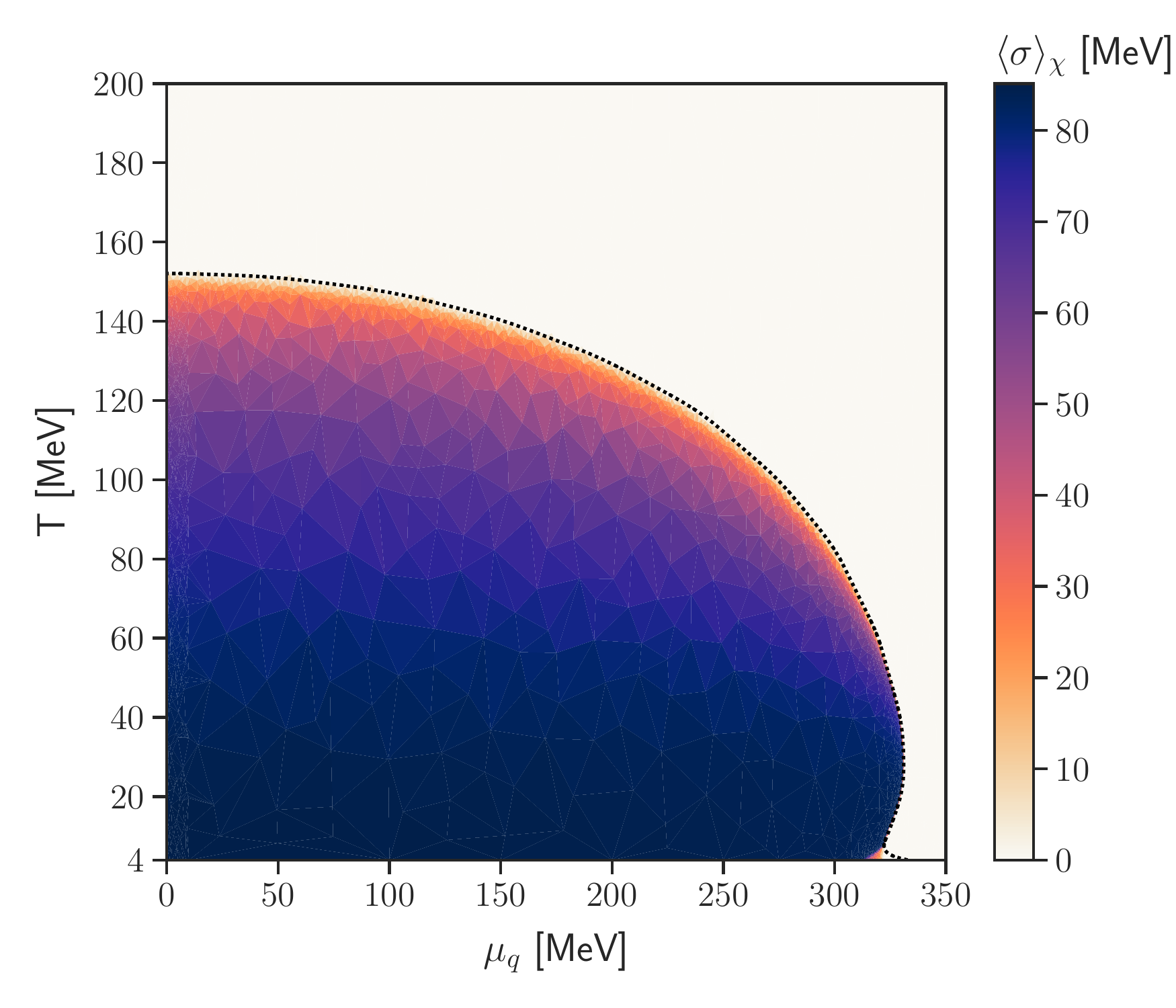}
	\end{subfigure}%
	\begin{subfigure}{0.5\linewidth}
		\centering
		\includegraphics[width=1\linewidth]{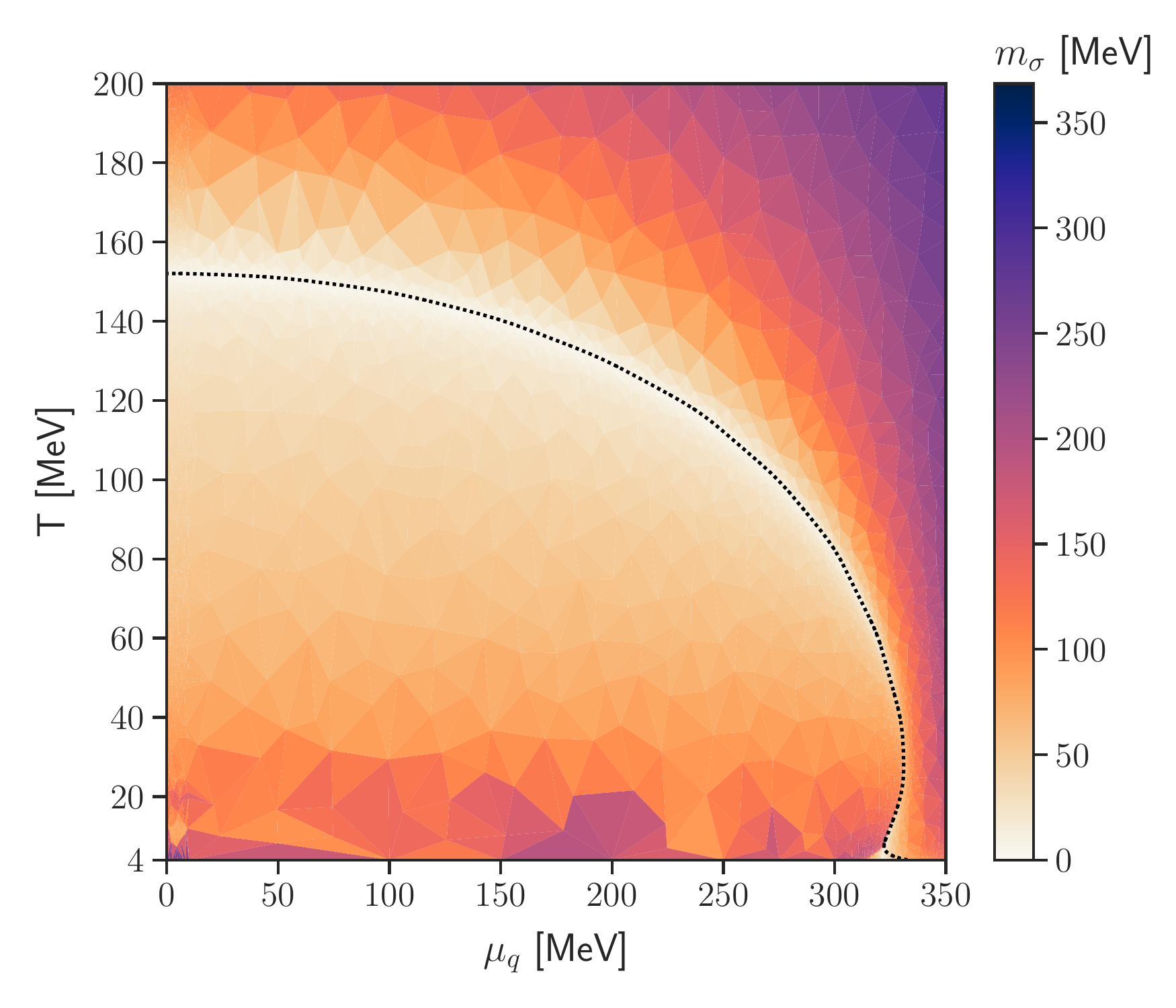}
	\end{subfigure}
	\begin{subfigure}{0.5\linewidth}
		\centering
		\includegraphics[width=1\linewidth]{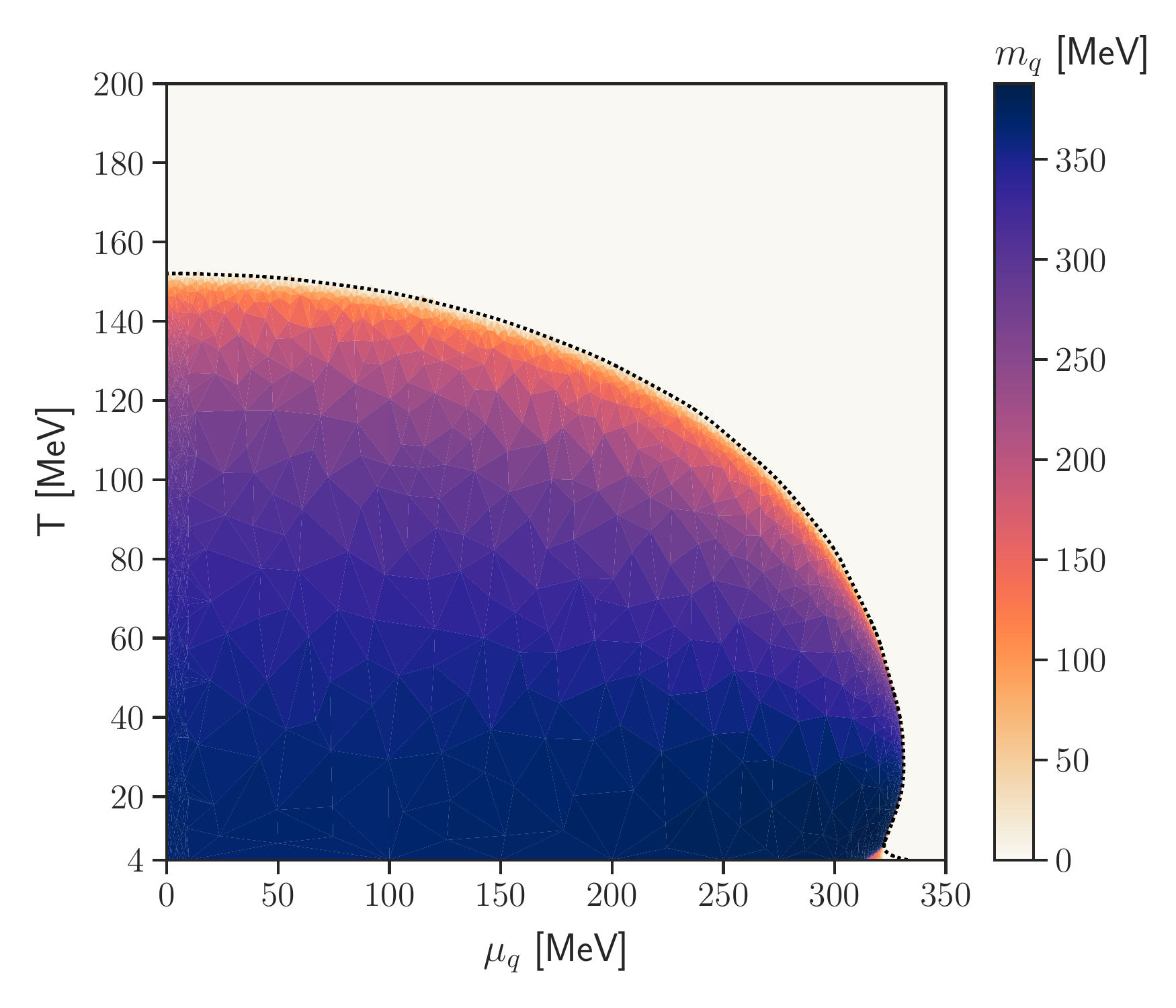}
	\end{subfigure}%
		\begin{subfigure}{0.5\linewidth}
		\centering
		\includegraphics[width=1\linewidth]{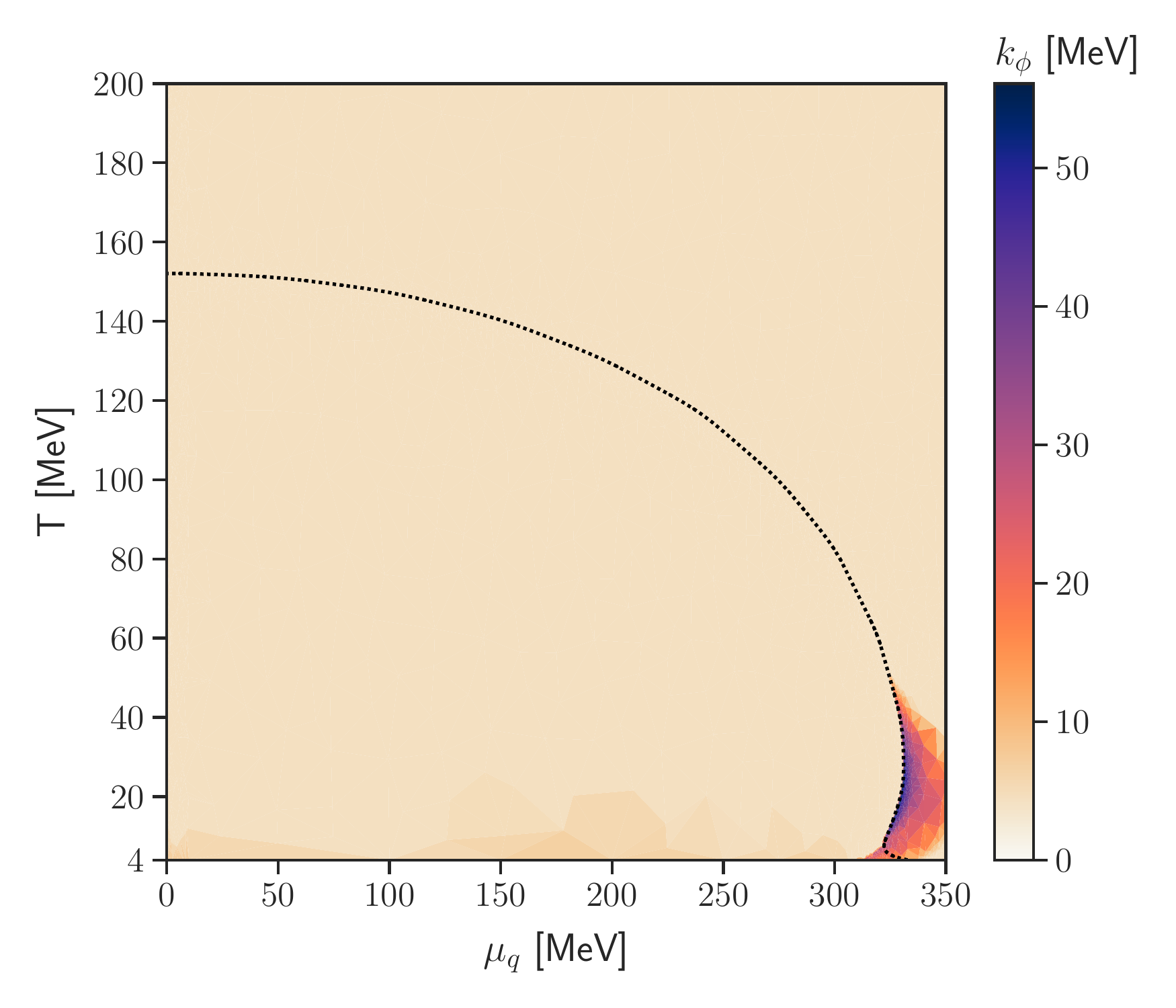}
	\end{subfigure}
	\caption{Order parameter $\langle\sigma\rangle$, sigma mass $m_\sigma$ and quark mass $m_q$ for the quark-meson model at $\gamma = 0.77$ and running $h_k(\rho)$ (scLPA). The bottom right plot shows a convergence diagram in $k_\phi$ (scLPA). n comparison to \Cref{fig:QM_PD_1} the splitting at low temperature and high density is significantly reduced, but it is still present, as is a CEP.}
	\label{fig:QMY_PD_13}
\end{figure}
\begin{figure}[h]
	\begin{subfigure}{0.5\linewidth}
		\centering
		\includegraphics[width=1\linewidth]{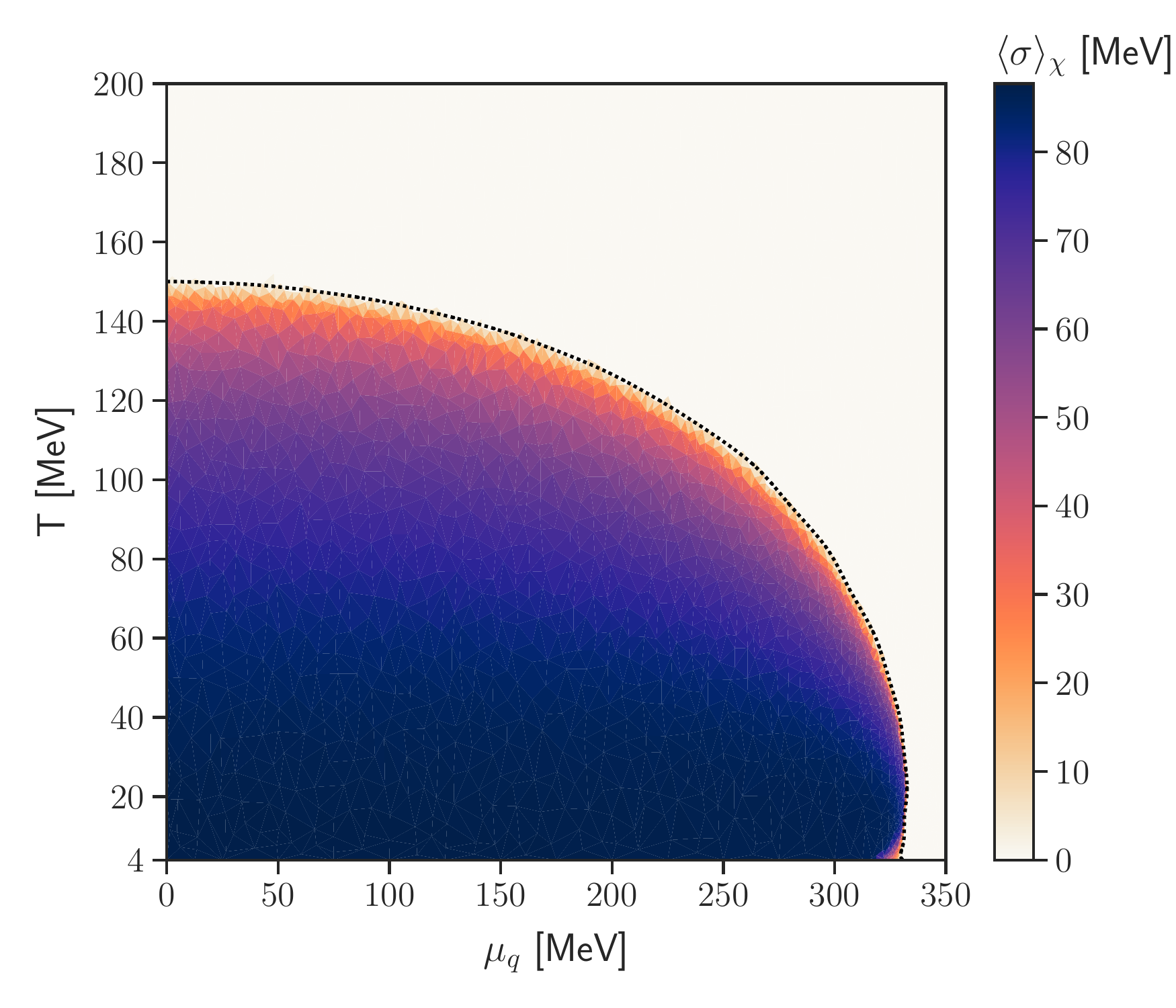}
	\end{subfigure}%
	\begin{subfigure}{0.5\linewidth}
		\centering
		\includegraphics[width=1\linewidth]{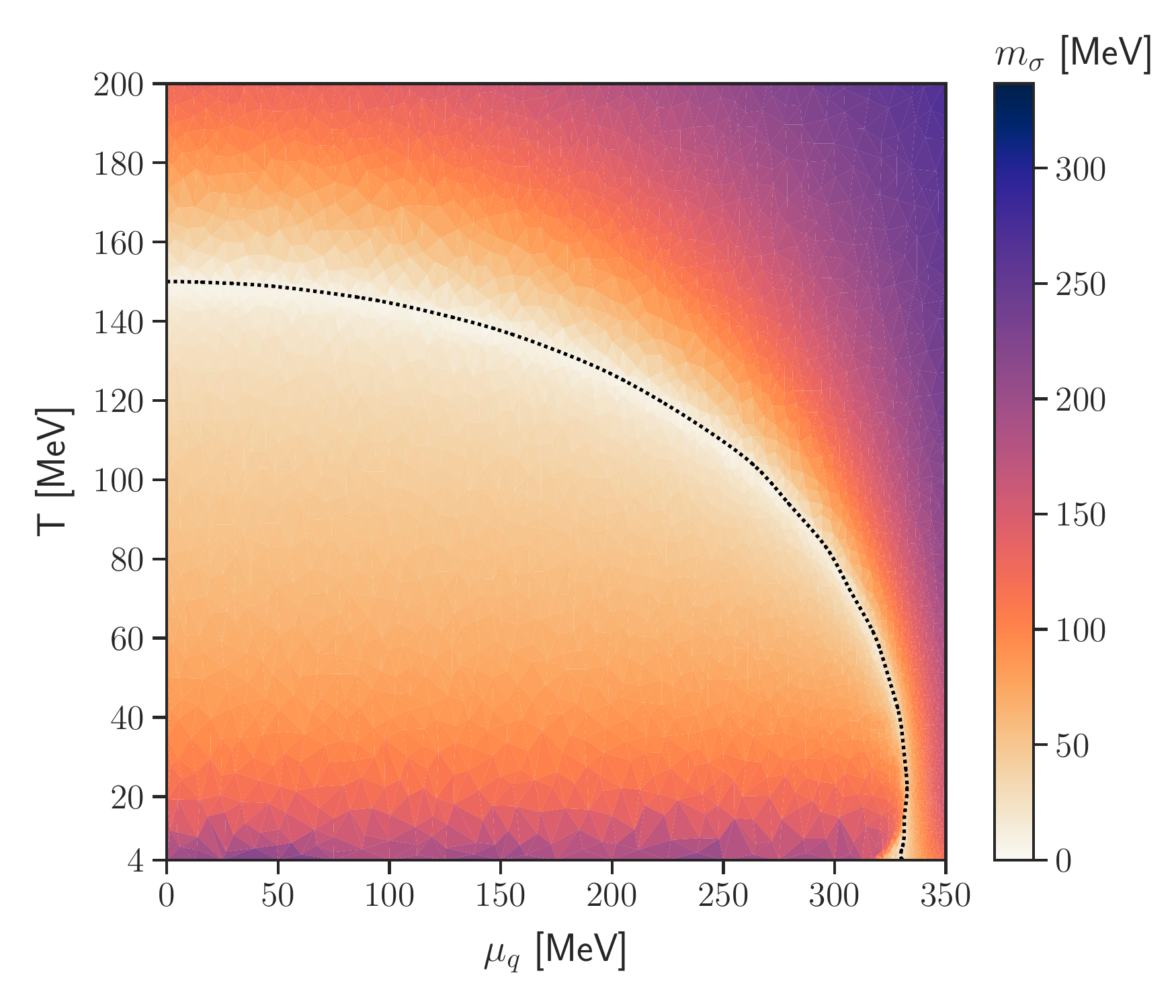}
	\end{subfigure}
	\begin{subfigure}{0.5\linewidth}
		\centering
		\includegraphics[width=1\linewidth]{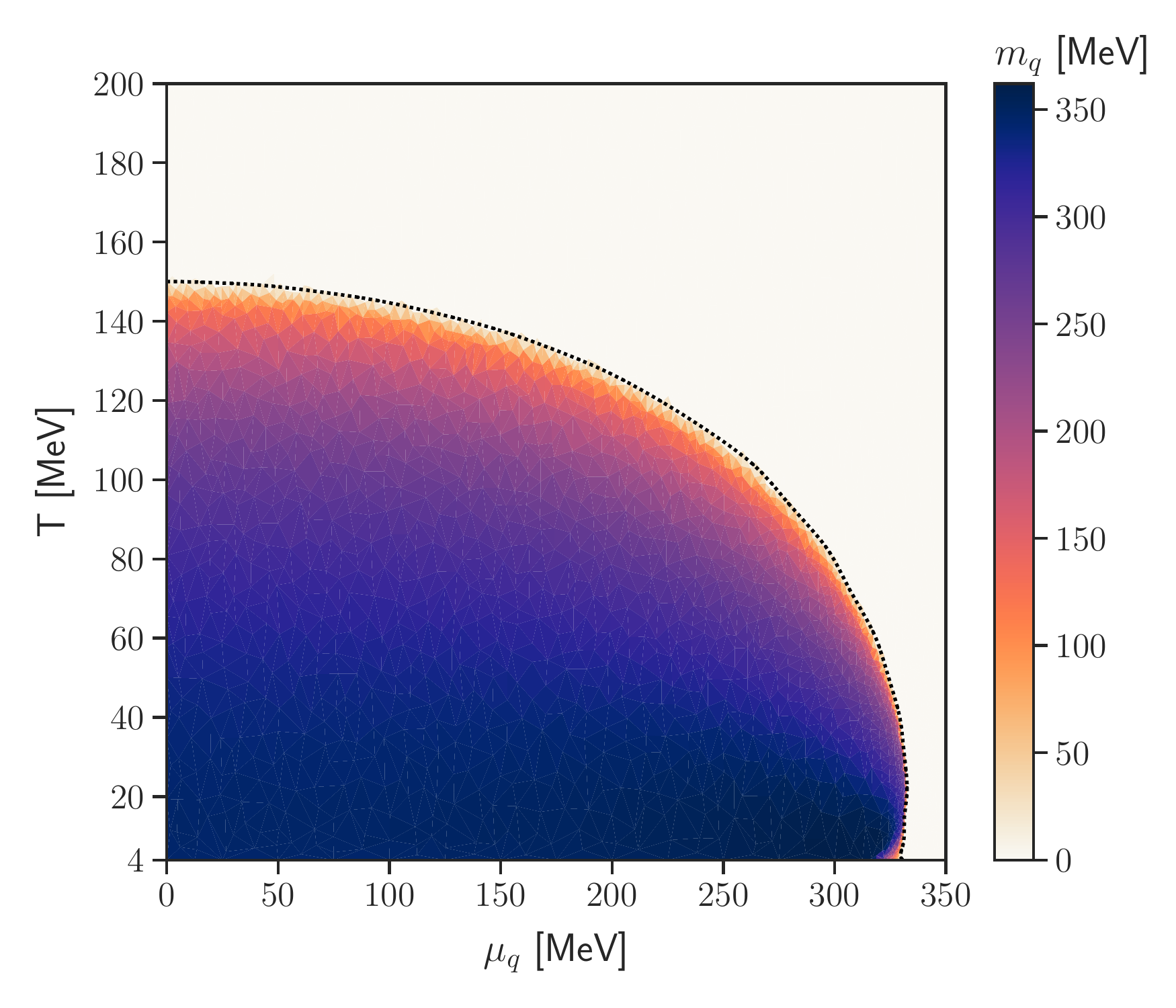}
	\end{subfigure}%
	\begin{subfigure}{0.5\linewidth}
		\centering
		\includegraphics[width=1\linewidth]{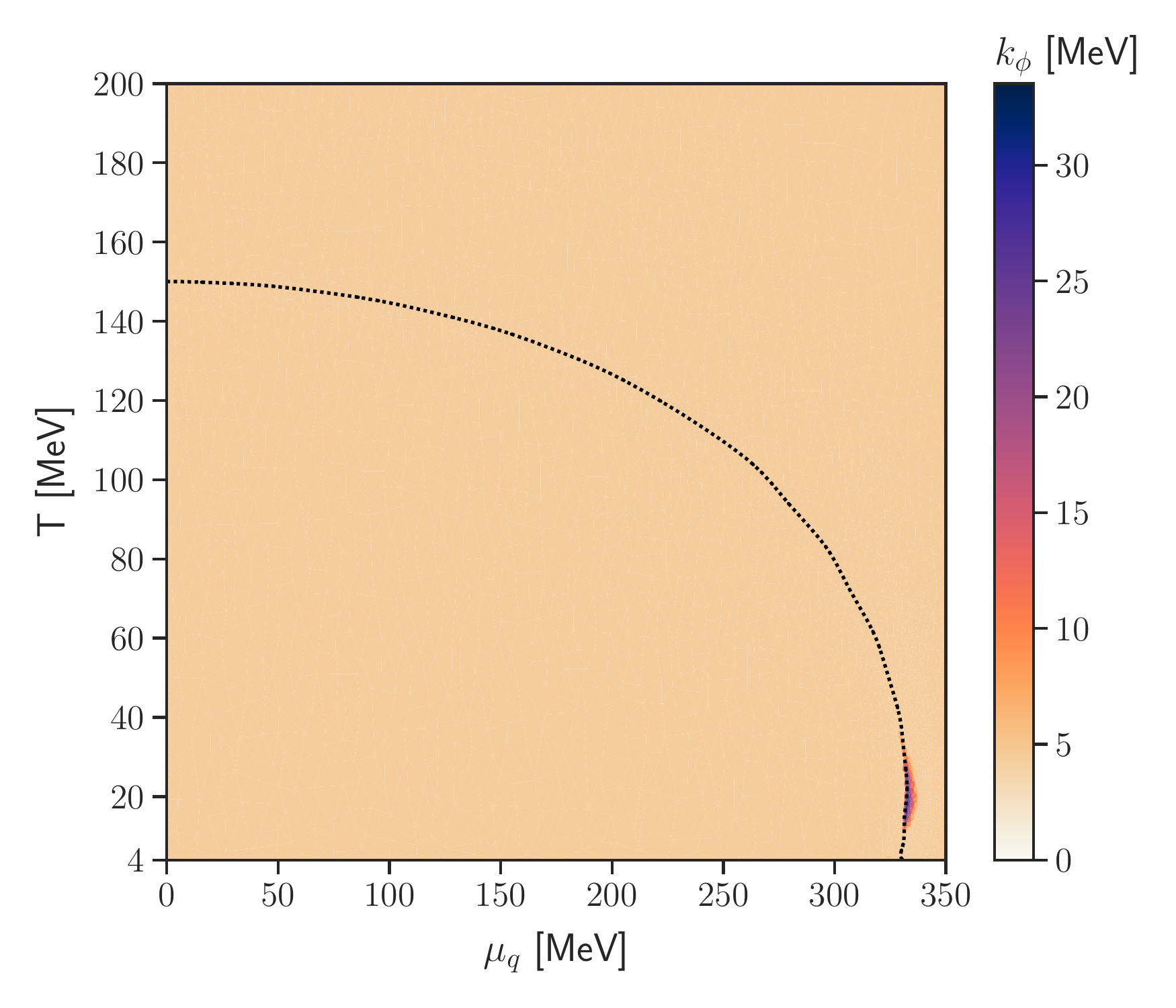}
	\end{subfigure}
	\caption{The order parameter $\langle\sigma\rangle$, sigma mass $m_\sigma$ and quark mass $m_q$ for the quark-meson model at $\gamma = 0.67$ and running $h_k(\rho)$. The bottom right plot shows a convergence diagram in $k_\phi$ (scLPA). The back-bending is reduces in comparison to the higher value of $\gamma=0.77$ in \Cref{fig:QMY_PD_13}. Moreover, the CEP is moving towards $T=0$ for smaller $\gamma$.}
	\label{fig:QMY_PD_15}
\end{figure}

\twocolumngrid

\endgroup
\clearpage
\bibliographystyle{apsrev4-2}
\bibliography{ref-lib} 

\end{document}